\documentstyle[12pt,graphicx]{article}

\def\mathfrak{\bf}

\def\be{\begin{equation}}
\def\ee{\end{equation}}
\def\bea{\begin{eqnarray}}
\def\eea{\end{eqnarray}}

\def\dt#1{\on{\hbox{\bf .}}{#1}}                
\def\Dot#1{\dt{#1}}
\def\IR{\relax{\rm I\kern-.18em R}}
\def\binomial#1#2{\left(\,{\buildrel 
{\raise4pt\hbox{$\displaystyle{#1}$}}\over
{\raise-6pt\hbox{$\displaystyle{#2}$}}}\,\right)}

\def\[{\lfloor{\hskip 0.35pt}\!\!\!\lceil}
\def\]{\rfloor{\hskip 0.35pt}\!\!\!\rceil}

\newcommand{\AmS}{{\protect\the\textfont2
  A\kern-.1667em\lower.5ex\hbox{M}\kern-.125emS}}

\catcode`@=11
\def\un#1{\relax\ifmmode\@@underline#1\else
        $\@@underline{\hbox{#1}}$\relax\fi}
\catcode`@=12

\def\fracm#1#2{\hbox{\large{${\frac{{#1}}{{#2}}}$}}}

\def\ad{{\kern0.5pt
                   \alpha \kern-5.05pt
\raise5.8pt\hbox{$\textstyle.$}\kern
0.5pt}}

\def\Dot#1{{\kern0.5pt
     {#1} \kern-5.05pt \raise5.8pt\hbox{$\textstyle.$}\kern
0.5pt}}

\def\a{\alpha}
\def\b{\beta}

\def\d{\delta}
\def\e{\epsilon}
\def\f{\phi}
\def\g{\gamma}
\def\h{\eta}

\def\j{\psi}

\def\l{\lambda}

\def\p{\pi}
\def\q{\theta}

\def\s{\sigma}
\def\t{\tau}

\def\x{\xi}
\def\z{\zeta}

\def\F{\Phi}
\def\G{\Gamma}
\def\J{\Psi}
\def\L{\Lambda}

\def\cv{{\cal V}}

\def\bo{{\raise.15ex\hbox{\large$\Box$}}}               
\def\pa{\partial}                                       
\def\TH{{\raise.2ex\hbox{$\displaystyle \bigodot$}\mskip-4.7mu \llap H
\;}}
\def\face{{\raise.2ex\hbox{$\displaystyle \bigodot$}\mskip-2.2mu \llap
{$\ddot
        \smile$}}}                                      

\def\bj#1{{}_{\rm #1}}                          
\def\Tilde#1{\widetilde{#1}}                    
\def\Hat#1{\widehat{#1}}                        
\def\Bar#1{\overline{#1}}                       
\def\leftrightarrowfill{$\mathsurround=0pt \mathord\leftarrow \mkern-6mu
        \cleaders\hbox{$\mkern-2mu \mathord- \mkern-2mu$}\hfill
        \mkern-6mu \mathord\rightarrow$}
\def\dvec#1{\vbox{\ialign{##\crcr
        \leftrightarrowfill\crcr\noalign{\kern-1pt\nointerlineskip}
        $\hfil\displaystyle{#1}\hfil$\crcr}}}           
\def\dt#1{{\buildrel {\hbox{\LARGE .}} \over {#1}}}     

\def\fracm#1#2{\hbox{\large{${\frac{{#1}}{{#2}}}$}}}
\def\frac#1#2{{\textstyle{#1\over\vphantom2\smash{\raise.20ex
        \hbox{$\scriptstyle{#2}$}}}}}                   
\def\sfrac#1#2{{\vphantom1\smash{\lower.5ex\hbox{\small$#1$}}\over
        \vphantom1\smash{\raise.4ex\hbox{\small$#2$}}}} 
\def\bfrac#1#2{{\vphantom1\smash{\lower.5ex\hbox{$#1$}}\over
        \vphantom1\smash{\raise.3ex\hbox{$#2$}}}}       
\def\afrac#1#2{{\vphantom1\smash{\lower.5ex\hbox{$#1$}}\over#2}}    
\def\on#1#2{\mathop{\null#2}\limits^{#1}}               

\newskip\humongous \humongous=0pt plus 1000pt minus 1000pt
\def\caja{\mathsurround=0pt}
\def\eqalign#1{\,\vcenter{\openup2\jot \caja
        \ialign{\strut \hfil$\displaystyle{##}$&$
        \displaystyle{{}##}$\hfil\crcr#1\crcr}}\,}
\newif\ifdtup

  \def\pp{{\mathchoice
          {
              \kern 1pt%
              \raise 1pt
              \vbox{\hrule width5pt height0.4pt depth0pt
                    \kern -2pt
                    \hbox{\kern 2.3pt
                          \vrule width0.4pt height6pt depth0pt
                          }
                    \kern -2pt
                    \hrule width5pt height0.4pt depth0pt}%
                    \kern 1pt
           }
            {
              \kern 1pt%
              \raise 1pt
              \vbox{\hrule width4.3pt height0.4pt depth0pt
                    \kern -1.8pt
                    \hbox{\kern 1.95pt
                          \vrule width0.4pt height5.4pt depth0pt
                          }
                    \kern -1.8pt
                    \hrule width4.3pt height0.4pt depth0pt}%
                    \kern 1pt
            }
            {
              \kern 0.5pt%
              \raise 1pt
              \vbox{\hrule width4.0pt height0.3pt depth0pt
                    \kern -1.9pt  
                    \hbox{\kern 1.85pt
                          \vrule width0.3pt height5.7pt depth0pt
                          }
                    \kern -1.9pt
                    \hrule width4.0pt height0.3pt depth0pt}%
                    \kern 0.5pt
            }
            {
              \kern 0.5pt%
              \raise 1pt
              \vbox{\hrule width3.6pt height0.3pt depth0pt
                    \kern -1.5pt
                    \hbox{\kern 1.65pt
                          \vrule width0.3pt height4.5pt depth0pt
                          }
                    \kern -1.5pt
                    \hrule width3.6pt height0.3pt depth0pt}%
                    \kern 0.5pt
            }
        }}

  \def\mm{{\mathchoice
                       {
                             \kern 1pt
               \raise 1pt    \vbox{\hrule width5pt height0.4pt depth0pt
                                  \kern 2pt
                                  \hrule width5pt height0.4pt depth0pt}
                             \kern 1pt}
                       {
                            \kern 1pt
               \raise 1pt \vbox{\hrule width4.3pt height0.4pt depth0pt
                                  \kern 1.8pt
                                  \hrule width4.3pt height0.4pt depth0pt}
                             \kern 1pt}
                       {
                            \kern 0.5pt
               \raise 1pt
                            \vbox{\hrule width4.0pt height0.3pt depth0pt
                                  \kern 1.9pt
                                  \hrule width4.0pt height0.3pt depth0pt}
                            \kern 1pt}
                       {
                           \kern 0.5pt
             \raise 1pt  \vbox{\hrule width3.6pt height0.3pt depth0pt
                                  \kern 1.5pt
                                  \hrule width3.6pt height0.3pt depth0pt}
                           \kern 0.5pt}
                       }}

\def\pd{{\kern0.5pt
                   + \kern-5.05pt \raise5.8pt\hbox{$\textstyle.$}\kern
0.5pt}}

\def\pmd{{\kern0.5pt
                  \pm \kern-5.05pt \raise6.3pt\hbox{$\textstyle.$}\kern1.5pt}}

\def\md{{\mathchoice
   {
      {{\kern 1pt - \kern-6.2pt \raise5pt\hbox{$\textstyle.$}\kern 1pt}}}
    {
      {{\kern 1pt - \kern-6.2pt \raise5pt\hbox{$\textstyle.$}\kern 1pt}}}
    {
      {\kern0.5pt - \kern-5.05pt \raise3.4pt\hbox{$\textstyle.$}\kern0.5pt}}
    {
      {\kern0.5pt - \kern-5.05pt \raise3.4pt\hbox{$\textstyle.$}\kern0.5pt}}}}

\def\ad{{\dot{\alpha}}}

\def\pp{{\mathchoice
          {
              \kern 1pt%
              \raise 1pt
              \vbox{\hrule width5pt height0.4pt depth0pt
                    \kern -2pt
                    \hbox{\kern 2.3pt
                          \vrule width0.4pt height6pt depth0pt
                          }
                    \kern -2pt
                    \hrule width5pt height0.4pt depth0pt}%
                    \kern 1pt
           }
            {
              \kern 1pt%
              \raise 1pt
              \vbox{\hrule width4.3pt height0.4pt depth0pt
                    \kern -1.8pt
                    \hbox{\kern 1.95pt
                          \vrule width0.4pt height5.4pt depth0pt
                          }
                    \kern -1.8pt
                    \hrule width4.3pt height0.4pt depth0pt}%
                    \kern 1pt
            }
            {
              \kern 0.5pt%
              \raise 1pt
              \vbox{\hrule width4.0pt height0.3pt depth0pt
                    \kern -1.9pt  
                    \hbox{\kern 1.85pt
                          \vrule width0.3pt height5.7pt depth0pt
                          }
                    \kern -1.9pt
                    \hrule width4.0pt height0.3pt depth0pt}%
                    \kern 0.5pt
            }
            {
              \kern 0.5pt%
              \raise 1pt
              \vbox{\hrule width3.6pt height0.3pt depth0pt
                    \kern -1.5pt
                    \hbox{\kern 1.65pt
                          \vrule width0.3pt height4.5pt depth0pt
                          }
                    \kern -1.5pt
                    \hrule width3.6pt height0.3pt depth0pt}%
                    \kern 0.5pt
            }
        }}

  \def\mm{{\mathchoice
                       {
                             \kern 1pt
               \raise 1pt    \vbox{\hrule width5pt height0.4pt depth0pt
                                  \kern 2pt
                                  \hrule width5pt height0.4pt depth0pt}
                             \kern 1pt}
                       {
                            \kern 1pt
               \raise 1pt \vbox{\hrule width4.3pt height0.4pt depth0pt
                                  \kern 1.8pt
                                  \hrule width4.3pt height0.4pt depth0pt}
                             \kern 1pt}
                       {
                            \kern 0.5pt
               \raise 1pt
                            \vbox{\hrule width4.0pt height0.3pt depth0pt
                                  \kern 1.9pt
                                  \hrule width4.0pt height0.3pt depth0pt}
                            \kern 1pt}
                       {
                           \kern 0.5pt
             \raise 1pt  \vbox{\hrule width3.6pt height0.3pt depth0pt
                                  \kern 1.5pt
                                  \hrule width3.6pt height0.3pt depth0pt}
                           \kern 0.5pt}
                       }}

\def\pd{{\kern0.5pt
                   + \kern-5.05pt \raise5.8pt\hbox{$\textstyle.$}\kern
0.5pt}}

\def\pmd{{\kern0.5pt
                  \pm \kern-5.05pt \raise6.3pt\hbox{$\textstyle.$}\kern1.5pt}}

\def\md{{\mathchoice
   {
      {{\kern 1pt - \kern-6.2pt \raise5pt\hbox{$\textstyle.$}\kern 1pt}}}
    {
      {{\kern 1pt - \kern-6.2pt \raise5pt\hbox{$\textstyle.$}\kern 1pt}}}
    {
      {\kern0.5pt - \kern-5.05pt \raise3.4pt\hbox{$\textstyle.$}\kern0.5pt}}
    {
      {\kern0.5pt - \kern-5.05pt \raise3.4pt\hbox{$\textstyle.$}\kern0.5pt}}}}

\def\dslash{\not{\hbox{\kern-2pt $\partial$}}}
\def\Dslash{\not{\hbox{\kern-4pt $D$}}}
\def\pslash{\not{\hbox{\kern-2.3pt $p$}}}
 \newtoks\slashfraction
 \slashfraction={.13}
 \def\slash#1{\setbox0\hbox{$ #1 $}
 \setbox0\hbox to \the\slashfraction\wd0{\hss \box0}/\box0 }
 
\font\ro=cmsy10                          
\def\kcr{{\hbox{\ro \char'170}}}                
\def\ktl{{\hbox{\ro \char'170}}}        
\def\ktr{{\hbox{\ro \char'170}}}        
\def\kbl{{\hbox{\ro \char'170}}}        
\def\kbr{{\hbox{\ro \char'170}}}        

\def\plpl{\raise-2pt\hbox{$\raise3pt\hbox{$_+$}\hskip-6.67pt\raise0.0pt
\hbox{$^+$}\hskip 0.01pt$}}
\def\mimi{\raise-2pt\hbox{$\raise3pt\hbox{$_-$}\hskip-6.67pt\raise0.0pt
\hbox{$^-$}\hskip 0.01pt$}} 

\def\bo{{\raise.15ex\hbox{\large$\Box$}}}               
\def\pa{\partial}                                       
\def\TH{{\raise.2ex\hbox{$\displaystyle \bigodot$}\mskip-4.7mu \llap H \;}}
\def\face{{\raise.2ex\hbox{$\displaystyle \bigodot$}\mskip-2.2mu \llap {$\ddot
        \smile$}}}                                      

\def\Tilde#1{\widetilde{#1}}                    
\def\Hat#1{\widehat{#1}}                        
\def\Bar#1{\overline{#1}}                       
\def\leftrightarrowfill{$\mathsurround=0pt \mathord\leftarrow \mkern-6mu
        \cleaders\hbox{$\mkern-2mu \mathord- \mkern-2mu$}\hfill
        \mkern-6mu \mathord\rightarrow$}
\def\dvec#1{\vbox{\ialign{##\crcr
        \leftrightarrowfill\crcr\noalign{\kern-1pt\nointerlineskip}
        $\hfil\displaystyle{#1}\hfil$\crcr}}}           
\def\dt#1{{\buildrel {\hbox{\LARGE .}} \over {#1}}}     

\def\fracm#1#2{\hbox{\large{${\frac{{#1}}{{#2}}}$}}}
\def\frac#1#2{{\textstyle{#1\over\vphantom2\smash{\raise.20ex
        \hbox{$\scriptstyle{#2}$}}}}}                   
\def\sfrac#1#2{{\vphantom1\smash{\lower.5ex\hbox{\small$#1$}}\over
        \vphantom1\smash{\raise.4ex\hbox{\small$#2$}}}} 
\def\bfrac#1#2{{\vphantom1\smash{\lower.5ex\hbox{$#1$}}\over
        \vphantom1\smash{\raise.3ex\hbox{$#2$}}}}       
\def\afrac#1#2{{\vphantom1\smash{\lower.5ex\hbox{$#1$}}\over#2}}    
\def\on#1#2{\mathop{\null#2}\limits^{#1}}               

\parskip=\medskipamount                 
\thicklines                         

\def\oldheadpic{                                
        \setlength{\unitlength}{.4mm}
        \thinlines
        \par
        \begin{picture}(349,16)
        \put(325,16){\line(1,0){4}}
        \put(330,16){\line(1,0){4}}
        \put(340,16){\line(1,0){4}}
        \put(335,0){\line(1,0){4}}
        \put(340,0){\line(1,0){4}}
        \put(345,0){\line(1,0){4}}
        \put(329,0){\line(0,1){16}}
        \put(330,0){\line(0,1){16}}
        \put(339,0){\line(0,1){16}}
        \put(340,0){\line(0,1){16}}
        \put(344,0){\line(0,1){16}}
        \put(345,0){\line(0,1){16}}
        \put(329,16){\oval(8,32)[bl]}
        \put(330,16){\oval(8,32)[br]}
        \put(339,0){\oval(8,32)[tl]}
        \put(345,0){\oval(8,32)[tr]}
        \end{picture}
        \par
        \thicklines
        \vskip.2in}
\def\oldtitle#1#2#3#4{\oldheadpic\begin{center}\vglue.5in{\large\bf #1}\\[.6in]
        {#2}\\[.1in] {\it Department of Physics and Astronomy}\\
        {\it University of Maryland, College Park, MD 20742}\\[.6in]
        Physics Publication \#{#3}\\ {#4}\\[1.5in] {\bf ABSTRACT}\\[.1in]
        \end{center} \begin{quotation}}                 
\def\oldTitle#1#2#3#4#5#6#7{\oldheadpic\begin{center} \vglue .4in
        {\large\bf #1}\\[.4in]
        {#2}\\[.1in] {\it Department of Physics and Astronomy}\\
        {\it University of Maryland, College Park, MD 20742}\\[.1in]
        {#3}\\[.1in] {\it {#4}}\\ {\it {#5}}\\[.4in]
        Physics Publication \#{#6}\\ {#7}\\[.5in] {\bf ABSTRACT}\\[.1in]
        \end{center} \begin{quotation}}                 
\def\border{                                            
        \setlength{\unitlength}{1mm}
        \newcount\xco
        \newcount\yco
        \xco=-21
        \yco=12
        \begin{picture}(140,0)
        \put(\xco,\yco){$\ktl$}
        \advance\yco by-1
        {\loop
        \put(\xco,\yco){$\kcr$}
        \advance\yco by-2
        \ifnum\yco>-240
        \repeat
        \put(\xco,\yco){$\kbl$}}
        \xco=158
        \yco=12
        \put(\xco,\yco){$\ktr$}
        \advance\yco by-1
        {\loop
        \put(\xco,\yco){$\kcr$}
        \advance\yco by-2
        \ifnum\yco>-240
        \repeat
        \put(\xco,\yco){$\kbr$}}
        \put(-20,13){\tiny **University of Maryland * Center for String and 
         Particle  Theory* Physics Department***University of Maryland *Center  
        for String and Particle  Theory** }
        \put(-20,-241.5){\tiny **University of Maryland * Center for String and 
         Particle  Theory* Physics Department***University of Maryland *Center  
        for String and Particle  Theory** }
        \end{picture}
        \par\vskip-8mm}
\def\bordero{                                           
        \setlength{\unitlength}{1mm}
        \newcount\xco
        \newcount\yco
        \xco=-31
        \yco=12
        \begin{picture}(140,0)
        \put(\xco,\yco){$\ktl$}
        \advance\yco by-1
        {\loop
        \put(\xco,\yco){$\kclr}
        \advance\yco by-2
        \ifnum\yco>-240
        \repeat
        \put(\xco,\yco){$\kbl$}}
        \xco=151
        \yco=12
        \put(\xco,\yco){$\ktr$}
        \advance\yco by-1
        {\loop
        \put(\xco,\yco){$\kcr$}
        \advance\yco by-2
        \ifnum\yco>-240
        \repeat
        \put(\xco,\yco){$\kbr$}}
        \put(-20,12){\ooo bacdefghidfghghdhededbihdgdfdfhhdheidhdhebaaahjhhdahba

hgdedge
   hgfdiehhgdigicba}
        \put(-20,-241.5){\ooo ababaighefdbfghgeahgdfgafagihdidihiidhiagfedhadbfd

ecdcdfa
   gdcbhaddhbgfchbgfdacfediacbabab}
        \end{picture}
        \par\vskip-8mm}
\def\headpic{                                           
        \indent
        \setlength{\unitlength}{.4mm}
        \thinlines
        \par
        \begin{picture}(29,16)
        \put(165,16){\line(1,0){4}}
        \put(170,16){\line(1,0){4}}
        \put(180,16){\line(1,0){4}}
        \put(175,0){\line(1,0){4}}
        \put(180,0){\line(1,0){4}}
        \put(185,0){\line(1,0){4}}
        \put(169,0){\line(0,1){16}}
        \put(170,0){\line(0,1){16}}
        \put(179,0){\line(0,1){16}}
        \put(180,0){\line(0,1){16}}
        \put(184,0){\line(0,1){16}}
        \put(185,0){\line(0,1){16}}
        \put(169,16){\oval(8,32)[bl]}
        \put(170,16){\oval(8,32)[br]}
        \put(179,0){\oval(8,32)[tl]}
        \put(185,0){\oval(8,32)[tr]}
        \end{picture}
        \par\vskip-6.5mm
        \thicklines}
\def\title#1#2#3#4{\border\headpic {\hbox to\hsize{#4 \hfill UMDEPP #3}}\par
        \begin{center} \vglue .5in {\large\bf #1}\\[.6in]
        {#2}\\[.1in] {\it Department of Physics and Astronomy}\\
        {\it University of Maryland, College Park, MD 20742}\\[1.5in]
        {\bf ABSTRACT}\\[.1in] \end{center} \begin{quotation}}  
\def\Title#1#2#3#4#5#6#7{\border\headpic
        {\hbox to\hsize{#7 \hfill UMDEPP #6}}\par
        \begin{center} \vglue .4in {\large\bf #1}\\[.4in]
        {#2}\\[.1in] {\it Department of Physics and Astronomy}\\
        {\it University of Maryland, College Park, MD 20742}\\[.1in]
        {#3}\\[.1in] {\it {#4}}\\ {\it {#5}}\\[.5in] {\bf ABSTRACT}\\[.1in]
        \end{center} \begin{quotation}}                 
\def\endtitle{\end{quotation}\newpage}                  

\def\qd{{\kern0.5pt
                   q \kern-5.05pt \raise5.8pt\hbox{$\textstyle.$}\kern
0.5pt}}

\begin{document}

\def\dt#1{\on{\hbox{\bf .}}{#1}}                
\def\Dot#1{\dt{#1}}

\def\gfrac#1#2{\frac {\scriptstyle{#1}}
        {\mbox{\raisebox{-.6ex}{$\scriptstyle{#2}$}}}}
\def\gg{{\hbox{\sc g}}}
\border\headpic {\hbox to\hsize{November 2002 \hfill
{UMDEPP 02-054}}}
\par
{\hbox to\hsize{$~$ \hfill
{CALT-68-2387}}}
\par
\setlength{\oddsidemargin}{0.3in}
\setlength{\evensidemargin}{-0.3in}
\begin{center}
\vglue .10in
{\large\bf When Superspace Is Not Enough\footnote
{Supported in part  by National Science Foundation Grant
PHY-0099544.}\  }
\\[.5in]
S. James Gates, Jr.\footnote{gatess@wam.umd.edu}, W. D. Linch,
III\footnote{ldw@physics.umd.edu} and J.
Phillips\footnote{ferrigno@physics.umd.edu}
\\[0.06in]
{\it Center for String and Particle Theory\\ 
Department of Physics, University of Maryland\\ 
College Park, MD 20742-4111 USA}\\[3in]

{\bf ABSTRACT}\\[.01in]
\end{center}
\begin{quotation}
{We give an expanded discussion of the proposal that spacetime
supersymmetry representations may be viewed as having their
origins in 1D theories that involve a special class of real
Clifford algebras.  These 1D  theories reproduce the supersymmetric 
structures of spacetime supersymmetric theories after the latter
are reduced on a 0-brane.  This leads us to propose that spacetime
appears as a bundle  in KO-theory. }

${~~~}$ \newline
PACS: 04.65.+e, 11.15.-q, 11.25.-w, 12.60.J

\endtitle

\noindent
\section{Introduction}

~~~~In previous work on the subject of $1D$, arbitrarily large
$N$-extended supersymmetry \cite{jim1,jim2}, it was shown how
to reformulate the usual notion of supersymmetry transformations 
on an NSR $D0$-brane, defined through the action of a set of derivations, 
in terms of an algebraic structure dubbed the ``general real algebra 
of extension $N$ and dimension $d$," or ${\cal GR}$(d, $N)$ for short. 
In many ways the matrices that are used to realize these constructs
behave as real-valued Euclidean space  analogues of the usual Pauli
matrices and hence our assigned moniker.   The  spinning particle is a
unique supersymmetrical system at this time, in that  we have complete
knowledge of its {\it {off-shell}} representation theory,  i.e., the
auxiliary  fields are known for all $N$.  The problem of finding the
off-shell linear representations of all supersymmetrical systems has {\it
{remained}} {\it {unsolved}} {\it {for}} {\it {almost}} {\it a} {\it
{quarter}}  {\it {of}} {\it a} {\it {century}} \cite{Fun}.

Insight into this general problem may arise from a more careful 
investigation of the spinning particle system.  In previous work, 
this same algebraic structure was found to occur in all on-shell
$N$-extended  scalar (and their dual vector) multiplets in three
dimensional superspaces.   Since this is the case, the way is open to
study the problem of the arbitrarily $N$-extended supersymmetric BF
theories.  This class of theories naturally leads to ``zero-curvature''
conditions on field strength and this means there  is the possibility of
a link to integrable systems.

Our main motivation is to understand these two problems.  In the following 
we will present the ${\cal GR}({\rm d}, N)$ algebras more explicitly than in
our previous work.  Most importantly, we will investigate how these algebras
organize the problem of component fields in conventional superspace and in
superfields.  Further, using 1D models constructed from these algebras we
will show that these algebras can encode the structure of supermultiplets
from higher dimensions.  This letter begins with an abstract definition of
${\cal GR}({\rm d},N)$, followed by some useful derivations using an
explicit realization of these algebras.  We then focus our attention on
writing field theories using ${\cal GR}({\rm d},N)$.

The form of this paper is as follows.  Formal arguments are
presented in Section 2.  These include geometrical
motivations for the ${\cal GR}$ algebra, and conventions for the
representation of these algebras.  In section 3, we show how
these algebras encode the supersymmetric structure of higher
dimensional supersymmetric theories.  Finally, section 4 shows how
we apply the ${\cal GR}$ algebras to write off-shell $N$-extended
spinning particle actions.

\section{Formal Arguments}
 
In this section we present a geometrical description for the
definition of the ${\cal GR}$(d, $N)$ algebra.  We also show the
explicit connection between these algebras and real Clifford
algebras.  The connection to real Clifford algebras is
then exploited in order to understand the structure of the
enveloping algebras of ${\cal GR}$(d, $N)$.
\subsection{Geometric Preliminaries}

~~~~Let d $ \geq 1$ be some fixed natural number and consider a collection of 
d + d diffeomorphisms $\f_i,\j_{\hat l}:{\mathfrak R}\longrightarrow
{\mathfrak R}{}^d$, with $i = 1, \, \, \cdots,d$ and ${\hat l}=1,\, \cdots,d$.
Define 
${\cal V}_L$ and ${\cal V}_R$ be the free vector spaces generated by
$\{\f_i\}_{i=1}^d$ and
$\{\j_{\hat l}\}_{{\hat l}=1}^d$ respectively. Note that ${\cal V}_L \cong
{\mathfrak R}^d\cong {\cal V}_R$, however, we do {\it not} wish to
identify the two. 

Next we consider the set of linear transformations acting on these spaces
in the following manner.  Let $\{ {\cal M}_L \}$ denote all linear maps
that send elements of ${\cal V}_L$ into elements of ${\cal V}_R$, $\{
{\cal M}_R \}$ denote  all linear maps that send elements of ${\cal V}_R$
into elements of 
${\cal V}_L$, $\{ {\cal U}_L \}$ denote all linear maps that send
elements of ${\cal V}_L$ into elements of ${\cal V}_L$ and $\{ {\cal U}_L 
\}$ denote  all linear maps that send elements of ${\cal V}_L$ into elements 
of ${\cal V}_{L}$.   Thus we have\footnote{Here it understood that ${\cal 
M}_L$ is an element of $\{ {\cal M}_L \}$.},
\be \eqalign{
{\cal M}_{L} &: {\cal V}_L ~\to~ {\cal V}_R  ~~~, \cr
{\cal M}_{R} &: {\cal V}_R ~\to~ {\cal V}_L  ~~~,  \cr
{\cal U}_{L} &: {\cal V}_L ~\to~ {\cal V}_L  ~~~, \cr
{\cal U}_{R} &: {\cal V}_R ~\to~ {\cal V}_R  ~~~.  
} \label{eq:2Chp1} \ee
Since the dimension of the vector spaces is $\rm d$, it follows that 
$\dim\{ {\cal M}_L\}$ = $\dim\{ {\cal M}_R\}$ = $\dim\{ {\cal U}_L\}$ = 
$\dim\{ {\cal U}_R\}$ = d${}^2$.

The definition of these maps implies that the compositions ${\cal M}_R
\circ {\cal M}_L$ and ${\cal M}_L\circ {\cal M}_R$ have the properties
\be \eqalign{
{\cal M}_R\circ {\cal M}_L &: {\cal V}_L ~\to~ {\cal V}_L  ~~~, \cr
{\cal M}_L\circ {\cal M}_R &: {\cal V}_R ~\to~  {\cal V}_R  ~~~,  \cr
} \label{eq:2Chp2} \ee
and are thus elements of $\{{\cal U}_L\}$ and $\{{\cal U}_R\}$ respectively.

All of these structures may be illustrated by means of a
``Placement-putting Graph'' that is presented on the following page.  
The d-dimensional  vector spaces are represented in a Venn diagram
containing two disjoint sets and the linear transformations are
represented by directed line segments that act between the sets
appropriately.  The line segments may be regarded as being composed of 
d${}^2$ ``fibers'' representing the d${}^2$-linearly independent linear 
maps acting between the vector spaces.

\begin{figure}[htbp]
\begin{center}
\centerline{{\includegraphics[width=6in]{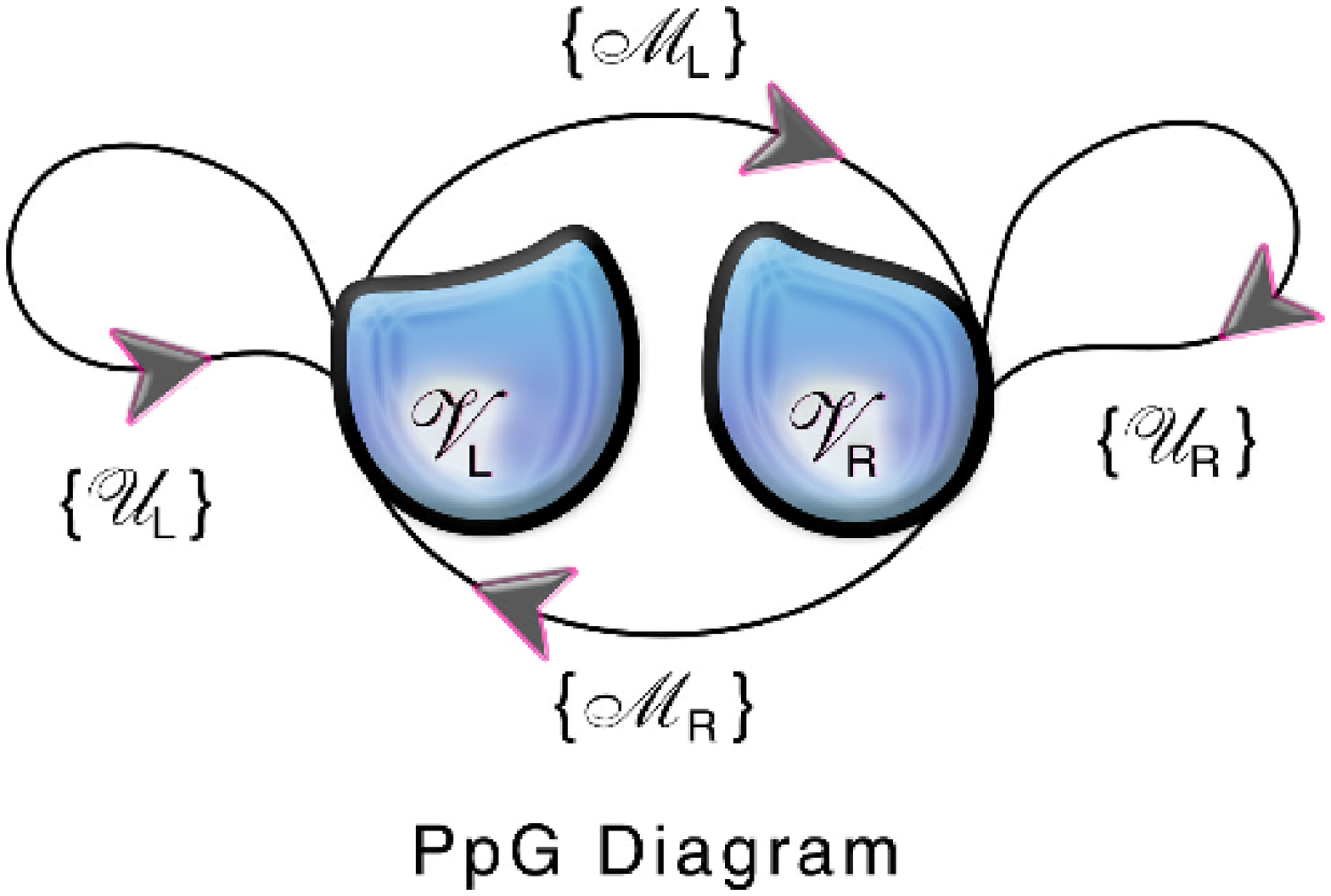}
}}
${~}$
\end{center}
\end{figure}

Two conditions are necessary to define the ${\cal GR}({\rm d},N)$ algebra.  
We will show how these conditions yield a representation of supersymmetry 
in 1D.  The first condition is a definition of the fundamental generators 
of ${\cal GR}({\rm d},~ N)$.  The second is a statement of how the generators
behave when an inner product is defined on the vector spaces ${\cal V}_L$ 
and ${\cal V}_R$.   Consider the family of $N+N$ (for any natural number 
$N$) linear maps $L_{\bj I}\in\,\{ {\cal M}_L \}$ and $R_{\bj I}\in\,\{ 
{\cal M}_R\} , {\rm I} =1,\cdots,N$ such that
\bea
L_{\bj I}\circ R_{\bj K} 
~+~ L_{\bj K}\circ R_{\bj I}
&=&-2\d_{{\bj I} \, {\bj K}}{\bf I}_{{\cal V}_L} ~~~, \cr 
R_{\bj I}\circ L_{\bj K} 
~+~ R_{\bj K} \circ L_{\bj I}
&=&-2\d_{{\bj I} \, {\bj K}}{\bf I}_{{\cal V}_R} 
~~~, ~~~\forall ~{\rm I},{\rm K}=1,\cdots, N ~~~.
\label{eq:GR1}
\eea
Here ${\bf I}_{{\cal V}_L}$ and ${\bf I}_{{\cal V}_R}$ denote the identity 
maps on ${\cal V}_L$ and ${\cal V}_R$ respectively.  Next we equip ${\cal 
V}_L$ and ${\cal V}_R$ with Euclidean inner products $\langle \cdot,\cdot 
\rangle_{{\cal V}_L}$ and $\langle \cdot,\cdot \rangle_{ {\cal V}_R}$ 
respectively. The second defining relation of the ${\cal {GR}}($d, $N)$ 
algebra is the condition that for any ${\rm I} =1,\cdots,N$
\be
\label{eq:GR2}
\langle \f, L_{\bj I} (\j) \rangle_{{\cal V}_L}~=~-\langle 
R_{\bj I} (\f) , \j \rangle_{{\cal V}_R} ~~~~~
\forall (\f,\j)\in {\cal V}_L \oplus {\cal V}_R ~~~.
\ee
This equation implies that the objects $R_{\bj I}$ equal to minus
the adjoints of $L_{\bj I}$ and the definitions in (\ref{eq:GR1}) can be solely
expressed in term of the sets of $L_{\bj I}$ and  their adjoints $L_{\bj I}^*$.
In all, then, we define the general real algebra of level $N$ and dimension 
d to be the sub-algebra of $\{ {\cal M}_R \} \oplus \{ {\cal M}_L \}$ generated
by the relations (\ref{eq:GR1}) and (\ref{eq:GR2}). We will occasionally resort 
to the notation ${\cal GR}(N)$ when referring to the abstract algebra or when 
the dimension d is otherwise unimportant.

We will now motivate the definition of the ${\cal GR}(N)$ algebra by using it 
to write a supersymmetric theory in 1D as follows.  Let $\{\f_k\}\subset
{\cal V}_L$ and $\{\j_{\hat l}\}\subset {\cal V}_R$ denote a set of d + d
real-valued fields, henceforth to be known as bosonic and fermionic
respectively.  We define an algebraic derivation $\d:{\cal
V}_L \oplus {\cal V}_R\rightarrow{\cal
V}_L \oplus {\cal V}_R$ by the relation
\be
\label{eq:algSUSY}
\d_\a:(\f,\j)\mapsto \lgroup i\a \cdot L(\j), \, - i \a \cdot R(D\f)
\rgroup,
\ee 
where $D:=i\partial$ is the 1D translation generator and $\a\cdot L=\a^{
\bj  I} L_{\bj I}$ and $\a\cdot R=\a^{\bj I} R_{\bj I}$ are elements of
the ${\cal GR}(N)$ algebra described above.  Owing to this we may easily
derive (using (\ref{eq:GR1}) and the facts that the parameter and fermion
are classical anti-commuting parameters)
\be
\label{eq:SUSY}
[ ~\d_\a~,~\d_\b~ ]~=~-2 \, \a\cdot \b \, D,
\ee
on $ {\cal V}_L \oplus {\cal V}_R$, which is nothing but the statement
that these algebraic derivations are, in fact, generators of
supersymmetry.  Next, consider the standard free lagrangian
\be
\label{eq:SUSYact}
{\cal L}~=~-{\fracm 12}(D\f)\cdot(D\f)~+~\j\cdot D\j ~=~{\fracm 12} \,
\langle \f,D^2 \f \rangle_{{\cal V}_L}~+~ {\fracm 12} \, \langle \j,D\j
\rangle_{{\cal V}_R},
\ee
up to total derivatives, which we neglect throughout this work. It is easy
to check that owing to the relation (\ref{eq:GR2}), the action constructed
from this lagrangian commutes with the derivation defined in
(\ref{eq:algSUSY}):
\be
\d_\a{\cal L}~=~-i\a\cdot \Big\{\langle D\f,L(D\j)\rangle_{{\cal
V}_L}~+~ \langle R(D\f),D\j\rangle_{{\cal V}_R} \Big\}~=~0.
\ee

Note that the relations (\ref{eq:GR1}) above are not consistent for all
values of $N$ and d. In particular, for any $N\in {\mathfrak N}$, there 
is a minimum value of d, which we will denote by d${}_N$, such that $N$
linearly independent quantities $L_{\bj I}$ and $N$ linearly independent 
quantities $R_{\bj I}$ can be constructed to satisfy (\ref{eq:GR1}) and
(\ref{eq:GR2}).  If we write $N$ as related to $n$ and $m$ through the
equations $N = n + 8 m$ where $1 \le n \le 8$ and further, if we use the
rule that  if $N = 8 k \, \Rightarrow \,  n = 8, \& \, m = k  - 1$ 
for $ k = 1, 2, ..., \infty$.  then the minimal value of d$_N$ can be 
expressed in the form of a functional relationship 
\be \eqalign{
{\rm d}_N ~=~ 2^{4m}\,F_{\cal {RH}}(N) ~=~ {16}^{m}\,F_{\cal {RH}}(n)  
~~~~ .
\label{eq:EX00}}
\ee  
where $F_{\cal {RH}}(N)$ is the Radon-Hurwitz function as was noted in 
\cite{PASH}.  In writing the second form of this relation, we have used 
the period eight property of this function.  In our previous work
\cite{jim1,jim2}, the Radon-Hurwitz function  appeared in tabular form. 
Our previous work  also presented an algorithm for the construction of an
explicit matrix representation for $L_{\bj I}$ and $R_{\bj I}$ for all
values of $N$ and  d$_N$. 

In what follows we will be interested not only in the ${\cal GR}({\rm 
d},N)$ algebra but two other related algebras.  The first is the
enveloping algebra  of ${\cal GR}({\rm d},~N)$ denoted ${\cal EGR}({\rm
d},~N)$.  This (${\cal EGR}({\rm d},~N)$) denotes the set of all linear 
maps on and between ${\cal V}_L$, and ${\cal V}_R$.  More simply stated:
\bea
{\cal EGR}:=\{ {\cal M}_L\}\oplus\{ {\cal M}_R\}
\oplus\{ {\cal U}_L\}\oplus\{ {\cal U}_R\}
\eea 
This new algebraic structure contains the ${\cal GR}({\rm d},N)$  sub-algebra.
Because of the structure of (\ref{eq:GR1}) it is natural to work
with objects that are antisymmetric combinations of $L_{\bj I}$ and
$R_{\bj I}$.  We therefore  make the following definitions.  Let $f_{\bj
{[n]}}$ denote the n-th  antisymmetric combination beginning with
$L_{\bj I}$, and ${\hat f}_{\bj {[n]}}$ the n-th antisymmetric
combination beginning with $R_{\bj I}$.  For example:
\bea
\label{wedging}
f_{\bj {[1]}}:=L_{\bj I}
~~&~~{\hat f}_{\bj {[1]}}:=R_{\bj I}\cr
f_{\bj {[2]}}:=L_{\bj {[I}}\circ R_{\bj {J]}}
~~&~~~{\hat f}_{\bj {[2]}}:=R_{\bj {[I}}\circ L_{\bj {J]}}
\eea
where ${\rm {[IJ]}}$ denotes antisymmetrization.  The algebra
formed by wedging 
$L$'s and $R$'s in this manner will be denoted by $\bigwedge {\cal GR}({\rm 
d},~ N)$. Although it sometimes happens that ${\cal EGR}({\rm d},~ N)\cong
\bigwedge {\cal GR}({\rm d},~ N)$, the generic situation is that $\bigwedge 
{\cal GR}({\rm d},~ N)$ forms a proper sub-algebra of ${\cal EGR}({\rm d},~ 
N)$. When this happens, $\bigwedge {\cal GR}({\rm d},~ N)$ does not span 
the space of all linear mapping on and between $\cv_L$ and $\cv_R$.  It is 
in this sense that ${\cal EGR}({\rm d},~ N)$ completes ${\cal GR}({\rm d},
~N)$. 

It is hoped that the discussion above has convinced the reader that by
allowing the elements of ${\cal V}_L$ and ${\cal V}_R$ to depend upon a 
real parameter for which the derivation $D$ is well defined, a natural
representation of 1D supersymmetry is induced among the elements.
However, this should also suggest to the reader that by permitting the
elements of ${\cal EGR}(N)$ to also depend on such a parameter implies
that the elements of ${\cal EGR}(N)$ must also carry a representation of
1D supersymmetry.  We will see that this is so in the next section.

With these preliminaries out of the way, we turn in the next section to
the representation theory of these ${\cal GR}(N)$ algebras. As explained
there, the representations of this algebra can be derived directly from
the known real representation theory of ordinary Clifford algebras. What
may, perhaps, come as a surprise to some readers is the fact that the
correspondence between the representations of the two algebras is not
entirely trivial. In fact, as will be explained below, the condition
(\ref{eq:GR2}) above amounts in the use of Clifford algebras to 
requiring the existence of an extra matrix which anti-commutes with the
algebra elements and squares to the identity.

\subsection{Enveloping Algebra Representation Theory}

~~~As mentioned in the previous section, it turns out that ${\cal GR}$(d
, $N)$ fits naturally into the context of real Clifford algebras.  We note
first that we need only consider $C(N,1)$ Clifford algebras, that is
(with $I$ = 1, $\dots$ , $N+1$):
\bea
\label{Cliff}
\g^{({}_{ I}}\g^{{}_{ J})}~=~- 2 \h^{{}_{I} {}_{ J}} \, {\bf 
I} \eea
Where $\h^{ {IJ}}=\rm{diag}(1\dots 1, -1)$.\footnote{Please note the different
indices.  $I$ corresponds to the adjoint Clifford algebra index that 
\newline ${~~~~\,}$ runs from $1$ to $N+1$, whereas ${\rm I}$ is the 
adjoint index corresponding to ${\cal GR}$(d,$N)$ taking \newline 
${~~~~\,}$ values $1$ to $N$.}  We find the irreducible representations
 of ${\cal GR}$(d,$N)$ within the context of this family of Clifford 
algebras.  It turns out that $C(N,1)$ Clifford algebras naturally
contain projection operators that lead to ${\cal GR}$(d,$N)$.  We can see 
this by constructing projection operators as follows:
\bea
P_{\pm}~=~{\fracm 12}(\, {\bf I}~\pm~\g^{N+1} \,) 
\eea
where $\g^{N+1}$ is the $(N+1)$-th $\g$-matrix in (12).
These operators satisfy the usual projection algebra, $P_iP_j=\d_{ij}P_i$, and
have the following property when acting on $\g^I$:
\bea
\label{goesto}
P_{\pm}\g^{\bj I} =\g^{\bj I} P_{\mp}
~~~,~~~P_{\pm}\g^{N+1}=\pm P_{\pm}
\eea
With these projectors we make the following identifications:
\be \eqalign{
L_{{}_{\rm I}}~\equiv~P_{+}\g_{{}_{\rm I}} \, P_{-} ~~~,~~~
R_{{}_{\rm I}}~\equiv~P_{-}\g_{{}_{\rm I}} \, P_{+} ~~~, } \label{eq:LR}
\ee
and we can now see that ${\cal GR}$(d, $N)$ arises naturally:
\bea
L_{{}_{( \rm I}}\, R_{{\,}_{{\rm J})}}~&=&~P_{+}\, \g_{{}_{( \rm
I}}\, P_{-} \, \g_{{}_{{\rm J})}} \, P_{+}~=~-2\d_{{{}_{\rm I} {}_{\rm 
J}}}P_{+} ~\equiv~-2\d_{{{}_{\rm I} {}_{\rm J}}}\, {\bf I}{}_+ ~~~, \cr 
R_{{}_{( \rm I}}\, L_{{}_{{\rm J})}}~&=&~P_{-}\g_{{}_{(\rm I}}\, P_{+}\,
\g_{{ }_{\rm J)}} \, P_{-}~=~-2\d_{{{}_{\rm I} {}_{\rm J}}}P_{-}  ~\equiv~
- 2 \d_{{{}_{\rm I} {}_{\rm J}}} \, {\Hat {\bf I}} {}_- ~~~.
\eea
We can also exhibit the skew symmetry relation of the L's to the R's.  
For the $C(N,1)$ Clifford algebras we can always find a basis such that:
\bea
\label{basis}
(\g^{N+1})^T&=&\g^{N+1}\cr
(\g^{\bj I})^T&=&-\g^{\bj I}
\eea
then it follows that:
\bea
L_{{}_{\rm I} }^T~=~(P_{+}\, \g_{{}_{\rm I} } \, P_{-})^T~=~P_{-}^T \, 
\g_{{}_{\rm I} }^T \, P_{+}^T ~=~-P_{-} \, \g_{{}_{\rm I} }
\,P_{+}~=~-R_{{}_{\rm I} }
\eea
Thus, we have shown how all of the defining properties of ${\cal GR}$(d,$N)$ can
be realized by considering only the real Clifford algebras $C(N, 1)$.

A natural expectation\footnote{However, this natural expectation is wrong
as will be discussed below.} is that ${\cal {EGR}}(N)$ = $\bigwedge {\cal 
GR}(N)$ i.\ e.\ represented:
\be \eqalign{ \label{spaces}
\{{\cal U}_L\}~&=~\{\, P_{+}, \, P_{+} \, \g_{{{}_{\rm I} {}_{\rm
J}}}P_{+}~,...,~P_{+}\g_{[N]} P_{+} \, \} ~~~, \cr
\{{\cal M}_L\}~&=~\{ P_{+}\g^{{}_{\rm I} }P_{-}~,...,~P_{+}\g_{[N-1]}P_{-} 
\, \} ~~~, \cr
\{{\cal U}_R\}~&=~\{ \, P_{-}, \, P_{-}\g^{{{}_{\rm I} {}_{\rm J}}}P_{-}
~,...,~ P_{-}\g_{[N]}P_{-} \, \} ~~~, \cr
\{{\cal M}_R\} ~&=~\{ P_{-}\g^{{}_{\rm I} }P_{+}~,...,~P_{-}\g_{[N-1]}P_{+} 
\, \} ~~~~~. } \label{eq:Exp} \ee
Where $\rm {[N]}$ means antisymmetrization over $\rm N$ indices.
This procedure realizes the four transformations between the two spaces 
and automatically obeys the left/right composition rule since $P_{\pm}
P_{\mp}~=~0$.  We can also see why each space of $\bigwedge {\cal GR}(N)$ 
contains half of the elements that $\bigwedge C(N,1)$ contains.  The Clifford
algebra element $\g^{N+1}$ does not appear in (\ref{spaces}) because of
(\ref{goesto}).

When dealing with Minkowski-space Dirac matrices, it is customary 
to use the analogs of ${\rm L}$ and ${\rm R}$  (i.e. the usual
$\s$-matrices) to generate  the analogs of $\{ {\cal M}_L \}$, $\{ {\cal
M}_R \}$, $\{ {\cal U}_L \}$ and $\{  {\cal U}_R \}$ by repeatedly taking
higher and higher order nested commutators.  The procedure generally
fails for the ${\cal {GR}}$(${\rm  d},~ N$) algebras as implied by a work
of Okubo \cite{okubo} who also gave the proper way to build the complete
enveloping algebras.  His classification of the enveloping algebras point
to the existence of three distinct types of real Clifford algebras\footnote{In
the mathematical literature, these are simply referred to as irreducible Clifford 
modules \newline ${~~\,~~}$ that are Real, Complex, or Quaternionic, respectively.}; 

 ${~~~~~~~~~}$(a.) N-type (normal) \newline \indent
 ${~~~~~~~~~}$(b.) AC-type (almost complex)  \newline 
\indent
 ${~~~~~~~~~}$(c.) Q-type (quaternionic) .

In the case of the normal enveloping algebras (a.), we can begin
with the $N+1$  representations $\g^I$ and use their wedge
products to form a set 
$\{ \G \}$.  This set is equivalent to the enveloping
algebra
\be \eqalign{
\{ \G \} &=~ \{~ {\bf I}, \, \g^{I} , \, \g^{JK} , \dots \, , \g^{[N+1]} ~\}
} \label{eq:3Chp8} 
\ee
In turn, this set can be split according to whether elements contain even 
or odd powers of the $\g^{\bj I}$-matrices.
\be \eqalign{
\{ \G_e \} ~=~ \{~ {\bf I}, \, \g^{N+1}, \, \g^{\bj {IJ}}, \, 
\g^{\bj {IJ}}\g^{N+1}, \, \g^{\bj {IJKL}}, \dots ~\} ~~~,}  
\label{eq:3Chp9} 
\ee
\be \eqalign{
\{ \G_o \} ~\equiv~ \{~  \g^{\bj I} , \, \g^{\bj I}\g^{N+1}, \, 
\g^{\bj {IJK}} , \, \g^{\bj {IJK}}\g^{N+1} \dots  ~\}
~~~. }  \label{eq:3Chp10} 
\ee
Finally, the elements of $\{ \G_e \}$ and $\{ \G_o \}$ are such that they can
be put into a two-to-one correspondence with $\{ {\cal M}_L \}$, $\{ {\cal
M}_R \}$, $\{ {\cal U}_L \}$ and $\{  {\cal U}_R \}$ according to (\ref{eq:Exp}).

For the almost complex enveloping algebra (b.), the wedge products span only
one half of the enveloping algebra.  These algebras contain one element ${\cal
J}$ that commutes with $\g^I$ and squares\footnote{Actually ${\cal J}=\pm\g^{1}
\g^2\cdots\g^{N+1}$} to $-{\bf I}$.   In order to completely identify all of
the elements, we must introduce one additional quantity (denoted by Okubo as 
${\cal D}$) that anti-commutes with $\g^I$.
${\cal D}$ is used to double the number of wedge products.  Thus, the enveloping
algebra of the almost complex representations takes the form:
\bea
\{ \G \} &=~ \{~ {\bf I}, \, {\cal D}, \, \g^{\bj I}, \, \g^{\bj I}{\cal D}, \,
\dots \, ,\g^{[N+1]}, \, \g^{[N+1]}{\cal D} ~\}
\eea
This enveloping algebra can now be split into even and odd parts relative to
$\g^{\bj I}$.  These even and odd algebras can then be put into a two-to-one
correspondence with ${\cal EGR}(N)$ in the same manner as the normal case above.

Finally, in the case of quaternionic enveloping algebras, (c.), the wedge
products yield one quarter of the enveloping algebra.  This is remedied
by introducing three objects that commute with the elements
$\g^I$ \cite{okubo}. We  denote these three objects by ${\cal E}^{\hat \a}$ with
$\hat
\a =  (1,2,3)$ and they satisfy
\be
[ \, {\cal E}^{\hat \a} ~,~ {\cal E}^{\hat \b} \,] ~    =~ 2 \e^{{\hat \a} 
\, {\hat \b} \, {\hat \g}} \, {\cal E}^{\hat \g} ~~~,~~~ {\cal E}^{\hat
\a} \,  {\cal E}^{\hat \b} ~    =~ - \d^{{\hat \a} \,
{\hat \b} }{\bf I} ~~~.
\label{eq:3Chp14} 
\ee
The enveloping algebra now takes the form:
\bea
\{ \G \} &=~ \{~ {\bf I}, \, {\cal E}^{\hat \a}, \, \g^{\bj I}, \, 
\g^{\bj I}{\cal E}^{\hat\a}, \,
\dots \, ,\g^{[N+1]}, \, \g^{[N+1]}{\cal E}^{\hat\a} ~\}
\eea
Splitting into odd and even as before, we again arrive at a two-to-one
correspondence between the elements of this enveloping algebra and ${\cal
EGR}(N)$.

The fact that the enveloping algebras have these three distinct
structures makes a great difference when considering either the left or 
right multiplication of the enveloping algebras by single $\g$-matrices.  
In the normal case such a multiplication will cause an orbit to move 
through the entirety of $\{ {\cal M}_L \}$, $\{ {\cal M}_R \}$, $\{ 
{\cal U}_L \}$ and $\{  {\cal U}_R \}$.  In the almost complex and 
quarternionic cases, the results of such a multiplication are very 
different.  In the almost complex case such a multiplication will
only move through half of $\{ {\cal M}_L \}$, $\{ {\cal M}_R \}$, $\{ 
{\cal U}_L \}$ and $\{  {\cal U}_R \}$ since such multiplications can 
never produce objects in the ``${\cal D}$-sector.''  Similarly, in 
the quaternionic case such a multiplication will only move through 
one-fourth of $\{ {\cal M}_L \}$, $\{ {\cal M}_R \}$, $\{ {\cal U}_L
 \}$ and $\{  {\cal U}_R \}$ since such multiplications can never 
produce objects in the ``${\cal E}{}^{\hat \a}$-sector.''  We will 
refer to the part of the enveloping algebras that is independent of 
either $\cal D$ or ${\cal E}{}^{\hat \a}$ as ``the normal part (NP) 
of the enveloping algebra.''  Thus the dimension of the normal part 
of ${\cal {EGR}}(N)$ (which is $\bigwedge {\cal GR}(N)$) is $2^N$.

In our previous work \cite{jim1,jim2}, we constructed explicit
representations of the objects in (\ref{eq:LR}) and also developed an
algorithm for constructing arbitrarily large $N$ representations.  The results
of this analysis are presented in the following\footnote{By dimension in
the following table we are referring the size of the matrices that  \newline
${~~~~\,}$ comprise $C(N,1)$, or the dimension of the spin representation 
it  acts  upon.} table for the cases $1\le N\le 8$.
 
\begin{center}
\renewcommand\arraystretch{1.2}
\begin{tabular}{|c|c|c|c| }\hline
$~~{\cal {GR }}$(${\rm d},~ N)~~$   & 
dim$C(N,1)$&
${\cal {EGR }}$ (${\rm d},~ N) ~ {\rm {generators}} $ & 
Type \\ \hline  \hline
$~~{\cal {GR }}(8, 8)~~$   & 16  & $f_{{\,}_{\rm {I}}}$ & 
N \\ \hline  
$~~{\cal {GR }}(8, 7)~~$   & 16  & $f_{{\,}_{\rm {I}}}$ 
& N \\ \hline 
$~~{\cal {GR }}(8, 6)~~$   & 16  & $f_{{\,}_{\rm {I}}} \,,
~ {\cal D}$  & AC \\ \hline  
$~~{\cal {GR }}(8, 5)~~$   & 16  & $f_{{\,}_{\rm {I}}} \,,
~ {\cal E}^{\hat \a}$  & Q \\ \hline 
$~~{\cal {GR }}(4, 4)~~$  & 8   & $f_{{\,}_{\rm {I}}} \,,
~ {\cal E}^{\hat \a}$  & Q \\ \hline 
$~~{\cal {GR }}(4, 3)~~$  & 8   & $f_{{\,}_{\rm {I}}} \,,
~ {\cal E}^{\hat \a}$  & Q \\
\hline  
$~~{\cal {GR }}(2, 2)~~$  & 4   & $f_{{\,}_{\rm {I}}} \,,
~ {\cal D}$  & AC \\ \hline
$~~{\cal {GR }}(1, 1)~~$  & 2   & $f  ~ $ 
& N \\ \hline
\end{tabular}
\end{center}
\vskip.2in
\centerline{{\bf Table I}} 
By using Bott periodicity this table can be extended to all values of $N$. 
We also see that the dimension of the irreducible representations of
$C(N, 1)$ are exactly twice that of ${\cal GR}$(d, $N)$.  Therefore, it makes
sense that ${\cal GR}$(d, $N)$ can be obtained by projection from $C(N,1)$.

This concludes our discussion of the overall mathematical structure
of ${\cal GR}(N)$.  In what follows we provide a more explicit
representation of $\bigwedge {\cal GR}$(d, $N)$ and
construction of ${\cal EGR}$(d, $N)$ for $N \leq 8$.  We
will use ${\cal GR}$(d, $N)$ notation and when necessary we will
revert to the Clifford algebra representations.

\subsection{${\cal GR}(8,8)$ and Dimensional Reductions}
~~~~The Clifford algebra $C(8,1)$ is 16 dimensional.  This means that
the associated ${\cal GR}$-algebra is ${\cal GR}$(8,$8)$.  Since $C(8,1)$ is
normal we will not have to introduce a new matrix to complete the enveloping
algebra.  Therefore, we can write ${\cal EGR}(8)$ using 
the elements of $\bigwedge {\cal GR}(8,8)$:
\bea
{\cal EGR}(8)_{ij}~\cong~\{ {\cal U}^8_L\} &=&~\{ {\bf I},~f_{\bj {IJ}},~
f_{\bj {IJKL}} \}\cr 
{\bf 64}~&=&~\{ {\bf 1},~{\bf 28},~{\bf 35} \} \cr 
{~~}  \cr 
{\cal EGR}(8)_{i{\hat j}}~\cong~\{ {\cal M}^8_L \}~&=&~\{ f_{\bj I},
~f_{\bj {IJK}} \} \cr 
{\bf 64}~&=&~\{ {\bf 8},~{\bf 56} \}
\eea
Naively, we would expect 5, 6, 7 and 8-forms in ${\cal EGR}(8)$ and that
the 4-form is a ${\bf 70}$.  However, if we write these forms using the
Clifford algebra basis we see that these forms are the duals of the 3, 
2, 1, and 0-forms respectively and that the 4-form has a definite duality. 
Taking:
\bea
\g^{[8]}~=~\ell\e^{[8]}\g^9~,
\eea
where $\e^{[8]}$ is the totally antisymmetric tensor with eight indices 
and $\ell=\pm 1$, we have:
\bea
f_{[8]}~&=&~P_{+}\g^{[8]}P_{+}~=~\ell\e^{[8]}P_{+}\g^9P_{+} ~=~\ell
\e^{[8]}P_{+}~\equiv~\ell\e^{[8]}\, {\bf I}
\eea
\bea
f_{[7]}~&=&~P_{+}\g^{[7]}P_{-}~=~-P_{+}\g^{[7]}\g^{{}_{\rm
I}}P_{+}\g^{{}_{\rm I}} ~=~-\ell\e^{[7]{\rm I}}P_{+}\g^9P_{+}\g^{{}_{\rm 
I}}\cr
~&=&~-\ell\e^{[7]I}P_{+}\g^{{}_{\rm I}}P_{-}~=~-\ell\e^{[7]{\rm I}}
f_{{}_{\rm I}} ~~~~ {\rm I} \notin [7]
\eea
\bea
f_{[6]}~&=&~P_{+}\g^{[6]}P_{+}~=~-{\fracm
1{2!}}P_{+}\g^{[6]}\g^{{}_{\rm I} {}_{\rm J}}\g^{{}_{\rm I} {}_{\rm J}}
P_{+}\cr
&=&-{\fracm {\ell}{2!}}\e^{[6]{\rm {{{}_{\rm I} {}_{\rm J}}}}}P_{+}\g^9
P_{+}\g^{{}_{\rm I} {}_{\rm J}} ~=~-{\fracm {\ell}{2!}} \e^{[6]{\rm 
{{{}_{\rm I} {}_{\rm J}}}}}f_{{}_{\rm I} {}_{\rm J}}~~~~{\rm I,J}\notin [6]
\eea
\bea
f_{[5]}~&=&~P_{+}\g^{[5]}P_{-}~=~{\fracm
1{3!}}P_{+}\g^{[5]}\g^{{}_{\rm I} {}_{\rm J} {}_{\rm K} }\g^{{}_{\rm I} 
{}_{\rm J} {}_{\rm K} }P_{-}\cr
&=&{\fracm {\ell}{3!}}\e^{[5]{\rm {{{}_{\rm I} {}_{\rm J}} {}_K}}}P_{+}\g^9
P_{+}\g^{{}_{\rm I} {}_{\rm J}  {}_{\rm K}} ~=~ {\fracm {\ell}{3!}}
\e^{[5]{\rm {{{}_{\rm I} {}_{\rm J}}{}_K}}}P_{+}\g^{{}_{\rm I}{}_{\rm 
J} {}_{\rm K} }P_{-}\cr ~&=&~{\fracm {\ell}{3!}}\e^{[5] {\rm {{{}_{\rm I} 
{}_{\rm J}}{}_K}}}f_{{}_{\rm I} {}_{\rm J} {}_{\rm  K} } 
~~~~{\rm I,J,K}\notin [5]
\eea
and:
\bea
f_{{}_{\rm I} {}_{\rm J} {}_{\rm K} {}_{\rm L}}~&=&~P_{+}\g^{{}_{\rm I} 
{}_{\rm J} {}_{\rm K} {}_{\rm L}}P_{+}~=~{\fracm 1{4!}} P_{+}\g^{{}_{\rm 
I}{}_{\rm J} {}_{\rm K} {}_{\rm L}}\g^{{}_{\rm {MNOP}}}\g^{{}_{\rm
{MNOP}}}P_{+}
\cr~&=&~{\fracm
{\ell}{4!}}\e^{{}_{\rm {IJKLMNOP}}}P_{+}\g^{{}_{\rm 
{MNOP}}}P_{+}~=~{\fracm {\ell}{4!}}\e^{{}_{\rm {IJKLMNOP}}}f_{{}_{\rm
{MNOP}}} \cr
&~&~~~{\rm M,N,O,P}\not= {\rm I,J,K,L} 
\eea
This explains the economy of ${\cal EGR}(8)$ written in terms of ${\cal
GR}(8,8)$.  Note that the ``hatted'' objects may have different minus signs
since, $P_{-}\g^9P_{-}~=~-P_{-}$.

It is a fact that the irreducible representations of $C(7,1)$, $C(6,1)$,
and $C(5,1)$ are 16 dimensional.  Because of this, we can obtain the enveloping
algebras of ${\cal GR}$(8,$7)$, ${\cal GR}$(8,$6)$, and ${\cal GR}$(8,$5)$ by
dimensional reduction from ${\cal EGR}$(8,$8)$.  From this perspective we will
see why ${\cal EGR}$(8,$7)$ is a normal algebra.  We will also see how the
almost complex and quaternionic structures appear in ${\cal EGR}$(8,$6)$
and ${\cal EGR}$(8,$5)$, respectively.

${\cal EGR}$(8,$7)$ takes the form:
\bea
{\cal EGR}(8, 7)_{ij}~\cong~\{ {\cal U }^7_L \} ~&=&~ \{ {\bf I},
f_{\bj {IJ}}, f_{\bj {IJKL}}, 
f_{\bj {[6]}} \}\cr
{\bf 64}&=&~ \{ {\bf 1},~{\bf 21},~{\bf 35},~{\bf 7} 
\}~\cr
&{~~}&  \cr
{\cal EGR}(8, 7)_{i{\hat j}}~\cong\{ {\cal M}^7_L \} ~&=&~ \{f_{\bj I},
f_{\bj {IJK}}, f_{\bj {[5]}}, f_{\bj {[7]}} \}\cr
{\bf 64}~&=&~ \{ {\bf 7},~{\bf 35},~{\bf 21},~{\bf 
1} \}
\eea
One may ask ``Where did the duality go?"  The answer to this question is 
that the dimensional reduction of $C(8,1)$ to $C(7,1)$ uses the duality 
to transform all of the ``eights" away.  For example:
\bea
f_{\bj {[6]}}~=~-{\fracm {\ell}{2!}}\e^{\bj {[6]I8}}f_{\bj {I8}}
\eea
From these arguments, we see that ${\cal EGR}$(8,$7)$ is normal.  Because
${\cal EGR}$(8,$7)$ is normal we can use it, instead of ${\cal EGR}$(8,$8)$, to
obtain ${\cal EGR}$(8,$6)$ and ${\cal EGR}$(5,$8)$ via dimensional reduction.
${\cal EGR}$(6,$8)$ looks like:
\bea
{\cal EGR}(8,6)_{ij}~&=&~\{ {\bf I}, \,
f_{\bj {I7}}, \, f_{\bj {IJ}}, \, f_{\bj {IJK7}}, \,
f_{\bj {IJKL}}, \, f_{\bj {[5]7}}, \, f_{\bj {[6]}} \}\cr
{\bf 64}~&=&~ \{ {\bf 1},~{\bf 6},~{\bf 15},~~~{\bf 20},
~~~{\bf 15},~~~{\bf 6}, ~~{\bf 1} ~\}\cr
&{~~}&  \cr
{\cal EGR}(8, 6)_{i{\hat j}}~&=&~ \{f_{\bj 7}, \,
f_{\bj I}, \, f_{\bj {IJ7}}, \, f_{\bj {IJK}}, \, f_{\bj {[4]7}}, \, 
f_{\bj {[5]}}, f_{\bj {[6]7}} \}\cr 
{\bf 64}~&=&~ \{ {\bf 1}, ~~{\bf 6},~\, {\bf 15},~~{\bf 20},~\, {\bf 15},~
~\, {\bf 6}, ~~{\bf 1}~ \}
\eea
Here we see that $f_{\bj 7}$ plays the role of the projected ${\cal D}$. 
${\cal D}$ should be in the mixed space since it anti-commutes with
$\g^I$.  Furthermore, ${\cal D}\hat {\cal D}=-{\bf I}$ necessarily.  Reducing
once more we have:
\bea
{\cal EGR}(8,5)_{ij}~&=&~\{ {\bf I}, \, f_{\bj {67}}, \,
f_{\bj {I6}}, \, f_{\bj {I7}}, \, f_{\bj {IJ}}, \, f_{\bj {IJ67}}, \, 
f_{\bj {IJK7}}, \, f_{\bj {IJK6}}, \, f_{\bj {IJKL}}, \, f_{\bj {IJKL67}}, \,
f_{\bj {[5]7}}, \, f_{\bj {[5]6}} \}~~~~\cr 
{\bf 64}~&=&~ \{ {\bf 1},~~{\bf 1},~~{\bf 5},~~{\bf 5}, ~{\bf 10},~~{\bf 10},
~~~{\bf 10}, ~~~{\bf 10}, ~~\,~~{\bf 5},~~~~~~\,{\bf 5}~~~~~ {\bf 1},
~~{\bf 1}~\}\cr  
&{~~}&  \cr
{\cal EGR}(8, 5)_{i{\hat j}}~&=&~ \{f_{\bj 7}, \, f_{\bj 6}, \, 
f_{\bj I}, \, f_{\bj {I67}}, \, f_{\bj {IJ7}}, \, f_{\bj {IJ6}}, \, 
f_{\bj {IJK}}, \, f_{\bj {IJK67}}, \, f_{\bj {[4]7}}, \,  f_{\bj {[4]6}}, \,
f_{\bj {[5]}}, f_{\bj {[5]67}} \}\cr  
{\bf 64}~&=&~ \{ {\bf 1}, ~~{\bf 1},~~{\bf 5},~~\,
{\bf 5},~\,{\bf 10},~\, {\bf 10},~~ {\bf 10}, ~~~~{\bf 10},
~~~{\bf 5},~~~\,{\bf 5},~~\,{\bf 1}, ~~~{\bf 1}~ \}
\eea
Here we see the quaternionic structure:
\bea
{\cal E}^{\hat\a}=\Big( f_{\bj {67}},~f_{\bj {[5]6}},~f_{\bj {[5]7}}
\Big)
\eea
In the following sections we will see representations of a different
quaternionic and almost complex enveloping algebras.  These cases can
not be understood as dimensional reductions of larger Clifford algebras 
of the form $C(N+A,1)$.
\subsection{${\cal GR}(4,4)$ and Dimensional Reduction to ${\cal
GR}(3,4)$}
~~~~The irreducible representation of $C(4,1)$ is 8 dimensional and
quaternionic.  To construct the enveloping algebra we take the wedge products
and the quaternions.  Thus
${\cal EGR}$(4,$4)$ takes the form:
\bea
{\cal EGR}(4)_{ij}~&=&~\{ ~ {\bf I}, \, f_{\bj {IJ}}, \, {\cal E}^{\hat \a}, \,
f_{\bj {IJ}} {\cal E}^{\hat \a}  ~ \}\cr  
{\bf 16}~&=&~\{ \,{\bf 1},~~\,{\bf 3},~~{\bf 3},~~~{\bf 9}~~~\,\}\cr
&{~~}&  \cr
{\cal EGR}(4)_{i{\hat j}}~&=& ~\{f_{{}_{\rm I}}, \, f_{{}_{\rm I}}{\hat 
{\cal E}}^{\hat \a}\} \cr 
{\bf 16}~&=&~\{\,{\bf 4},~~{\bf 12}~\}
\eea
As in the case of ${\cal EGR}$(8), there is some duality that must be taken care
of in order to write the enveloping algebra in this form.  In this case we have
\bea
\g^{\bj {IJKL}}~=~\ell \e^{\bj {IJKL}}\g^{5}~,
\eea
and upon projection
\bea
f^{\bj {IJKL}}~\equiv~P_{+}\g^{\bj {IJKL}}P_{+}
~=~\ell\e^{\bj {IJKL}}P_{+}\g^5P_{+}
~\equiv~\ell\e^{\bj {IJKL}} \, {\bf I}
\eea
\bea
f^{\bj {IJK}}~\equiv~P_{+}\g^{\bj {IJK}} P_{-}
~=~-P_{+} \, \g^{\bj {IJK}}
\g^{\bj L} \, P_{+}\, \g^{\bj L} ~\equiv~-\ell\e^{\bj {IJKL}}f_{\bj L}
~~~{\rm L}\not={\rm I,J,K}
\eea
\bea
f^{\bj {IJ}}~\equiv~P_{+}\g^{\bj {IJ}} \, P_{+}
~=~- {\fracm {\ell}{2!}}\e^{\bj {IJKL}}
P_{+}\, \g^5 \, P_{+}\g^{\bj {KL}} 
~\equiv~-{\fracm {\ell}{2!}}\e^{\bj {IJKL}}f_{\bj {KL}}~.
\eea
Here we see that the two-form has definite duality, and the 3 and 4 forms are
the duals of the 1 and 0 forms.  We have also projected the quaternionic
generators:
\bea
{\cal E}^{\hat \a}~:=~P_{+} \,{\cal E}^{\hat \a} \, P_{+}~,~{\Hat {\cal
E}}^{\hat \a}~:=~P_{-} \, {\cal E}^{\hat \a} \, P_{-}
\eea

The Clifford algebra $C(3,1)$ is 8 dimensional and quaternionic just like 
$C(4,1)$.  This similarity means that we can just dimensionally reduce ${\cal
EGR}$(4,$4)$ to get ${\cal EGR}$(4,$3)$:
\bea
{\cal EGR}(3)_{ij}~&=&~\{~ {\bf I}, \, f_{\bj {IJ}}, \, {\cal E}^{\hat
\a},f_{\bj {IJ}} {\cal E}^{\hat \a}\}\cr
{\bf 16}~&=&~\{~{\bf 1},~~{\bf 3},~~{\bf 3},~~{\bf 9}~~~\}\cr
&{~~}&\cr
{\cal EGR}(3)_{i{\hat j}}~&=& ~\{~f_{\bj I}, \, f_{\bj {IJK}}, \,
f_{\bj {IJK}}{\hat {\cal E}}^{\hat \a} ,\, 
f_{\bj I}{\hat {\cal E}}^{\hat \a} ~ \}\cr    
{\bf 16}~&=&~\{~\,{\bf 3},~~~~{\bf 1},~~~~\,~~{\bf 3},~~{\bf 9}~~~~\}
\eea
Note, as in the case of ${\cal GR}(7,8)$, we have used the duality to
transform all of the ``fours" away.
\subsection{${\cal GR}(2,2)$ and ${\cal GR}(1,1)$}
~~~~$C(2,1)$ is an almost complex, 4 dimensional Clifford algebra.  So the
enveloping algebra of ${\cal GR}$(2,$2)$ looks like:
\bea
{\cal EGR}(2,2)_{ij}~&=&~\{ {\bf I}, \, f_{\bj {IJ}}
, \, {\cal D}, \, f_{\bj {IJ}}{\cal D}~\}\cr
{\bf 4}~&=&~\{ {\bf 1}, ~\,{\bf 1}, ~\,{\bf 1}, ~~{\bf 1}~~\,~ \}\cr
&{~~}&\cr
{\cal EGR}(2,2)_{i\hat j}~&=&~\{ f_{\bj I}
, \, f_{\bj I}{\cal D}~\}\cr
{\bf 4}~&=&~\{~{\bf 2}, ~\,{\bf 2}~~~ \}
\eea
In this case ${\cal D}\hat{\cal D}={\bf I}$.  The final algebra to consider 
is ${\cal GR}(1,1)$.  The Clifford algebra $C(1,1)$ is two dimensional and 
normal.  A one dimensional algebra may not be worth noting, but these
structures obey Bott periodicity.  So even though ${\cal GR}(1,1)$ is trivial,
the form of ${\cal EGR}(1,1)$ will be the same for $N=9, 17\dots$.  It is just
the wedge products:
\bea
{\cal EGR}(1,1)_{ij}&=&~\{ {\bf I} \}\cr
{\bf 1}~&=&~\{ {\bf 1}\}\cr
&{~~}&\cr
{\cal EGR}(1,1)_{i\hat j}&=&~\{ f \}\cr
{\bf 1}~&=&~\{ {\bf 1}\}
\eea
We note that there is no duality in these cases.  This is because $N+1=2n$ 
in this case.  If $N+1$ had been odd, as in the case $N=8$ above , there 
would have been duality.  This concludes the mathematical discussion of $\cal
GR$-algebras.  In what follows we will show how the $\cal GR$-algebras can lead
to a better understanding of supersymmetric theories in higher dimensions.
\section{Connections to Higher Dimensional Supersymmetric Systems}

~~~~The set of objects $\phi_i$ and $\psi_{\hat l}$ with which we
began our discussion in the introduction provide a purely algebraic 
definition of a ``superfield\footnote{As we will see this definition 
is equivalent to the usual Salam-Strathdee definition of a superfield.
}.''  So we may call the set $(\phi_i, \,\psi_{\hat l})$ a superfield.
In particular, the elements of ${\cal V}_L$ may be taken to be a set 
of bosonic commuting 1D fields and the elements of ${\cal V}_L$ may 
be taken to be a set of fermionic anti-commuting 1D fields.  When the  
parameters $\a^{{}_{\rm I}}$ that first appear below (\ref{eq:algSUSY})
are also considered to be anti-commuting, we obtain a description of a
system that realizes $N$-extended supersymmetry.  The simplest invariant
action for this system appears in (\ref{eq:SUSYact}).

However, as already pointed out, the superfields defined above are not 
sufficient to describe spinning particle systems in Minkowski space.  
The basic problem is that the bosonic components that correspond to the 
``$x$-space'' coordinate of the spinning particle within the framework 
of these representations necessarily increase with $N$.   Thus, the 
superfields defined above are appropriate for ``iso-spinning particle'' 
systems but not spinning particle systems.  To overcome this,  it was 
suggested that there is another\footnote{In fact, there are an infinite 
number of such definitions.}  Clifford-algebraic definition of a superfield.  
It was observed that the elements of $\{{\cal U}_L\} \oplus \{{\cal M}_R\}$ 
(as well as $\{{\cal U}_R\} \oplus\{ {\cal M}_L\}$) can be used to define
superfields in precisely the same way as the elements ${\cal V}_L$ and
${\cal V}_R$.  This second Clifford-algebraic definition of a superfield
appears to be compatible with the structure of spinning particle models 
in Minkowski space.

To this end, introduce a set of maps ($\F_{i} {}^{j}$, $\J_{\hat l}
{}^i$)$:{\mathfrak R}\rightarrow \{{\cal U}_L\} \oplus \{{\cal M}_R\}$
and define an algebraic derivation  $\d_{\a} : \{{\cal U}_L\} \oplus
\{{\cal  M}_R\} \rightarrow\{{\cal U}_L\} \oplus \{{\cal M}_R\} $ 
by the relations
\be
\d_\a:(\Phi,\Psi)\mapsto \lgroup i\a \cdot L(\Psi)\, ,~ - i \a \cdot
R(D\Phi) ~ \rgroup  ~~~.
\label{eq:4.1}   \ee 
Above we mean that $L$ and $R$ act upon $\Phi$ and $\Psi$ by left 
composition.  Acting again with $\d_\b$ we find that the condition in 
(\ref{eq:SUSY}) is satisfied on this representation of fields.  We will see that
this representation of ${\cal GR}$(d, $N)$ plays a key role in connection 
with higher dimensional theories.

The constructions above also naturally engender the realization of 
``twisted'' superfield representations.  If we concentrate only on 
the field representations and neglect the question of an invariant 
action, then there exists a potentially interesting ambiguity.  Let us 
focus upon two sets of objects $(\phi_i^{(1)}, \,\psi_{\hat l}^{(1)})$
and $(\phi_{\hat l}^{(2)}, \, \psi_i^{(2)})$ where it should be noted
that these two sets of fields possess different markings of their index 
types.  Both of these sets will necessarily define supermultiplets
that satisfy a supersymmetry algebra.  If they are inequivalent, we may
call one the twisted version of the other. There is also a second 
potential source of twisted multiplets.  If there exist several
inequivalent representations of a given ${\cal GR}(N)$ 
algebra\footnote{From the mathematical literature, the is known
to occur for quaternionic case.}, then the fields defined with respect to 
these inequivalent representations may also be referred to as the twisted 
versions of one another.  Finally, one can consider a set of mapping 
operations whose definition is given by the exchange among pairs of 
twisted multiplets.  We have long called such maps, ``mirror maps.''

The concepts of twisted multiplets and mirror maps can also be defined 
to act upon sets of objects that are defined to lie in $\{{\cal M}_R
\}\oplus\{{\cal U}_L\}$ (as well as $\{{\cal M}_L\} \oplus\{{\cal U}_R
\}$).  So for these larger superfield representations very similar 
ambiguities may also be realized.  In particular, the fact that normal, 
almost complex and quaternionic cases exist leads directly the 
occurrence of mirror representations here.  Consider a set of objects 
($\Hat \F_{\hat i} {}^{\hat j}$, $\Hat \J_{i}{}^{\hat j}$) with $\Hat 
\F_{\hat i} {}^{\hat j} \in \{{\cal U}_R\}$ and $\Hat \J_{i}{}^{\hat j} 
\in \{{\cal M}_L\}$.  Next choose only the cases where the enveloping
algebra contains an almost complex structure ${\cal J}$.  Finally,
we define define an algebraic derivation $\d_{\a} : \{{\cal U}_R\}
\oplus \{{\cal M}_L\}\rightarrow\{{\cal U}_R\} \oplus \{{\cal M}_L\} $ 
by the relations
\be
\d_\a:({\Hat \Phi}, \, {\Hat \Psi})\mapsto \lgroup i\a \cdot
{\cal J} L({\Hat \Psi})\, ,~ i \a \cdot {\cal J}R(D {\Hat \Phi}) 
~ \rgroup ~~~,
\label{eq:4.2}   \ee 
The set $({\Hat \Phi}, \, {\Hat \Psi})$ is a ``twisted multiplet''
relative to the set $(\Phi, \,\Psi)$.  Similar generalizations can
occur for quaternionic representations\footnote{It may be the
case that there occur a full $S^2$ of such twistings resulting from
the \newline ${~~~~\,}$ distinct ways of choosing a complex  structure on the quaternions.}.

Due to the structure of the enveloping algebras discussed previously,
the superfields that appear in (\ref{eq:4.1}) are in general reducible
representations.  Only in the cases where the enveloping algebra is
normal, do these superfields form irreducible representations and this
only occurs for the cases of $N$ = 1, 7 and 8 mod(8).  For all other 
values of $N$, only the normal parts of the almost complex and
quaternionic enveloping algebras form irreducible superfields.

However, even in the case of either normal enveloping algebras or
restricting to the normal part of enveloping algebras, there is a
remarkable property of these irreducible representations.  In these
cases there is a freedom to transmute ``auxiliary fields'' into
``physical fields'' and vice-versa.  In the following, we display
this property and show some connections between
${\cal GR}$(d, $N)$ and higher dimensional theories.

\subsection{Auxiliary/Physical Field Duality in ${\cal GR}$(d,
$N)$}

~~~~Since reduction on a cylinder is a universal procedure, it can
be applied to any field theory in any dimension greater than one.
Let us begin with the scalar and spinor multiplets of the well known
heterotic string.  The starting point includes the (1,0) ``scalar''
superfields and (1,0) ``heterotic fermion'' superfields,
\be \eqalign{
{\bf X}{}^{\un m}(\z^+ , \, \t, \, \s) \, &=~ X{}^{\un m}(\t, \, \s)  
~+~ \z^+ \, \psi_+ {}^{\un m}(\t, \, \s)  ~~~, \cr
\L {}_- {}^{\Hat I}(\z^+ , \, \t, \, \s) \,  &=~ \eta {}_-
{}^{\Hat I}(\t, \, \s)  ~+~ i \, \z^+ \, F{}^{\Hat I}(\t, \, \s)  
~~~.} \label{eq:5.1}    \ee
The well-known supersymmetrically invariant actions for these take
the forms
\be \eqalign{  {~~~~~~~}
{\cal S}_{scalar} ~&=~ \int d^2 \s \, d \z^+ ~ \Big[ ~ i \fracm 12 \,
\eta_{\un m \, \un n} \, {\bf X}{}^{\un m} \, \pa_{\mm} \, D_+ 
 \,{\bf X}{}^{\un n} ~ \Big] 
 ~~~, \cr
~&=~ \int d^2 \s ~ \Big[ ~ - \fracm 12 \,
\eta_{\un m \, \un n} \, X{}^{\un m} \, \pa_{\mm} \, \pa_{\pp}
 \, X{}^{\un n} ~+~ i \fracm 12 \,
\eta_{\un m \, \un n} \, \psi_+ {}^{\un m} \, \pa_{\mm} \, 
 \, \psi_+ {}^{\un n}  ~ \Big]   ~~~, \cr
{\cal S}_{spinor} ~&=~  \int d^2 \s \, d \z^+ ~ \Big[ ~ - \, 
\fracm 12 \, \L{}_- {}^{\Hat I} \, D_+ \, \L{}_- 
{}^{\Hat I} ~ \Big]  ~~~,  \cr 
~&=~ \int d^2 \s ~ \Big[ ~ i \, \fracm 12 \, \eta {}_- {}^{\Hat I} \,
\pa_{\pp} \, \eta{}_-  {}^{\Hat I} ~+~ \fracm 12 \, F {}^{\Hat I} \, 
F {}^{\Hat I} ~ \Big]   ~~~.
} \label{eq:5.2}  \ee
For our purposes, the target space 10D vector and SO(32) indices
may be suppressed.

Reducing these fields on a cylinder amounts to simply dropping the 
dependence on $\s$ as well as dropping the ``vectorial'' superscripts 
($\pp$ and $\mm$) and ``spinorial'' subscripts $\pm$ that in 2D keep 
track of helicity (we ignore higher KK modes throughout):
\be \eqalign{
{\bf X}(\z^+ , \, \t, \, \s) \, &\to  {~~} {\bf X}(\z , \, \t)
~=~ X  (\t)   ~+~ i \,\z \, \psi (\t)  ~~~, \cr
\L{}_- (\z^+ , \, \t, \, \s) \, &\to  ~~ \L (\z , \, \t) \, ~=~ {\bf
\eta} (\t)  ~ ~+~ \z \, F(\t )  
~~~\,~.}    \label{eq:5.3}  \ee
In writing the 1D superfields, we have made certain re-definitions so that
the final appearance of the components insure their reality under superspace
conjugation.  
\be \eqalign{
X ~=~ [\, X \, ]^* ~~~,~~~ \psi ~=~ [\, \psi \,]^* ~~~,~~~ \eta ~=~
[\, \eta \,]^* ~~~, ~~~ F ~=~ [\, F \, ]^* 
~~~\,~.}    \label{eq:5.3b}  \ee
The 2D actions in (\ref{eq:5.2}), under the action of the
reduction, become
\be \eqalign{  {~~~~}
{\cal S}_{scalar} ~&=~ \int d \t \, d \z ~ \Big[ ~ i \fracm 12 \,
{\bf X} \, \pa_{\t} \, D_{\z}  \,{\bf X} ~ \Big] ~=~ \int d \t ~ 
\Big[ ~ - \fracm 12 \, X \, \pa_{\t}^2 \, X ~-~ i \fracm 12 \, \psi  
\, \pa_{\t} \, \psi   ~ \Big]   ~~~, \cr 
&=~ \int d \t ~ \Big[ ~  \fracm 12 \, (\pa_{\t} X ) \, (\pa_{\t} X ) 
~-~ i \fracm 12 \, \psi  \, \pa_{\t} \, \psi ~ \Big]   
~~~, } \label{eq:5.4a}  \ee
\be \eqalign{  {~~\,~}
{\cal S}_{spinor} ~&=~  \int d \t \, d \z ~ \Big[ ~  \, 
\fracm 12 \, \L \, D_{\z} \, \L ~ \Big]  ~~~,  
{~~~~~~~~~~~~~~~~~~~~~~~~~~~~~~~~~~~~~~~~~~~~~~~~~~~~~~~~~~}\cr 
~&=~ \int d^2 \s ~ \Big[ ~ -  i \, \fracm 12 \, \eta \, \pa_{\t} \, \eta 
~+~ \fracm 12 \, F \, F  ~ \Big]   ~~~.
} \label{eq:5.4b}  \ee
The $(X, \, \psi)$ multiplet and the $(\eta, \, F)$ multiplet are both
representations of ${\cal GR}$(1, $1)$.  For the first of these we may
make the identifications $X \in {\cal V}_L$ and $\psi \in {\cal V}_R$ and
for the second $\eta \in {\cal V}_L$ and $F \in {\cal V}_R$.  

There is a transformation that maps one of these representation into the
other.  This can be seen by simply making the following field 
re-definitions
\be \eqalign{  {~~~~}
(X, \, \psi) ~ \leftrightarrow ~ (\,  (\pa_{\t}^{-1} F) , \, \eta \,)  
~~~,
} \label{eq:5.5}  \ee
which has the effect of transforming the two actions one into the other, 
${\cal S}_{scalar}  \leftrightarrow {\cal S}_{spinor} $.  An interesting 
feature of the re-definition in (\ref{eq:5.5}) is that although it involves 
a formally non-local transformation, there is no sign of the non-locality 
(after implementing the map) in either the transformation laws or
the actions.  A more remarkable feature of the map defined by (\ref{eq:5.5})
is that it acts to map the ``physical field'' $X$ in the 1D scalar
multiplet into the ``auxiliary field'' $F$ of the 1D spinor multiplet
and vice-versa.  So within the representation theory of ${\cal GR}$(d, 
$N)$ bosonic fields which in higher dimensions correspond to propagating
and auxiliary fields are accorded a unified treatment and a ``duality''
map (\ref{eq:5.5}) exists between them.  I

One other fact to note about the map in (\ref{eq:5.5}) is that from the
point of view of the superfields it corresponds to
\be \eqalign{ 
\L  ~ \leftrightarrow ~ - i \,D_{\z} {\bf X} ~~~.
} \label{eq:5.6}  \ee
Thus implementing the map also has the effect upon a superfield construction
of ``changing'' where the component fields appear in the $\z$-expansion
of superfields.  It should noted that the equation in (\ref{eq:5.6}) is
particular to 1D.  In the original 2D theory (\ref{eq:5.1}), (\ref{eq:5.2})
the 2D Lorentz invariance forbids the possibility to write (\ref{eq:5.6}).
It should also be clear that only in the case of equal numbers of scalar
multiplets and spinor multiplets can such a transformation be implemented.

\subsection{``Root Superfields'' in ${\cal GR}$(d, $N)$}

~~~~The observation that such non-local transformations exist for ${\cal 
GR}$(d, $N)$ allows for an interesting generalization for higher values 
of d and $N$ when we start with superfields that are defined by Clifford 
algebra expansions as described in (\ref{eq:4.1}).  Due to their 
definitions, it follows that we may write
\be \eqalign{ 
\F &=~ \phi(\t) \, {\bf I} ~+~ \phi_{ {\,}_{{\rm I}_1}{\,}_{{\rm I}_2}}
(\t)  \, f^{ {\,}_{{\rm I}_1}{\,}_{{\rm I}_2} }~+~ \phi_{ {\,}_{{\rm
I}_1}{\,}_{{\rm I}_2} {\,}_{{\rm I}_3}{\,}_{{\rm I}_4} } (\t) \, f^{ 
{\,}_{{\rm I}_1}{\,}_{{\rm I}_2} {\,}_{{\rm I}_3}{\,}_{{\rm I}_4} }
~+~ \dots ~~~, 
\cr
\J &=~ \psi_{ {\,}_{{\rm I}_1}} (\t)  \, \hat f^{ {\,}_{{\rm
I}_1}}~+~ 
\psi_{ {\,}_{{\rm I}_1}{\,}_{{\rm I}_2} {\,}_{{\rm I}_3} } (\t)
 \, \hat f^{ {\,}_{{\rm I}_1}{\,}_{{\rm I}_2} {\,}_{{\rm I}_3} }
~+~ \dots ~~~. 
} \label{eq:5.7}  \ee
Written with these definitions, the component fields are analogous to
components of ${\bf X}$.  However, we observe that there exist many
distinct ``dualities'' like that defined in (\ref{eq:5.5}) that may be 
implemented on these components.  The ``dualized'' superfields 
have the forms
\be \eqalign{ {~~~}
{\Tilde \F} &=~ [(\pa_{\t})^{a_0} \phi] \, {\bf I} ~+~ [(\pa_{\t})^{a_2} 
\phi_{ {\,}_{{\rm I}_1}{\,}_{{\rm I}_2} } ] \, f^{ {\,}_{{\rm I}_1}
{\,}_{{\rm I}_2} } ~+~ [(\pa_{\t})^{a_4} \phi_{ {\,}_{{\rm I}_1}{\,}_{
{\rm I}_2} {\,}_{{\rm I}_3}{\,}_{{\rm I}_4} } ] \, f^{ {\,}_{{\rm
I}_1}{\,}_{{\rm I}_2} {\,}_{{\rm I}_3}{\,}_{{\rm I}_4} } ~+~ \dots ~~~, 
\cr
{\Tilde \J} &=~ [(\pa_{\t})^{a_1} \psi_{ {\,}_{{\rm I}_1}} ] \,
\hat f^{  {\,}_{{\rm I}_1}}~+~ [(\pa_{\t})^{a_3} \psi_{ {\,}_{{\rm
I}_1}{\,}_{ {\rm I}_2} {\,}_{{\rm I}_3} } ] \, \hat f^{ {\,}_{{\rm
I}_1}{\,}_{{\rm I}_2} {\,}_{{\rm I}_3} } ~+~ \dots ~~~.  }
\label{eq:5.8}  \ee There are many different choices for the
non-positive integer exponents
$a_0, \, a_1, \, \dots$ such that using (\ref{eq:4.1}) leads to purely
local transformation laws among the component fields.  However, just as 
in our ``toy'' example of looking at the effect of (\ref{eq:5.5}) on the
$\z$-expansion of the equivalent superfields, here the different choices 
of the exponents ``shift'' the various component fields among the
different $\z$-levels of Salam-Strathdee superfields.  In the previous
work of \cite{jim1,jim2} this  freedom was exploited to define a set of
component fields that have been suggested to provide an off-shell
representation for the arbitrary $N$-extended spinning particle.  In the
chapter on applications, we will return to this proposed description.

It might at first appear that the 1D duality between physical bosons and
auxiliary bosons is peculiar to the case of the heterotic starting point
of our discussion.  The next point we wish to make is that this is {\em
{not }} the case and the simplest context for giving this demonstration
is to consider the 2D, $N$ = 1 scalar superfield
\be \eqalign{ {~~~}
{\bf X}(\z^+ , \, \z^- , \, \t, \, \s) \, &=~ X(\t, \,
\s)   ~+~ \z^+ \, \psi_+ (\t, \, \s)  
~+~ \z^- \, \psi_- (\t, \, \s) \cr
  &~~~~  ~+~ i \, \z^+ \, \z^- \, F(\t, \, \s)  
~~~.} \label{eq:5.9}    \ee
After reduction to 1D on a cylinder this takes the form
\be \eqalign{ {~~~~~~~~}
{\bf X}(\z^+ , \, \z^- , \, \t) \, &=~ X(\t) ~+~ \z^+ \, \psi_+ (\t)  
~+~ \z^- \, \psi_- (\t) +~ i \, \z^+ \, \z^- \, F(\t)  
~~~.} \label{eq:5.10z}    \ee
Note that since the two spinors $\psi_+$ and $\psi_-$ as well as $\e^+$ 
and $\e^-$ are independent quantities, we retain the indices from 2D
even though they no longer carry a representation of helicity.  

The presence of the indices in 1D allows us to use superspace conjugation
so that
\be \eqalign{ {~~~~~~~}
X ~=~ [\, X \, ]^* ~~,~~ \psi_+ ~=~ - [\, \psi_+ \,]^* ~~,~~ \psi_- ~=~
- [\, \psi_- \,]^* ~~, ~~ F ~=~ [\, F \, ]^* 
~~~\,~.}    \label{eq:5.3bb}  \ee

The supersymmetry variation of the superfield in (\ref{eq:5.9}) can also 
be reduced on a cylinder and afterward takes the form
\be \eqalign{ {~~~}
\d_Q X &=~  \e^+ \, \psi_+  ~+~  \e^- \, \psi_- ~~~, \cr
\d_Q \psi_+  &=~ i  \e^+ \, \pa_{\t} \, X ~+~ i \e^- \, F  ~~~, \cr
\d_Q \psi_- &=~  i  \e^- \, \pa_{\t} \, X~-~ i \e^+ \, F   ~~~, \cr
\d_Q F &=~ - \, [ ~ \e^+ \, \pa_{\t} \, \psi_-  ~-~  \e^- \,
\pa_{\t} \,\psi_+ ~ ]    ~~~.} \label{eq:5.10}    \ee
so that when evaluated on any of the fields we have
\be
[\, \d_{Q_1} ~,~ \d_{Q_1} \, ] ~=~ - i 2  \, [ ~ \e^+_1 \, \e^+_2
\, \pa_{\t} ~+~ \e^-_1 \, \e^-_2 \, \pa_{\t} ~ ] ~~~.
\label{eq:5.10a}    \ee

We next take a Clifford algebraic superfield of the form given in
(\ref{eq:5.8}) expanded over the normal part of ${\cal U}_L$(2, $2)$
$\oplus$ ${\cal M}_L$(2, $2)$
\be \eqalign{ {~~~}
{\Tilde \F} &=~ [(\pa_{\t})^{a_0} \phi] \, {\bf I} ~+~ [(\pa_{\t})^{a_2} 
\phi_{ {\,}_{{\rm I}_1}{\,}_{{\rm I}_2} } ] \, f^{ {\,}_{{\rm I}_1}
{\,}_{{\rm I}_2} }  ~~~, 
\cr
{\Tilde \J} &=~ [(\pa_{\t})^{a_1} \psi_{ {\,}_{{\rm I}_1}} ] \,
\hat f^{  {\,}_{{\rm I}_1}} ~~~.  } \label{eq:5.11}  \ee
Upon making the identifications $a_0 \,=\, a_1 \,=\, 0 , ~ a_2 \,=\, -1$
and 
\be \eqalign{ {~~~}
\a{}^{{}_{\rm I}} & =~ (\e {}^+ , \, \e {}^- ) ~~~,~~~ \psi{}^{{}_{\rm I}}
~ =~ (\psi {}_+ , \, \psi {}_- )  
~~~,~~~ \phi ~=~ X  ~~~,~~~ \phi{}_{ {\,}_{{\rm I}_1}{\,}_{{\rm I}_2} }
~=~ \e {}_{ {\,}_{{\rm I}_1}{\,}_{{\rm I}_2} } \, F~~~.  }
\label{eq:5.12}  \ee it is seen that the transformation law in (\ref{eq:4.1}) 
exactly reproduces the results in (\ref{eq:5.10}).  An interesting feature to 
note is that the exponent for $F$ takes on a different value from those of 
$X$ and $ \psi{}^{{}_{\rm I}}$.  Since the former is known to be an ``auxiliary'' 
field while the latter two are ``physical'' fields, this suggest that the presence
of the operator $\pa^{-1}$ is associated with this distinction.

We have seen that there is a sense in which the 1D Clifford-algebraic superfield
in (\ref{eq:5.11}) ``encodes'' the structure of the 2D, $N$ = 1 superfield 
in (\ref{eq:5.9}).  We will refer to the former of these superfields as the
``root'' superfield for the latter.  This brings us to a conjecture

${~~~~~}$ {\it {All superfields that provide a {\underline 
{linear}} representation of spacetime super- \newline ${~~~~~~~~~}$
symmetry in all dimensions can be represented as Clifford-algebraic
\newline ${~~~~~~~~~}$ root superfields.}} 

It is our eventual hope that the root superfield concept will prove useful
in understanding off-shell representation theory in higher dimensions as
the concepts of                 ``roots and weights'' play a similar role in Lie algebra theory.

\subsection{Higher Dimensional Off-shell v.s. On-shell SUSY and 
Embedding in ${\cal GR}$(d, $N)$ Representations}

~~~~The fact that higher dimensional superfields can be related to 1D
root superfields has other interesting implications for making a 
Clifford-algebraic distinction between on-shell superfields and off-shell
superfields.  This can be seen by re-considering the 2D, $N$ = 1 
scalar multiplet in (\ref{eq:5.9}).  The 2D, $N$ = 1 off-shell
supersymmetry  variations take the form
\be \eqalign{ {~~~}
\d_Q X &=~  \e^+ \, \psi_+  ~+~  \e^- \, \psi_- ~~~, \cr
\d_Q \psi_+  &=~ i  \e^+ \, \pa_{\pp} \,X ~+~ i \e^- \, F  ~~~, \cr
\d_Q \psi_- &=~  i  \e^- \, \pa_{\mm} \, X~-~ i \e^+ \, F   ~~~, \cr
\d_Q F &=~ -  \, [ ~ \e^+ \, \pa_{\pp} \, \psi_-  ~-~  \e^- \,
\pa_{\mm} \,\psi_+ ~ ]    ~~~.} \label{eq:6.1}    \ee
so that when evaluated on any of the fields we have
\be
[\, \d_{Q_1} ~,~ \d_{Q_2} \, ] ~=~ - i 2  \, [ ~ \e^+_1 \, \e^+_2
\, \pa_{\pp} ~+~ \e^-_1 \, \e^-_2 \, \pa_{\mm} ~ ] ~~~.
\label{eq:6.1a}    \ee

In the previous section, we established that the reduction of these
off-shell variations yields the 1D variations that appear in (\ref{eq:5.10}).
In turn these variations could be embedded into the transformation law
of (\ref{eq:4.1}) if the Clifford-algebraic superfield was expanded
over the normal part of ${\cal U}_L$(2, $2)$ $\oplus$ ${\cal M}_L$(2, $2)$
embedded in ${\cal EGR}$(2, 2).

Let us consider the on-shell massless limit of (\ref{eq:6.1}) which begins
by imposing the condition that the usual auxiliary component field
$F$ should be subjected to the algebraic condition $F$ = 0.  From the
last line in (\ref{eq:6.1}), if $F$ = 0, it follows that 
\be \eqalign{ {~~~}
\pa_{\pp} \, \psi_-  ~=~ 0 ~~~,~~~   \pa_{\mm} \, \psi_+   
~=~ 0 ~~~.} \label{eq:6.2}    \ee
The solutions to these equations are very simple, namely
\be \eqalign{ {~~~}
 \psi_-  ~=~  \psi_- (\s^{\mm} ) ~~~,~~~   
 \psi_+  ~=~  \psi_+ (\s^{\pp} )
~~~.} \label{eq:6.3}    \ee
Subject to these restrictions and as well using the restriction $F$ = 0,
we next apply $\pa_{\mm}$ to the second line of (\ref{eq:6.1}) (or 
alternately applying $\pa_{\pp}$ to the third line of (\ref{eq:6.1})).
We find
\be \eqalign{ {~~~}
 \pa_{\pp} \,\pa_{\mm} \,X ~=~ 0 ~~~,
~~~.} \label{eq:6.4}    \ee
whose solution is given by
\be \eqalign{ {~~~}
X ~=~ X_L(\s^{\pp}) ~+~ X_R (\s^{\mm})  ~~~,
~~~.} \label{eq:6.5}    \ee

The solutions in (\ref{eq:6.3}) and (\ref{eq:6.5}) can be substituted
back into (\ref{eq:6.1}) and since the 2D light-cone coordinates
$\s^{\pp}$ and $\s^{\mm}$ are independent, it follows that the equations in
(\ref{eq:6.5}) can actually be separated into {\em {two}} distinct sets of
equations so  that we can write,
\be \eqalign{ {~~~~~~~~}
\d_Q \, \left[ \, X_L(\s^{\pp}) \, \right] &=~  
\e^+ \, \psi_+ (\s^{\pp} )  ~~~, ~~~
\d_Q \left[ \, \psi_+ (\s^{\pp}) \, \right] ~=~ i \e^+ \, \pa_{\pp} \,
X_L(\s^{\pp})  ~~~, \cr
\d_Q \, \left[ \,  X_R (\s^{\mm}) \, \right] &=~  
 \e^- \, \psi_- (\s^{\mm} ) ~~~, ~~~
\d_Q \left[ \, \psi_- (\s^{\mm}) \, \right] ~=~  i \e^- \, \pa_{\mm} \,
X_R(\s^{\mm})    ~~~.} \label{eq:6.7}    \ee

Now we can reduce these results on a cylinder and then ask,  ``What
${\cal GR}$(d, $N)$ Clifford-algebraic superfields can re-produce these
results?''  The answer turns out the be rather interesting.  As can
easily be seen in (\ref{eq:6.7}), the component fields there actually
form two distinct representations; $X_L$ and $\psi_+$ form one
representation and $X_R$ and $\psi_-$ form another.  The 
Clifford-algebraic superfields that produces these transformation
laws are valued in ${\cal EGR}$(1, $1)$.  Stated another way, the
results in (\ref{eq:6.7}) constitute a reducible representation of
${\cal EGR}$(1, $1)$.

The lesson from the example is starkly clear.  The ${\cal GR}$(d, $N)$
characterization of the same multiplet changes depending on whether the 
multiplet is on-shell or off-shell.   Stated another way, for a given 
supermultiplet, there is an algebraic way to distinguish between its 
on-shell versus off-shell representation.  From this explicit example, 
we are led to a second conjecture regarding the off-shell versus on-shell 
distinction for spacetime supersymmetric representations when viewed 
from their embedding into ${\cal GR}$(d, $N)$

${~~~~~}$ {\it {If an on-shell supermultiplet is embedded into a
representation of}} \newline ${~~~~~~~~~}$ ${\cal EGR}$(d${}_N, \,N)$,
{\it {then an off-shell representation of this supermultiplet 
\newline ${~~~~~~~~~}$ is embedded into}} ${\cal EGR}$(d${}_{2N}, \,2N)$.

\subsection{4D Chiral Superfield on a D0-Brane}

~~~~In light of our presentation in the previous chapter, it seems 
plausible that we might gain insight into the possibility of embedding 
the superfield representations of 4D spacetime supersymmetry.  Let us 
begin with a 1D representation that bears a striking
resemblance to the 4D, $N$ = 1 chiral superfield.  We introduce
a multiplet $\left({\cal Z} (\t), \varphi{}_{ {\,}_{I} } (\t), F 
(\t) \right)$ that forms an 1D, $N$ = 4 scalar 
multiplet.  The global supersymmetry variations of these fields when 
reduced to 1D read 
\be \eqalign{
\d_{Q} {\cal Z} &= ~ \e^{ {\,}_{I}} \varphi {}_{ {\,}_{I}} ~~~, \cr
\d_{Q} \varphi { {\,}_{I}} &=~ i \, {\Bar{\e}} { {\,}_{I}} \, (\pa_{\t}
{\cal Z}) ~+~ i \e {}^{ {\,}_{K}} \e_{ {\,}_{K} {\,}_{I}}\, F ~~~,
\cr
\d_{Q} F &=~   {\Bar{\e}}{}_{ {\,}_{I}} \, ( \pa_{\t} \varphi {}_{
{\,}_{K}} ) \, \e^{ {\,}_{I} {\,}_{K}}  ~~~.  }  
\label{eq:A6.1}\ee
A few words about notation are in order. These results are directly
obtained from the reduction.  The fields $\left({\cal Z} (\t), 
\varphi{}_{ {\,}_{I} } (\t), F (\t) \right)$ are all complex and functions 
of $\t$.  The indices $I$ on the supersymmetry parameter $\e$ and the
fermion field $\varphi$ in 1D correspond to an isospin index that
takes on two values.  This is not to be identified with the index I
which takes on four values.  

When evaluated on the fields that appear in (\ref{eq:A6.1}), the
commutator algebra takes the form
\be
[ \, \d_{Q_1} ~,~ \d_{Q_2} \, ] ~=~ - i \, (\, \e_1{}^I \, {\bar 
\e} {}_2{}_I ~+~ {\bar \e}{}_1{}_I \, \e_2{}^I \,) \, \pa_{\t}
~~~, \label{eq:A6.4} \ee
but this is not the basis for extracting the ${\cal GR}$(d, $N)$
structure of the theory.  For this purpose it is necessary to 
express the supersymmetry parameter and fields in terms of real 
quantities by writing
\be  \eqalign{
\e{}^I &\equiv~ (\, \e{}^{1(1)} \,+\, i\e{}^{1(2)} , \,
\e{}^{2(1)} \,+\, i\e{}^{2(2)}  \, )
~~~, \cr
{\cal Z} &\equiv~ {\rm A} ~+~ i {\rm B} ~~~, ~~~
F ~\equiv~ {\rm F} ~+~ i {\rm G} ~~~, \cr
\varphi{}_I &\equiv~ (\, \varphi{}_{1(1)} \,+\, i \varphi{}_{1(2)} , \,
\varphi{}_{2(1)} \,+\, i \varphi{}_{2(2)}  \, )
~~~, \cr
}\label{eq:A6.5} \ee
Once these definitions are made, then it is possible to introduce
four component quantities $\a{}^{{}_{\rm I}}$ and $\j{}^{{}_{\rm
I}}$  via the definitions
\be \eqalign{
\a{}^{{}_{\rm I}} &=~ (\, \e{}^{1(1)} , \, \e{}^{1(2)} , \,
\e{}^{2(1)} , \, \e{}^{2(2)}  \, ) ~~~, \cr
\j{}^{{}_{\rm I}} &=~ - i \, (\, \varphi{}_{1(1)} , \,
\varphi{}_{1(2)} , \,
\varphi{}_{2(1)} , \, \varphi{}_{2(2)}  \, )      ~~~.
}\label{eq:A6.6} \ee
It is a simple matter to verify that (\ref{eq:A6.4}) using the definitions
in (\ref{eq:A6.5}) and (\ref{eq:A6.6}) takes the form of (\ref{eq:SUSY}).

Finally, the transformation laws in (\ref{eq:A6.1}) must be expressed
in terms of A, B, F, G, $\j{}^{{}_{\rm I}}$ and $\e{}^{{}_{\rm I}}$
in order to uncover how the 4D, $N$ = 1 chiral multiplet is embedded
into representations of ${\cal GR}$(d, $N)$.  In terms of these,
we find
\be \eqalign{
\d_Q {\rm A} &=~  i \, [~ \a^1 \, \psi^2 \,+\, \a^2 \, \psi^1 \,+\, 
\a^3 \, \psi^4 \,+\, \a^4 \, \psi^3  ~]   ~~~, \cr
\d_Q {\rm B} &=~  i \, [~ - \a^1 \, \psi^1 \,+\, \a^2 \, \psi^2 \,-\, 
\a^3 \, \psi^3 \,+\, \a^4 \, \psi^4   ~]  ~~~,    \cr
\d_Q \j{}^1 &=~   [~ - \a^1 \, (\pa_{\t} {\rm B}) \,+\, \a^2 \, (
\pa_{\t} {\rm A}) \,+\, \a^3 \, {\rm G} \,+\, \a^4 \, {\rm F} ~] 
~~~, \cr
\d_Q \j{}^2 &=~   [~ \a^1 \, (\pa_{\t} {\rm A}) \, + \, \a^2 \, 
(\pa_{\t} {\rm B}) \,-\, \a^3 \, {\rm F} \,+\, \a^4 \, {\rm G}   
~]  ~~~,    \cr
\d_Q \j{}^3 &=~   [~ - \a^1 \, {\rm G}  \,-\, \a^2 \,  {\rm F}  
\,-\, \a^3 \, (\pa_{\t} {\rm B}) \,+\, \a^4 \, (\pa_{\t} {\rm A})
 ~] ~~~, \cr
\d_Q \j{}^4 &=~ [~ \a^1 \, {\rm F} \,-\, \a^2 \, {\rm G} \,+\, 
\a^3 \, (\pa_{\t} {\rm A}) \,+\, \a^4 \, (\pa_{\t} {\rm B})  ~]     
~~~, \cr
\d_Q {\rm F} &=~  i \, \pa_{\t} [~ \a^1 \, \psi^4 \,-\, \a^2 \, 
\psi^3 \,-\, \a^3 \, \psi^2 \,+\, \a^4 \, \psi^1  ~]   ~~~, \cr
\d_Q {\rm G} &=~  i \,  \pa_{\t} [~  - \a^1 \, \psi^3 \,-\, \a^2 
\, \psi^4 \,+\, \a^3 \, \psi^1 \,+\, \a^4 \, \psi^2   ~]  
~~~. }\label{eq:A6.5a} \ee
It is clearly the case that these variations can be written in the 
forms
\be \eqalign{ {~~~~~~~~}
\d_Q {\rm A} &=~  i \, \a{}^{{}_{\rm I}}\, {\rm L{}^{(\rm A)}}{}_{
{}_{\rm I} {}_{\rm K}} \, \psi{}^{{}_{\rm K}}  ~~~, ~~~
\d_Q {\rm B} ~=~ i \, \a{}^{{}_{\rm I}}\, {\rm L{}^{(\rm B)}}{}_{
{}_{\rm I} {}_{\rm K}} \, \psi{}^{{}_{\rm K}}  ~~~,    \cr
\d_Q \j {}^{{}_{\rm K}} &=~ - \, [~ {\rm R{}^{(\rm A)
}}{}^{{}^{\rm K} {}^{\rm L}} \,(\pa_{\t} {\rm A}) \,+\, {\rm R{}^{
(\rm B)}}{}^{{}^{\rm K} {}^{\rm L}} \,(\pa_{\t} {\rm B}) \,+\,
{\rm R{}^{(\rm F)}}{}^{{}^{\rm K} {}^{\rm L}} \, {\rm F} \,+\,
{\rm R{}^{(\rm G)}}{}^{{}^{\rm K} {}^{\rm L}} \, {\rm G} ~] \,
\a{}^{{}_{\rm L}}~~~, \cr
\d_Q {\rm F} &=~  i \, \a{}^{{}_{\rm I}}\, {\rm L{}^{(\rm F)}}{}_{
{}_{\rm I} {}_{\rm K}} \, \pa_{\t} \psi{}^{{}_{\rm K}}  ~~~, ~~~
\d_Q {\rm G} ~=~   i \, \a{}^{{}_{\rm I}}\, {\rm L{}^{(\rm G)}}{}_{
{}_{\rm I} {}_{\rm K}} \, \pa_{\t} \psi{}^{{}_{\rm K}} 
~~~, }\label{eq:A6.6a} \ee
in terms of some constant coefficients (the ${\rm L}$'s and ${\rm R}$'s).
These variations satisfy (\ref{eq:SUSY}) when evaluated on any of the
real fields.  The explicit forms of the  ${\rm L}$ and ${\rm R}$  
quantities are given by
\be \eqalign{ 
\Big( \, {\rm L{}^{(\rm A)}} , \, {\rm L{}^{(\rm B)}}, \, {\rm L{}^{
(\rm F)}} , \, {\rm L{}^{(\rm G)}}   \, \Big) ~&=~   \Big( \,  {{\bf{\rm 
I}}}   \otimes \s^1 , \,  - {{\bf{\rm  I}}}   \otimes \s^3 , \, - 
\s^2\otimes\s^2  , \, - i \s^2 \otimes {{\bf{\rm  I}}}      \, \Big)  
\cr
&=~   \Big( \,  - \, {\rm R{}^{(\rm A)}} , \, - \, {\rm R{}^{(\rm B)}}, 
\, - \, {\rm R{}^{(\rm F)}} , \, {\rm R{}^{(\rm G)}}   \, \Big)  
~~~. }\label{eq:A6.7a} \ee
If we use the symbol $\cal F$ where $\cal F = \{  {(\rm A)} , \, {(\rm B)}, 
\, {(\rm F)}, \, {(\rm G)} \}$ to denote the different bosonic fields 
then the ${\rm L}$'s and ${\rm R}$'s satisfy the algebra of ${\cal G}{\cal 
R}$(4);
\be \eqalign{ 
 {\rm L{}^{\cal F}}{}_{{}_{\rm I} {}_{\rm K}}  {\rm R{}^{{\cal F}'}}
{}_{{}_{\rm K} {}_{\rm L}} ~+~ {\rm L{}^{{\cal F}'}}{}_{{}_{\rm I} 
{}_{\rm K}}  {\rm R{}^{{\cal F}}}{}_{{}_{\rm K} {}_{\rm L}} ~=~ - 2 
\d^{{\cal F} \, {\cal F}'} \, \d{}_{ {}_{\rm I} \, {}_{\rm L}}
 ~~~. }\label{eq:A6.8a} \ee
Consequently, we are forced to conclude that the bosonic fields 
A, B, F and G constitute the same representation of ${\cal GR}$(d, $N)$
as the four components of $\psi{}^{{}_{\rm K}}$ or $\a{}^{{}_{\rm K}}$.  
This is true in spite of the fact that the engineering dimensions of the 
first two bosonic fields are different from that of the later two bosonic 
fields.  The condition (\ref{eq:A6.8a}) insures that the usual supersymmetry 
algebra is obeyed on all the bosonic fields.  

The closure of the algebra on the fermionic fields, however, requires 
something quite different.  In particular, the Fierz identity
\be \eqalign{ 
\sum_{\cal F} \, \Big( \, {\rm L{}^{\cal F}}{}_{{}_{\rm I} {}_{\rm K}}  
{\rm R{}^{{\cal F}}}{}_{{}_{\rm L} {}_{\rm M}} ~+~  {\rm L{}^{\cal F}}
{}_{{}_{\rm M} {}_{\rm K}}  {\rm R{}^{{\cal F}}} {}_{{}_{\rm L} {}_{\rm 
I}}  \, \Big) ~=~ - 2  \, \d{}_{ {}_{\rm I} \, {}_{\rm M}} \, \d{}_{ 
{}_{\rm J} \, {}_{\rm K}}
 ~~~, }\label{eq:A6.9a} \ee
must be satisfied.  Direct calculation using the representation in 
(\ref{eq:A6.7a}) shows that it is.  We emphasize that {\em {not}} all
${\rm L}$'s and ${\rm R}$'s that provide representations of
${\cal GR}$(d, $N)$ satisfy this Fierz condition.

\subsection{4D Chiral Superfield Alternate Embedding}

~~~~In our just concluded discussion, we have seen that the fundamental 
representation of 4D, $N$ = 1 supersymmetry, the chiral multiplet, does 
indeed provide a realization of structures associated with the geometry 
indicated in figure one for ${\cal GR}(4,4)$.   There is also a second 
way to interpret the chiral multiplet, namely it is also a representation 
of  ${\cal {EGR}}(2,2)$.  To begin this demonstration we first introduce 
four $\t$-dependent ``fields'' $\hat\J_{k \hat l}$, $\J_{\hat k l} $, 
$\F_{k l} $ and $\F_{\hat k \hat l} $ so that
\be \eqalign{
\hat\J_{k \, \hat l} &\in \{ {\cal M}_L\} ~~,~~ \J_{\hat k \, l} \in 
\{ {\cal M}_R\} ~~, \cr
\F_{k \, l} &\in \{ {\cal U}_L\} ~\,~~,~\,~ \F_{\hat k \, \hat l} \in 
\{ {\cal U}_R\} ~~. } 
\label{eq:A8.1}   \ee 
so that collectively\footnote{The fields do not, however, completely
saturate these spaces.} these are valued in the entirety of ${\cal {EGR
}}(2,2)$ as indicated in (10).  We next propose as their transformation 
laws,
\be \eqalign{
\d_Q \F_{k \, l} \, &=~ i\a^{{\scriptsize {\rm I}}} \, ({\rm L}^{
{\scriptsize {\rm I}}}){}_{k}{}^{\hat \ell} \, \J_{\hat \ell \, l} 
~+~ i\bar\a^{\scriptsize {\rm I}} \, ({\rm L}^{{\scriptsize {\rm 
I}}}){}_{l}{}^{\hat \ell}\, \Hat \J_{k \, \hat \ell} ~~~,  \cr
\d_Q \J_{\hat k \, l} \, &=~ -\a^{\scriptsize {\rm I}} \, ({\rm R}^{{
\scriptsize {\rm I}}})_{\hat k}{}^{\ell} \, \pa_{\t}  \F_{\ell \,  l}  
~+~  \bar \a^{\scriptsize {\rm I}}\, ({\rm L}^{{\scriptsize {\rm I}}}
)_{l}{}^{\hat \ell} \, \pa_{\t} \Hat \F_{\hat  k \, \hat \ell}   
~~~,   \cr
 \d_Q \Hat\J_{k \,  \hat l} \, &=~ -\a^{\scriptsize {\rm I}} \, ({\rm L
}^{{\scriptsize {\rm I}}})_{k}{}^{\hat \ell} \, \pa_{\t} \Hat\F_{\hat 
\ell \, \hat l} ~-~ \bar\a^{\scriptsize {\rm I}} \, ({\rm R}^{{
\scriptsize {\rm I}}})_{\hat l}{}^{\ell}  \, \pa_{\t}  \F_{k \, 
\ell} ~~~, \cr
\d_Q \Hat\F_{\hat k \, \hat l} \, &=~ i\a^{\scriptsize {\rm I}} \, ({\rm 
R}^{{\scriptsize {\rm I}}})_{\hat k}{}^{\ell} \, \Hat\J_{\ell \, \hat l} 
~-~ i\bar \a^{\scriptsize {\rm I}} \, ({\rm R}^{{\scriptsize {\rm I}}}
)_{\hat l}{}^{\ell} \,  \J_{\hat k \, \ell} } \label{eq:A8.2}   \ee
These variations close under commutation to:
\bea
[\d_1,\d_2] ~=~ -i 2\, (\a^{\scriptsize {\rm I}}_1\a^{\scriptsize 
{\rm I}}_2 ~+~ \bar\a^{\scriptsize {\rm I}}_1 \, \bar\a^{\scriptsize 
{\rm I}}_2) \, \pa_{\t} ~~~,
\label{eq:A8.3}  \eea
In ${\cal GR}(2,2)$ we have the following conventions and identities:
\bea
X^{[{\scriptsize {\rm I}}{\scriptsize {\rm J}}]} ~=~ \e^{{\scriptsize 
{\rm I}}{\scriptsize {\rm J}}} \, \e^{{\scriptsize {\rm M}}{\scriptsize 
{\rm N}}} \, X^{{\scriptsize {\rm M}}{\scriptsize {\rm N}}} ~\equiv~ 
\e^{{\scriptsize {\rm I}}{\scriptsize {\rm J}}}\e\cdot X ~~~,   \cr
f^{{\scriptsize {\rm I}}{\scriptsize {\rm J}}} ~\equiv~ {\fracm 12}
{\rm L}^{[{\scriptsize {\rm I}}}{\rm R}^{{\scriptsize {\rm J}}]} ~=~ 
{\fracm 12}\e^{{\scriptsize {\rm I}}{\scriptsize {\rm J}}}\, \e\cdot 
f~~~, \cr
{\rm L}^{\scriptsize {\rm I}}{\rm R}^{\scriptsize {\rm J}} ~=~ -\d^{
{\scriptsize {\rm I}}{\scriptsize {\rm J}}} \,+\, {\fracm 12}\e^{{
\scriptsize {\rm I}}{\scriptsize {\rm J}}} \, \e\cdot f~~~,~\cr
Tr[\e\cdot f] ~=~ 0~~~,~~~(\e\cdot f)^2 ~=~ -4 {\bf {\rm I}} ~~~.
\label{eq:A8.4}  \eea
We expand the ${\cal EGR}(2,2)$ fields in the following manner:
\be \eqalign{
\F_{k \, l} &=~{\fracm 12}\d_{k \, l} B \,+\, {\fracm 14}(\e\cdot 
f)_{k \, l} \, (\pa_{\t})^{-1}G ~~~, \cr
\Hat\F_{\hat k \, \hat l} &=~ {\fracm 12}\hat\d_{\hat k \hat l}A
\,+\, {\fracm 14}(\e\cdot \hat f)_{\hat k \hat l} (\pa_{\t} )^{-1}
F ~~~,   \cr
\Psi_{\hat k \, l} &=~ -{\fracm 12}{\rm R}^{{\scriptsize {\rm I}}
}_{\hat k  l} \j^{\scriptsize {\rm I}}~~~,~~~~~~~\cr
\Hat\Psi_{k \, \hat l} &=~ -{\fracm 12}{\rm L}^{{\scriptsize {\rm 
I}}}_{k \, \hat l} \hat\j^{\scriptsize {\rm I}}~~~.~~~~~~~
} \label{eq:A8.5}  \ee
We propose that these Clifford algebraic superfields may be regarded as 
``root'' superfields for the 4D, $N$ = 1 chiral multiplet.

Using ${\rm L}^{\scriptsize {\rm I}}$ $=$ $-({\rm R}^{\scriptsize {\rm 
I}})^T$ to do proper matrix multiplication when necessary, we can 
extract the ``component" results from the above variations. 
\be \eqalign{
\label{tada1}
\d_Q A &=~ i \, \a^{\scriptsize {\rm I}}\hat\j^{\scriptsize {\rm I}} 
~+~ i\, \bar\a^{\scriptsize {\rm I}} \j^{\scriptsize {\rm I}}
~~~~~~~~~~~~~~\cr
\d_Q B &= ~ i\, \a^{\scriptsize {\rm I}} \j^{\scriptsize {\rm I}} ~-~ 
i \bar\a^{\scriptsize {\rm I}}\hat\j^{\scriptsize {\rm I}}
~~~~~~~~~~~~~~\cr 
\d_Q F &= ~ -i\a^{\scriptsize {\rm I}}\e^{{\scriptsize {\rm I}}{
\scriptsize {\rm J}}}\pa_{\t} \, \hat\j^{\scriptsize {\rm J}} 
~+~ i\bar\a^{\scriptsize {\rm I}}\e^{{\scriptsize {\rm I}}
{\scriptsize {\rm J}}}\pa_{\t} \,\j^{\scriptsize {\rm J}}
~~~~~~~~\cr
\d_Q G &=~ -i\a^{\scriptsize {\rm I}}\e^{{\scriptsize {\rm I}}
{\scriptsize {\rm J}}}\pa_{\t} \, \j^{\scriptsize {\rm J}} 
~-~ i\bar\a^{\scriptsize {\rm I}}\e^{{\scriptsize {\rm I}}
{\scriptsize {\rm J}}}\pa_{\t} \, \hat\j^{\scriptsize {\rm 
J}}~~~~~~~~\cr
\d_Q \j^{\scriptsize {\rm I}} &=~ \a^{\scriptsize {\rm I}}\pa_{\t} 
\, B ~+~ \bar\a^{\scriptsize {\rm I}}\pa_{\t} \, A ~-~ \a^{
\scriptsize {\rm J}} \e^{{\scriptsize {\rm J}}{\scriptsize {\rm 
I}}}G ~+~ \bar\a^{\scriptsize {\rm J}}\e^{{\scriptsize {\rm J}}
{\scriptsize {\rm I}}} \,F  \cr
\d_Q\hat\j^{\scriptsize {\rm I}} &=~  \a^{\scriptsize {\rm I}}\pa_{\t} \, A 
~-~ \bar\a^{\scriptsize {\rm I}} \, \pa_{\t} B ~-~ \a^{\scriptsize {\rm J}}
\e^{{\scriptsize {\rm J}}{\scriptsize {\rm I}}} \, F ~-~ \bar\a^{
\scriptsize {\rm J}} \e^{{\scriptsize {\rm J}}{\scriptsize {\rm I}}}
 \, G
 } \label{eq:A8.6}  \ee
These variations are exactly those that appear in (\ref{eq:A6.5a}) after 
the following redefinitions:
\bea
&~~~~~~~\,\a^1\rightarrow\bar\a^1~~~,~~~\, \a^2\rightarrow\a^1 \,~~~,~~~
~ \a^3\rightarrow\bar\a^2~~~,~~~ \a^4\rightarrow\a^2
~~~,~~~~\cr
&~~~~\j^1\rightarrow\hat\j^1~~~,~~~\j^2\rightarrow\j^1 ~~~~,~~~\,
\j^3\rightarrow\hat\j^2 ~~~,~~~\j^4\rightarrow\j^2  ~~~.
\label{eq:A8.7}  \eea

One of the interesting points about this embedding of the transformations 
in (\ref{eq:A6.5a}) into representations of ${\cal {EGR}}(2,2)$ is 
that it suggests that a non-trivial role may be played by the $\cal 
D$-element associated with this algebra.  In particular, the expansions 
in (\ref{eq:A8.5}) are not {\em {not}} unique.  Referring back to the 
results in Table I, we see that ${\cal {EGR}}(2,2)$ is one of the almost 
complex cases.  This implies that we can utilize the alternate expansion 
given by
\be \eqalign{
\F_{k \hat l}^{\prime} &=~ {\fracm 12}{\cal D}_{k \hat l} \, B{}^{\prime} 
\,+\, {\fracm 14}(\e\cdot f \, {\cal D})_{k \hat l} \, (\pa_{\t})^{-1}
G{}^{\prime} ~~~, \cr
\Hat\F_{\hat k l}^{\prime} &=~ {\fracm 12} {\Hat {\cal D}}_{\hat k l} \,
A{}^{\prime} \,+\, {\fracm 14}(\e\cdot \hat f \, {\Hat {\cal D}})_{
\hat k l} (\pa_{\t} )^{-1}F{}^{\prime} ~~~,   \cr
\Psi_{k l}^{\prime} &=~ -{\fracm 12}({\rm L}^{ {\scriptsize {\rm I}}}
\, {\Hat {\cal D}} )_{k l}\,  \j^{\prime  \, \scriptsize {\rm I}}~~~,
~~~~~~~ \cr
\Hat \Psi_{\hat k \hat l}^{\prime} &=~ -{\fracm 12}({\rm R}^{ {\scriptsize 
{\rm I}}} \, {\cal D})_{\hat k \hat l} \, {\hat \j}^{\prime  \, \scriptsize 
{\rm I}}~~~,~~~~~~ 
} \label{eq:A8.8}  \ee
where we use the prime superscript to distinguish these fields from
those that appear in (\ref{eq:A8.5}).

In order to find the form of the supersymmetry transformation laws
for this representation, we propose the ansatz below.
\be \eqalign{
\d_Q \Phi_{k \, \hat l}^{\prime} \, &=~ i \a^{\scriptsize {\rm I}} \, ({\rm 
L}^{{\scriptsize {\rm I}}})_{~k}{}^{\hat \ell} \, \Hat\J_{\hat \ell \,
\hat l}^{\prime} ~+~ i \bar\a^{\scriptsize {\rm I}} \, ({\rm R}^{{
\scriptsize {\rm I}}})_{\hat k} {}^{\ell} \, \J_{k \ell}^{\prime} 
~~~, \cr
\d_Q \J_{k \, l}^{\prime} \, &=~ \a^{{\scriptsize {\rm I}}} \, 
({\rm L}^{\scriptsize {\rm I}})_{k} {}^{\hat \ell} \, 
\pa_{\t} {\Hat \Phi}{}_{\hat \ell \, l}^{\prime} ~-~ 
\bar\a^{\scriptsize {\rm I}} \, 
({\rm L}^{\scriptsize {\rm I}})_{l}{}^{\hat \ell} \, 
\pa_{\t}  \Phi_{k \, \hat \ell}^{\prime} ~~~, \cr
\d_Q \Hat\J_{\hat k \, \hat l}^{\prime} \, &=~ - \a^{\scriptsize {\rm 
I}} \, ({\rm R}^{{\scriptsize {\rm I}}})_{\hat k} {}^{\ell}\, \pa_{\t} 
\Phi_{\ell \, \hat l}^{\prime} ~-~ \bar \a^{\scriptsize {\rm I}} \, ({\rm 
R}^{{\scriptsize {\rm I}}})_{\hat l} {}^{\ell} \, \pa_{\t} \Hat \Phi_{
\hat k \, \ell}^{\prime} ~~~,  \cr
\d_Q  \Hat \Phi_{\hat k \, l}^{\prime} &= -\, i \a^{\scriptsize {\rm I}} \, 
({\rm R}^{{\scriptsize {\rm I}}})_{\hat k} {}^{\ell} \,  \J_{\ell \, l}^{\prime}  
~+~ i \bar \a^{\scriptsize {\rm I}}\, ({\rm L}^{{\scriptsize {\rm I}}
})_{l} {}^{\hat \ell} \, \Hat \J_{\hat k \, \hat \ell}^{\prime}  ~~~.
} \label{eq:A8.9}   \ee
Now it is a fact that when the expansions in (\ref{eq:A8.8}) are 
substituted into the variations in (\ref{eq:A8.9}), the component fields in
(\ref{eq:A8.8}) are found to obey {\em {exactly}} the same transformation
laws as appear in  (\ref{eq:A8.6}).  In other words, this is an alternate
description of the same multiplet.

This leads to a minor dilemma.  If the descriptions obtained previously
describe the chiral multiplet, how then is the ``anti-chiral multiplet'' to
be described?   A simple solution is provided by using local and
 non-local re-definitions as appear in (\ref{eq:5.5}).  In particular we
introduce new expansions 
\be \eqalign{
\F_{k l}^c  &=~{\fracm 12}\d_{k l} \, (\pa_{\t} )^{-1}G^c \,+\, {\fracm 14}
(\e\cdot f)_{k l} \, B^c ~~~, \cr
\hat\F_{\hat k \hat l}^c &=~ {\fracm 12}\hat\d_{\hat k \hat l} \, (\pa_{\t} 
)^{-1}F^c \,+\, {\fracm 14}(\e\cdot \hat f)_{\hat k \hat l} \, A^c ~~~,   \cr
\Psi_{\hat k l}^c &=~ -{\fracm 12}{\rm R}^{{\scriptsize {\rm I}}}_{\hat k l}
\j^{c ~ \scriptsize {\rm I}}~~~,~~~~~~~\cr
\hat\Psi_{l  \hat k}^c &=~ -{\fracm 12}{\rm L}^{{\scriptsize {\rm I}}}_{l 
\hat k} \hat\j^{c ~ \scriptsize {\rm I}}~~~.~~~~~~~
} \label{eq:A8.10}  \ee
and use the variations from (\ref{eq:A8.2})  The motivation for this
can be seen by a review of the relation between chiral and anti-chiral
multiplets in 4D.

In 4D, $N$ = 1 superspace, the component fields ($A$, $B$, $\psi_{\a}$,
$F$ and $G$) of a chiral multiplet may be defined by the $D$-expansion
of a chiral superfield
\be \eqalign{
A(x) ~+~ i B(x) ~\equiv~  \Phi {\Huge |} ~~,~~ \psi_{\a}(x)
 ~\equiv~  D_{\a}\Phi {\Huge |} ~~,~~  F(x) ~+~ i G(x)  ~\equiv~  
D{}^2 \Phi {\Huge |} ~~~.
} \label{eq:A8.10z}  \ee
But the component fields ($A^c$, $B^c$, ${\Bar \psi}{}^c_{\Dot \a}$,
$F^c$ and $G^c$) of an {\em {anti}}-{\em {chiral}} multiplet can also
be defined from the same superfield
\be \eqalign{
A^c(x) ~+~ i B^c(x)  ~\equiv~  D{}^2 \Phi {\Huge |}  ~~,~~
{\Bar \psi}{}^c_{\Dot \a}(x) ~\equiv~  i \, \pa_{\un a} {D}{}^{\a}\Phi {\Huge |} 
~~,~~  F^c(x) ~+~ i G^c(x) ~\equiv~  {}_\bo \, \Phi {\Huge |} ~~~
} \label{eq:A8.10y}  \ee
In particular, it can be seen that the spinor in the anti-chiral multiplet
is related to the spinor in the chiral multiplet by
\be \eqalign{
{\Bar \psi}{}^c_{\Dot \a}(x) ~=~  i \, \pa_{\un a} \, \psi^{\a}(x)  ~~~.
} \label{eq:A8.10x}  \ee

Using the expansions in (\ref{eq:A8.10}) and the variations in (\ref{eq:A8.3})
leads to the component results
\be \eqalign{
\label{tada2}
{~~~~~~} \d A^{c} &= ~ -i \a^{\scriptsize {\rm I}}\e^{{\scriptsize {\rm I
}}{\scriptsize {\rm J}}} \, \hat\j^{c~ \scriptsize {\rm J}} 
~+~ i \bar\a^{\scriptsize {\rm I}}\e^{{\scriptsize {\rm I}}{\scriptsize 
{\rm J}}} \,\j^{c~ \scriptsize {\rm J}}~~~~~~~~\cr
\d B^{c} &=~ -i \a^{\scriptsize {\rm I}}\e^{{\scriptsize {\rm I}}
{\scriptsize {\rm J}}} \, \j^{c~ \scriptsize {\rm J}} ~-~ i
\bar\a^{\scriptsize {\rm I}}\e^{{\scriptsize {\rm I}}{\scriptsize {\rm 
J}}} \, \hat\j^{c~ \scriptsize {\rm J}}~~~~~~~~\cr
\d F^{c} &=~ i \, \a^{\scriptsize {\rm I}} \, \pa_{\t}  \hat\j^{c~
 \scriptsize {\rm I}} ~+~  i\, \bar\a^{\scriptsize {\rm I}} \,
\pa_{\t} \j^{c~ \scriptsize {\rm I}}~~~~~~~~~~~~~~\cr
\d G^{c} &= ~  i\, \a^{\scriptsize {\rm I}} \, \pa_{\t} \j^{c~ \scriptsize 
{\rm I}}~-~ i\bar\a^{\scriptsize {\rm I}} \, \pa_{\t} \hat\j^{c~ \scriptsize 
{\rm I}}~~~~~~~~~~~~~~\cr 
\d \j^{c~ \scriptsize {\rm I}} &=~  -\,  \a^{\scriptsize {\rm J}} 
\e^{{\scriptsize {\rm J}}{\scriptsize {\rm I}}} \, \pa_{\t} B^{c} ~+~ 
 \bar\a^{\scriptsize {\rm J}}\e^{{\scriptsize {\rm J}}{\scriptsize {\rm 
I}}} \, \pa_{\t} A^{c} ~+~\a^{\scriptsize {\rm I}} \, G^{c} ~+~ 
 \bar\a^{\scriptsize {\rm I}} \, F^{c}  \cr
\d\hat\j^{c~ \scriptsize {\rm I}} &=~  -\,  \a^{\scriptsize {\rm J}}\e^{{
\scriptsize {\rm J}} {\scriptsize {\rm I}}} \, \pa_{\t} A^{c} ~-~  \bar\a^{ 
\scriptsize {\rm J}} \e^{{\scriptsize {\rm J}} {\scriptsize {\rm I}}} \, 
\pa_{\t} B^{c} ~+~  \a^{\scriptsize {\rm I}}  \, F^{c}  ~-~  \bar\a^{\scriptsize 
{\rm I}} \, G^{c} 
} \label{eq:A8.10ww}  \ee
Now the transformation laws in (\ref{eq:A8.10ww})  are related to those
in (\ref{eq:A8.2}) via the following field re-definitions.
\be \eqalign{
\pa_{\t} \, A \, &=~  F^c   ~~,~~  \pa_{\t} \, \psi{}^{\scriptsize {\rm 
I}} ~=~ \psi{}^{c \, \scriptsize {\rm I}} ~~,~~ F \, ~=~ \pa_{\t} \, A^c  
~~~,  \cr
\pa_{\t} \, B \, &=~ G^c ~~,~~  { \pa}{}_{\t} \, {\hat \psi}{}^{\scriptsize 
{\rm I}} ~=~  {\hat \psi}{}^{c \, \scriptsize {\rm I}} ~~,~~ G \, ~=~ \pa_{
\t} \, B^c  ~~~. \cr
} \label{eq:A8.10wv}  \ee
It is obvious now that spinors in our proposed 1D, $N$ = 4 ``chiral multiplet''
are related to the spinors in our proposed 1D, $N$ = 4 ``anti-chiral multiplet''
in a manner similar to the 4D, $N$ = 1 theory.  All of this suggests that the 
representation in (\ref{eq:A8.10}) is the root superfield representation of the
usual anti-chiral multiplet familiar from 4D, $N$ = 1 supersymmetry.

\subsection{Higher Rank ${\cal {EGR}}(2,2)$ Representations}

~~~~Let us call the quartet of fields
\be \eqalign{
&~~~~~~~~~ \F_{k_1 \, k_2}   \cr
&~~\, \J_{{\hat k}_1 \, k_2} ~~~~~~~ \Hat \J_{k_1 {\hat k}_2 }  \cr
&~~~~~~~~~\Hat\F_{{\hat k}_1 \, {\hat k}_2 } }
\label{eq:A8.11}   \ee
a rank two representation.  Having seen evidence this quartet of fields 
forms a 1D, $N$ = 4 representation that is closely related to the 4D, $N$ = 1
chiral multiplet suggests that there should exist even more complicated
1D, $N$ = 4 ${\cal {EGR}}(2,2)$ representations related to other 4D, $N$ = 1 
supermultiplets.  Let us discuss the general outline of this possible relation.   
As a first step let us introduce an alternate geometrical interpretation
of the quartet above.
We can consider a cartesian product space of the form
\be \eqalign{
{\cal P}_2 ~&\equiv~  {\cal V}_L^{(1)} \, \times \, {\cal V}_L^{(2)}  \,
\times \,  {\cal V}_R^{(1)} \, \times \, {\cal V}_R^{(2)} 
~~~.}
\label{eq:A8.11a}   \ee
Next we introduce a set of maps that act on rank-two projections of this space
into the real numbers defined by
\be \eqalign{
\F_{k_1 \, k_2}: \, {\cal V}_L^{(1)} \, \times \, {\cal V}_L^{(2)} \,& \,
\longrightarrow \, {\mathfrak R}{}^1  ~~~,    \cr
\J_{{\hat k}_1 \, k_2}: \,  {\cal V}_R^{(1)} \, \times \, {\cal V}_L^{(2)}  
\,& \, \longrightarrow \, {\mathfrak R}{}^1 ~~~,   \cr
\Hat \J_{k_1 {\hat k}_2 }: \, {\cal V}_L^{(1)} \, \times \, {\cal V}_R^{(2)}     
\,& \, \longrightarrow \, {\mathfrak R}{}^1 ~~~,    \cr
\Hat\F_{{\hat k}_1 \, {\hat k}_2 }:\,  {\cal V}_R^{(1)} \, \times \, {\cal 
V}_R^{(2)}  \,& \, \longrightarrow \, {\mathfrak R}{}^1   
~~~.}
\label{eq:A8.11b}   \ee
These coefficients are precisely the quartet that appears in (\ref{eq:A8.11}).

The generalization of this construction is obvious.  These begin by the
introduction of 
\be \eqalign{
{\cal P}_n ~&\equiv~  {\cal V}_L^{(1)} \, \times \, \dots \, \times 
\, {\cal V}_L^{(n)}  \, \times \, {\cal V}_R^{(1)} \, \times \, \dots \,
\times \, {\cal V}_R^{(n)} 
~~~.}
\label{eq:A8.11c}   \ee
Next one can consider all rank $n$ tensors that map rank $n$ projections of
${\cal P}_n$ into ${\mathfrak R}{}^1$.  For $n$ = 3, this leads to the fields
\be \eqalign{
&~~~~~~~~~~~~~~~~~ \F_{k_1 \, k_2 \, k_3 }   \cr
&~~\, \J_{{\hat k}_1  \, k_2 \, k_3  } ~~~~~ \J_{k_1 \, {\hat k}_2 
\, k_3}  ~~~~~  \J_{k_1 \, k_2 \, {\hat k}_3}  \cr
&~~\, \J_{{\hat k}_1  \, {\hat k}_2 \, k_3  } ~~~~~ \J_{{\hat k}_1 \, k_2 
\, {\hat k}_3}  ~~~~~  \J_{k_1 \, {\hat k}_2 \, {\hat k}_3}  \cr
&~~~~~~~~~~~~~~~~~ \F_{{\hat k}_1 \, {\hat k}_2 \, {\hat k}_3 }  ~~~,}
\label{eq:A8.12}   \ee
or the $n$ = 4 representation of the form\footnote{Since the representation in 
(\ref{eq:A8.11}) seems to correspond to a chiral multiplet, the representation 
in \newline ${~~~~~}$ (\ref{eq:A8.13})
seems appropriate to describe the usual 4D, $N$ = 1 vector multiplet.
}
\be \eqalign{
&~~~~~~~~~~~~~~~~~~~~~~~~~~~~~~~~~~~~~~~~~~~~~~~~~ \F_{k_1 \, k_2 \, k_3 \, 
k_4 }   \cr
&~~~~~~~~~~~~~~~~~~~\, \J_{{\hat k}_1  \, k_2 \, k_3  \, k_4 } ~~~~~ \J_{k_1 
\, {\hat k}_2 \, k_3  \, k_4}  ~~~~~  \J_{k_1 \, k_2 \, {\hat k}_3  \, k_4} 
 ~~~~~  \J_{k_1 \, k_2 \, k_3 \, {\hat k}_4}  \cr 
&~~~~~~~~\, \F_{{\hat k}_1  \, {\hat k}_2 \, k_3 \, k_4  } ~~~ \F_{{\hat k}_1 
\, k_2 \, {\hat k}_3  \, k_4 }  ~~~  \F_{{\hat k}_1 \, k_2 \, k_3 \, {\hat 
k}_4  }~~~ \F_{k_1 \, {\hat k}_2 \, {\hat k}_3 \, k_4 }  ~~~  \F_{k_1 \, 
{\hat k}_2 \, k_3  \, {\hat k}_4 } ~~~ \F_{k_1 \, k_2 \, {\hat k}_3 \, 
{\hat k}_4 }  \cr
&~~~~~~~~~~~~~~~~~~~\, \J_{{\hat k}_1  \, {\hat k}_2 \, {\hat k}_3  \, k_4 } 
~~~~~ \J_{{\hat k}_1 \, {\hat k}_2 \, k_3  \, {\hat k}_4}  ~~~~~  
\J_{{\hat k}_1 \, k_2 \, {\hat k}_3  \, {\hat k}_4} 
 ~~~~~  \J_{k_1 \, {\hat k}_2 \, {\hat k}_3 \, {\hat k}_4}  \cr
&~~~~~~~~~~~~~~~~~~~~~~~~~~~~~~~~~~~~~~~~~~~~~~~~~ \F_{{\hat k}_1 \, {\hat k}_2 \, 
{\hat k}_3 \, {\hat k}_4 }  }
\label{eq:A8.13}   \ee
(On each row, the number of fields is simply determined by the binomial coefficients.)
These simple considerations lead us to suggests that {\em {only}} the {\em {even}} 
rank representations can correspond to 4D, $N$ = 1 superfields.  The next question 
becomes whether there exist for (\ref{eq:A8.12}) and (\ref{eq:A8.13}) a set of variations 
analogous to (\ref{eq:A8.2}) that satisfy, for these representations, the condition 
in (\ref{eq:A8.3})?  We will not attempt to prove this but are confident that this
is the case.

Our arguments suggest that the vector multiplet appears as a root superfield
within the rank four structure associated with various representations 
of ${\cal {EGR}}(2,2)$.   The next highest even-rank structures, the rank six
tensors, would therefore be associated with matter gravitino multiplets
\cite{MGM}.  Finally the rank eight structures must contain the root representation
of the 4D, $N$ = 1 supergravity multiplet.  However, the supersymmetry variations 
of the root superfield representation of both the matter gravitino multiplet and super
gravity multiplet must be of the type that was suggested for the spinning
particle models of \cite{jim1,jim2}.

One possible use for these higher rank root superfields is for counting the degrees 
of freedom of higher dimensional supersymmetric irreducible multiplets.  Since 
dimensional reduction on a cylinder is well defined, any higher dimensional multiplet 
must be related to some ${\cal GR}$ representation.  In order to do this counting, 
we must know the total number of components of the supersymmetry parameters.  
For example, in $D=4, N=1$ there are four components in the supersymmetry parameter.  
This means that we are looking for a $N=4$ one dimensional representation.  This 
can be obtained by using ${\cal GR}(4,4)$, or by using ${\cal EGR}(2,2)$.  Both 
of these descriptions had four supersymmetry parameters.  This also tells us that 
there are no smaller representations then the chiral field in four dimensions.  

Could we do this for $N=2$ in 4D or even higher 4D, $N$-extended representations?  
At present this is not clear.  However, a hopeful starting point would be to
begin with structures associated with ${\cal EGR}(2N,2N)$.

\section{Other Applications}

~~~~In the following sections, we will present some concrete applications
of the results discussed in this work. We should point out that these 
examples have appeared before in very brief discussions.  So part of the 
benefit of our return to these is to give a much fuller description.

\subsection{Off-shell 1D, $N$-extended Spinning Particles}

~~~~We turn now to the discussion of free spinning particle (which was 
our original motivation for introducing these algebraic structures).  
Utilizing the concise ${\cal {GR}}$(d, $N)$  notation this action takes 
the form
\be \eqalign{
{\cal S}_{Spng~Part.}~=~ \int d \t \, [ &- \, i \fracm 12 {\rm d}^{-1} 
\, {({\cal X}{}^t)}{}_{\hat k} {}^{i} \, {\pa}_{\t} {\cal X}_i {}^{\hat 
k} \,-\, i  \fracm 12 \, \pi_{{\,}_{\rm I}} \, {\pa }_{\t} \pi_{{\,}_{
\rm I}}  \,-\, \frac 12 {\rm P}^2 \,-\, \frac 12 {\rm d}^{-1} \, ( ({
\cal G}{}^t)_i{}^j {\cal  G}_i {}^j )  \cr 
&- i \, \Psi_{{\,}_{\rm I}} (\, {\pa}_{\t} \pi_{{\,}_{\rm I}} \, )
  \,+\, {\rm P} (\, {\pa}_{\t} {\rm X} \, ) \,+\, {\rm d}^{-1} {\cal 
G}_i {}^j {\cal F}_j {}^i \,+\, i {\rm d}^{-1} {\cal X}_i {}^{\hat k} 
\L_{\hat k} {}^i  ~ ]
 ~~~~.   \cr}
\label{eq:8.1}
\ee
As described in \cite{jim1,jim2}, this action is actually constructed 
from two distinct multiplets; the USPM (the Universal Spinning Particle
Multiplet; ${\rm X},\, \Psi_{{\,}_{\rm I}}, \, {\cal F}_i{}^j,\, \L_{\hat
k}{}^i)$  and its canonically conjugate momentum multiplet $(\p_{{\,
}_{\rm I}},\,  {\cal X}_i {}^{\hat k},\, {\rm P}, \, {\cal G}_i{}^j
)$.  The fields (${\rm X},\, {\cal F}_i{}^j)$ $\oplus (\Psi_{{\,}_{\rm 
I}},\, \L_{\hat k}{}^i)$ form a representation of the normal part of
${\cal U}_L \oplus {\cal M}_R$ and the fields $({\rm P}, \, {\cal 
G}_i{}^j) \oplus ( \p_{{\, }_{\rm I}},\,  {\cal X}_i{}^{\hat k} )$ 
form a representation of the normal part of ${\cal U}_R 
\oplus {\cal M}_L$. 

We have also used the result described in the fifth chapter.  In our
original presentation of these results \cite{jim1,jim2} we described how 
it was necessary to begin with a Cartesian product of the ``iso-spinning
particle'' with a space isomorphic to ${\cal V}_L$.  The iso-spinning 
particle is geometrically described by the superfield and transformation 
laws that appear in (\ref{eq:algSUSY}).   After taking the Cartesian 
product, we arrive at the superfield and and transformation laws that 
appear in (\ref{eq:4.1}).   Finally in our earlier works \cite{jim1,
jim2}, we wrote a theory that is described by a the superfield  of 
the form that appears in (\ref{eq:5.8}).  The exponents were chosen in 
such a way that only the 0-brane coordinate $X(\t)$ and NSR fermions 
$\Psi {}_{{}_{\rm I}}(\t)$ are propagating degrees of freedom.

The supersymmetry transformation laws of the spinning particle
supermultiplets that leave  the action invariant are given by
\bea
{\d}_{Q} \, {\rm X} &=& i \a^{{\,}_{\rm I}} \, \Psi_{{\,}_{\rm I}} 
 ~~~, \cr 
{\d}_{Q} \, \Psi_{{\,}_{\rm I}} &=& - \, [~ \a_{{\,}_{\rm I}} \, 
(\pa_{\t} {\rm X}  ) ~+~ d^{-1} \a^{{\,}_{\rm J}} (f_{{\,}_{{\rm I}\, 
{\rm J}}})_i {}^j {\cal F}_j  {}^i  ~]  ~~~, \cr
{\d}_{Q} \,  {\cal F}_i {}^{\, j}  &=& i  \a^{{\,}_{\rm I}} \, (f_{{\,
}_{{\rm I}\, {\rm K}}})_i {}^j  (\pa_{\t} \Psi_{{\,}_{\rm K}}  )
~+~ i \a^{{\,}_{\rm K}} \, ({\rm  L}_{{\,}_{\rm K}})_i {}^{\hat k} 
\L_{\hat k} {}^j    ~~~ , \cr  
{\d}_{Q} \, \L_{\hat k} {}^j  &=&  \a^{{\,}_{\rm K}} \,\pa_{\t} 
\, [~ ({\rm   R}_{{\,}_{\rm K}})_{\hat k} {}^l {\cal F}_l {}^{\, j}
  ~+~ d^{-1} ({\rm R}^{{\,}_{\rm  I}})_{\hat k}{}^j (f_{{\,
}_{{\rm I}\, {\rm K}}})_j {}^l  {\cal F}_l {}^{\, k}  ~]  ~~~,
\cr    
{\d}_{Q} \, \p_{{\,}_{\rm I}}   &=&   \a_{{\,}_{\rm I}} \, {\rm
P}  ~+~ {\rm d}^{-1} \a_{{\,}_{\rm K}} \left( f_{{\,}_{{\rm K}\,
{\rm I}}} \right)_j {}^{i} \, {\cal G}_{i} {}^{j}  ~~~~, \cr 
{\d}_{Q} \, {\cal X}_i {}^{\hat k}  &=& -  \, \a_{{\,}_{{\rm K}}}
\left({\rm L }_{{\,}_{\rm K}} \right)_{k} {}^{\hat k}  \, {\cal G}_{i}
{}^{k} \, + \,  {\rm d}^{-1} \a_{{\,}_{\rm K}} \left( {\rm L}_{{\,
}_{\rm I}} \right)_i {}^{\hat k} \, \left( f_{{\,}_{{\rm I}\, {\rm K}}}
\right)_k {}^{l} \, {\cal G}_{l} {}^k   ~~~~, \cr 
{\d}_{Q} \, {\rm P}  &=& -i \,  \a_{{\,}_{\rm I}} \, {\pa}_{\t}
\p_{{\,}_{\rm I}}    ~~~~, \cr
{\d}_{Q} \, {\cal G}_{i} {}^{j}  &=& - i   \,  {\pa}_{\t} \, [~
\a_{{\,}_{ \rm J}} \left(  f_{{\,}_{{\rm I}\,{\rm J}}} \right)_{i} 
{}^{j} \, \p_{{\,}_{\rm I}} ~+~ \a_{{\,}_{\rm K}} \left( {\rm
R}_{{\,}_{\rm  K}} \right)_{\hat k} {}^{j} \, {\cal X}_i  {}^{\hat k} 
~]  ~~~~,
\label{eq:8.2}
\eea
and where ${\cal F}_{i} {}^i = \left( {\rm L}_{{\,}_{\rm I}} \right)_j 
{}^{\hat k} \L_{\hat k} {}^j = {\cal G}_{i} {}^i = \left({\rm
R}_{{\,}_{\rm I}} \right)_{\hat  k} {}^i {\cal X}_i {}^{\hat k} = 0$.  
These are equivalent to the transformation laws given in (\ref{eq:4.1}).
A more geometrical way to understand these transformation laws is to
note that the quantities $\F_1$, $\J_1$, $\F_2$, $\J_2$ defined by
\be \eqalign{ {~~~}
(\F_1) {}_{i k}  &\equiv~  {\rm X} \, {\bf I}{}_{i k} ~+~ [ \,
(\pa_{\t})^{-1} \, {\cal F}  {}_{ i \hat k} \, ]  ~~~, ~~~
(\J_1 ) {}_{\hat k i } ~\equiv~  \Psi_{{\,}_{\rm I}} \,
(\Hat f{}_{{\,}_{\rm I}} ){}_{\hat k i}  ~+~ [ \,
(\pa_{\t})^{-1} \L{}_{\hat k i}
\, ]  ~~~,  \cr
(\F_2) {}_{i k}  &\equiv~  {\rm P} \, {\bf I}{}_{i k} ~+~
 {\cal G}  {}_{i k}   ~~~~~~~~~~~~~~~,  ~~~ 
(\J_2 ) {}_{i \hat k} ~\equiv~  \pi_{{\,}_{\rm I}} \,
(f{}_{{\,}_{\rm I}} ){}_{i \hat k}  ~+~ {\cal X} {}_{i \hat k}
  ~~~, 
}\label{eq:8.2a}
\ee
are such that $\F_2$ and $\J_2$ have exactly the transformation law
given in (\ref{eq:4.1}).  Similarly $\F_1$ and $(\J_1){}^t$ also have
exactly the transformation law given in (\ref{eq:4.1}).  

The action in (\ref{eq:8.1}) can be somewhat simplified by making three 
field redefinitions
\bea
{\cal F}_i {}^j ~\to~ {\cal F}_i {}^j ~+~ \fracm 12 {\cal G}_i {}^j ~~~,
~~~ (\L^t){}_i {}^{\hat k} ~\to~ (\L^t){}_i {}^{\hat k} ~+~ \pa_{\t} {\cal
X}_i {}^{\hat k}  ~~~,~~~ \Psi_{{\,}_{\rm I}} ~\to~ \Psi_{{\,}_{\rm I}}
\,-\,
\pi_{{\,}_{\rm I}} ~~~,
\label{eq:8.3}
\eea
so that the redefined action is
\be \eqalign{
{\cal S}_{Spng~Part.} ~=~  \int d \t \, [ &-~ \frac 12 {\rm P}^2  ~+~ 
{\rm P}(\, {\pa}_{\t} {\rm X} \, ) ~-~ i \, \Psi_{{\,}_{\rm I}} \, (\,
{\pa}_{\t} \pi_{{\,}_{\rm  I}} \, ) ~{~~~~~~~~~~} \cr
&+~ {\rm d}^{-1}{\cal G}_i {}^j {\cal F}_j {}^i ~+~ i {\rm d}^{-1}{\cal
X}_i {}^{\hat  k} \L_{\hat k}{}^i ~ ]
 ~~~~.  }
\label{eq:8.4}
\ee
The bosonic fields (${\cal F}_i {}^j , \, {\cal
G}_i {}^j$) and fermionic fields ($\L_{\hat k}{}^i , \, {\cal X}_i{}^{
\hat k }$) are to be expanded as elements {\em {only}} of the normal 
parts of the enveloping algebras\footnote{This restriction to the normal
parts of the enveloping algebra has not been stated in our \newline
${~~~~\,}$ previous works on these models.},  subject to the algebraic 
constraints given below
(\ref{eq:8.2}).
\bea
{\cal F}_i {}^j  &=&  \fracm 1{2} (f^{\rm {{\,}_{I_1 I_2}}})_i {}^j \, 
{F}_{{\,}_{\rm {I_1 I_2}}}  ~+~ \fracm 1{4!} (f^{{\,}_{\rm {I_1 I_2 I_3
I_4}}})_i {}^j \,  {F}_{{\,}_{\rm {I_1 I_2 I_3 I_4}}}  ~+~ \cdots \cr 
&\,&~+~ \fracm 1{~(2{\rm p})!}(f^{{\,}_{\rm {I_1 \dots I_{2p}}}})_i {}^j 
\, {F}_{{\,}_{\rm {I_1 \dots I_{2p}}}} ~~~, \cr
{\L}_{\hat k} {}^{j}  &=& \fracm 1{3!} (f^{{\,}_{\rm {I_1 I_2  I_3}}}
)_{\hat k}{}^{j} \, \l_{{\,}_{{\rm {I_1 I_2 I_3 }}}} ~+~ \cdots ~+~
\fracm 1{~(2{\rm q} +1) !}(f^{{\,}_{\rm {I_1 \dots I_{2q + 1}}}})_{\hat
k} {}^{j} \, \l_{{\,}_{{\rm {I_1 \dots I_{2q + 1}}}}} ~~~, \cr 
{\cal G}_i {}^j  &=& \fracm 12 (f^{{\,}_{\rm {I_1 I_2}}})_i {}^j \,
{G}_{{\,}_{\rm  {I_1 I_2}}}  ~+~ \fracm 1{4!} (f^{{\,}_{\rm {I_1 I_2 I_3
I_4}}})_i {}^j \, {G}_{{\,}_{\rm {I_1 I_2 I_3 I_4}}}  ~+~ \cdots \cr 
&\,&~+~ \fracm 1{~(2{\rm p})!}(f^{{\,}_{\rm {I_1 \dots I_{2p}}}})_i {}^j 
\, {G}_{{\,}_{\rm  {I_1 \dots I_{2p}}}} ~~~, \cr
{{\cal X}}_{i} {}^{\hat k}  &=& \fracm 1{3!} (f^{{\,}_{\rm {I_1 I_2  
I_3}}})_i {}^{\hat k} \, {{\chi}}_{{\,}_{{\rm {I_1 I_2 I_3}}}} ~+~
\cdots ~+~ \fracm  1{~(2{\rm q} +1) !}(f^{{\,}_{\rm {I_1 \dots I_{2q +
1}}}})_{i} {}^{\hat k} \, {{\chi}}_{{\,}_{{\rm {I_1 \dots I_{2q + 1}}}}}
~~~. 
\label{eq:8.5}
\eea

When these expansions are inserted into the spinning particle action it 
is seen to lead to
\bea
{\cal S}_{Spng~Part.} ~=~  &\int d \t \, \Big[ ~-\, \frac 12 {\rm P}^2 
\,+\, {\rm P} (\, {\pa}_{\t} {\rm X} \, ) \,-\, i \, \Psi_{{\,}_{\rm I}} 
\, (\, {\pa}_{\t} \pi_{{\,}_{\rm  I}} \, ) \,+\, \sum_{\ell}  c_{2 \ell} 
\, G_{{\,}_{\rm {I_1 \dots I_{2 \ell}}}} \, F_{{\,}_{\rm {I_1 \dots I_{2
\ell}}}} \cr  
&{}+\, i \,\sum_{\ell}  c_{2 \ell + 1} \, \chi_{{\,}_{\rm 
{I_1 \dots I_{2 \ell + 1}}}} \, \l_{{\,}_{\rm {I_1 \dots I_{2 \ell + 
1}}}} \,~ \Big] ~~~~,
\label{eq:8.6}
\eea
here the values of the coefficients $c_{2\ell}$ and $c_{2\ell +1}$ can
be found by using  the trace relations (\ref{eq:traces}) given in the
appendix.  Likewise, the expansion of (\ref{eq:8.5}) can be substituted
into (\ref{eq:8.2}) in order to derive the explicit transformation laws
of all of the component fields.  For a fixed value of $N$ there are
2${}^{N}$ bosons and 2${}^{N}$ fermions in the action of (\ref{eq:8.6}).

For the values $N$ = 1,2 and 4, the action and theory described above is
reducible.  In these cases there is a truncation that may be performed
to obtain a smaller representation.  The reason for the exceptional
nature of these cases can be traced back to the representation theory
of the $\cal {EGR}$ algebras.  For these cases, the condition
\be
\dim(\{ f_{\rm I} \}) ~=~ 1 ~+~ \sum_{p = 1}^{p_{max}} \dim( f_{\rm I_1
\cdots \rm I_{2p}})   ~~~.
\label{eq:8.7}
\ee
is satisfied for some integer $p_{max}$. As a consequence the number of
NSR fermions is equal to the number of propagating and auxiliary bosons. 
Whenever this condition is satisfied, the theory described above is
reducible and the only cases of which we are aware are precisely
$N$ = 1,2, and 4 which we discuss below.

For the exceptional $N $ = 4 case, the action for the spinning particle
is of the form
\bea
{\cal S}_{Spng~Part.}^{N=4ex} ~=~ \int d \t ~ \Big[ ~ \fracm 12 \, (\, 
{\pa}_{\t} {\rm X} \,) (\, {\pa}_{\t}{\rm X} \,) ~+~ i \frac 12\,
\Psi_{{\,}_{\rm I}}  {\pa}_{\t} \Psi_{{\,}_{\rm I}}  ~+~ \fracm 14 \,
F_{{\,}_{{\rm I} {\rm J}}} \, F_{{\,}_{{\rm I} {\rm  J}}} ~\Big] ~~~, 
\label{eq:8.8}
\eea
where the auxiliary field satisfies ${F}_{{\,}_{{\rm I} {\rm J}}} = 
\fracm 12 \x \e_{{\,}_{{\rm I} \, {\rm J} \, {\rm K} \, {\rm L}}}{F}_{
{\,}_{{\rm  K} {\rm L}}} $ for $\x = \pm 1$.  The $N$ = 4  action is
invariant under the supersymmetry variations
\bea
{\d}_{Q} \, {\rm X}{}^{} &=& i \, \a_{{\,}_{\rm I}} \, \Psi_{{\,}_{\rm I}} 
~~~~, ~~~~ {\d}_{Q} \, \Psi_{{\,}_{\rm I}} {}^{} ~=~  - \a_{{\,}_{\rm I}} 
\, (\,{\pa }_{\t}{\rm X}{}^{} \, ) ~+~ \a_{{\,}_{\rm K}} {F}_{{\,}_{ {\rm 
K} {\rm I}}}   ~~~~, \cr
{\d}_{Q} \, {F}_{{\,}_{{\rm I} {\rm J}}} {}^{} &=& - \, i \, \fracm 12 \,
[~ \a_{ {\,}_{\rm I}} {\pa}_{\t} \Psi_{{\,}_{ {\rm J}} }{}^{} ~-~
\a_{{\,}_{\rm J}} {\pa}_{\t} \Psi_{{\,}_{ {\rm I}} }{}^{} ~+~ \x
\a_{{\,}_{\rm K}} \e_{{\,}_{{\rm I} {\rm J}   {\rm K}{\rm L}}} (\, 
{\pa}_{\t} \Psi_{{\,}_{{\rm  L}}} {}^{}\,) ~] ~~~~~. 
\label{eq:8.9}
\eea
The $N$ = 2 exceptional truncation of this is given by
\bea
{\cal S}_{Spng~Part.}^{N=2ex} ~=~ \int d \t ~ \Big[ ~ \fracm 12 \, (\, {\pa
}_{\t} {\rm X} \,) (\, {\pa}_{\t}{\rm X} \,) ~+~ i \frac 12 \, \Psi_{{\,}_{
\rm I}} {\pa}_{\t} \Psi_{{\,}_{\rm I}} ~+~ \fracm 12 \, F \, F ~\Big] ~~~, 
\label{eq:8.10}
\eea
with transformation laws given by
\bea
{\d}_{Q} \, {\rm X}{}^{} &=& i \, \a_{{\,}_{\rm I}} \, \Psi_{{\,}_{\rm 
I}} {}^{} ~~~~, ~~~~
{\d}_{Q} \, \Psi_{{\,}_{\rm I}} {}^{} ~=~  - \a_{{\,}_{\rm I}} \, (\,
{\pa}_{\t}{\rm  X} {}^{} \, ) ~+~ \a_{{\,}_{\rm I}} \, F ~~~~, \cr
{\d}_{Q} \, F &=& - i \, \a_{{\,}_{  {\rm I}}} \, (\, {\pa}_{\t} \Psi_{
{\,}_{\rm I}} \,)  ~~~~.
\label{eq:8.10b}
\eea
Finally there is the $N$ = 1 theory
\bea
{\cal S}_{Spng~Part.}^{N=1ex} ~=~ \int d \t ~ \Big[ ~ \fracm 12 \, (\, {\pa
}_{\t} {\rm X} \,) (\, {\pa}_{\t}{\rm X} \,) ~+~ i \frac 12 \, \Psi {\pa
}_{\t} \Psi ~\Big] ~~~, 
\label{eq:8.11}
\eea
with transformation laws given by
\bea
{\d}_{Q} \, {\rm X}{}^{} &=~ i \, \a \, \Psi ~~~~, ~~~~
{\d}_{Q} \, \Psi ~=~  -\, \a \, (\, {\pa}_{\t}{\rm  X} \, ) ~~~~.
\label{eq:8.12}
\eea

Up to this point in our discussion, all dynamical variables under 
consideration were ordinary functions of $\t$, our time-like variable.  
We now wish to consider a superspace formalism that is compatible 
with our previous discussion. We extend our one dimensional world-line
parametrized by $\t$ to a one dimensional super world-line parametrized 
by $(\t, \,\z^{{\,}_{\rm I}})$ where $\z^{{\,}_{\rm I}}$ is a Grassmann
coordinate\footnote{It has long been our convention to reserve the symbol
$\q$ for the Grassmann coordinate \newline ${~~~~~}$ 
associated with the target manifold.} with ${\rm I} = 1,\,
\dots , \, N$.

A standard rule for identifying the Salam-Strathdee superfields associated
with a component-level description is that the component fields with 
the highest engineering dimension and which transform solely into derivative 
terms under a supersymmetry transformation must occur as the last field in 
the Grassmann coordinate expansion of the superfield. This rule is applied 
in the following discussion.

The analysis of the exceptional irreducible cases of $N $ = 1, 2 and 4 is 
simplest, so we carry this out first.  Beginning our considerations in the 
$N$ = 4 case, we see that there is only a bosonic prepotential ${\bf U}
{}^{{\,}^{\rm {I\, J}}} (\z, \, \t),~ {\bf U}_{{\,}_{{\rm I} {\rm J}}} = 
\fracm 12 \x \e_{{\,}_{{\rm I} {\rm J} {\rm K}  {\rm L}}} {\bf U}_{{\,}_{
{\rm K} {\rm L}}} $ which describes the entire theory.  The manner in 
which the component fields appear is rather obvious.  We can define these
through projection.  For the $N$ = 4 case these are 
\bea
{\rm X} &=& D_{{\,}_{\rm K}} \, D_{{\,}_{\rm L}} \, {\bf U}^{{\,}^{{\rm 
K} {\rm L}}} \, {\Big |}~~~, \cr
\Psi_{{\,}_{\rm I}}\,  &=& i D_{{\,}_{\rm I}} \, D_{{\,}_{\rm K}} \, D_{
{\,}_{\rm L}} \, {\bf U}^{ {\,}^{{\rm K} {\rm L}}} \, {\Big |}~~~, \cr 
F^{{\,}^{{\rm K}{\rm L}}} \, &=& \fracm 1{4!} \, \e^{{\,}_{{\rm I} \,{\rm 
J} \, {\rm M} \, {\rm N} }}  D_{{\,}_{\rm I}} \, D_{{\,}_{\rm J}} \, D_{
{\,}_{\rm M}} \,  D_{{\,}_{\rm  N}} \, {\bf U}^{{\,}^{{\rm K} {\rm L}}} 
\, {\Big |}~~~.
\label{eq:8.13}
\eea

and the superfield action for the multiplet is simply 
\bea
{\cal S}_{Spng~Part.}^{N=4ex} &=~ \int d \t ~ d^4 \z ~ \Big[ \, \fracm 
1{2} \, {\bf U}^{{\,}^{{\rm I} {\rm J}}} \,  D_{{\,}_{\rm I}} \, D_{{\,
}_{\rm J}} \, D_{{\,}_{\rm K}} \, D_{{\,}_{\rm L}} \, {\bf U}^{{\,}^{
{\rm K} {\rm L}}} \, \Big] {~~~~~~}\cr 
&=~ \int d \t \, d^4 \z ~ \Big[ \, \fracm 12 \, {\bf U}^{{\,}^{{\rm K} 
{\rm L}}} \, {\bf F}{}^{{\,}^{{\rm K}  {\rm L}}} \, \Big] ~~~. 
\label{eq:8.14}
\eea

In the remaining exceptional cases, both theories are described by a scalar
pre-potential superfield ${\bf U}(\z , \, \t)$.  While in the $N$ = 2 case 
the components can be defined by 
\bea
{\rm X} &=~  {\bf U} \, {\Big |}~~~, ~~~ \Psi_{\rm I}\, ~=~ i D_{{\,}_{
\rm I}} \, {\bf U} \, {\Big |} ~~~, ~~~ F ~=~ i \, \fracm 1{2} \, \e^{
{\,}_{{\rm I} \,{\rm J}}}  D_{{\,}_{\rm I}} \, D_{{\,}_{\rm J}} \, {\bf
U} \,  {\Big |}~~~.
\label{eq:8.15}
\eea 
and the superfield action for the multiplet is simply 
\bea
{\cal S}_{Spng~Part.}^{N=2ex} &=~ \int d \t ~ d^2 \z ~ \Big[ \, i \, 
\fracm 1{4} \, {\bf U} \, \e^{{\,}_{{\rm I} \,{\rm J} \,}}  D_{{\,}_{
\rm I}} \, D_{{\,}_{\rm J}}  \, {\bf U} \, \Big] ~=~ \int d \t \, d^2 
\z ~ \Big[ \, \fracm 12 \, {\bf U} \, {\bf F} \, \Big] ~~~. 
\label{eq:8.16}
\eea

Finally in the $N$ = 1 case the components can be defined by 
\bea
{\rm X} &=~  {\bf U} \, {\Big |}~~~, ~~~ \Psi\, ~=~ i D \, {\bf 
U}  \, {\Big |}~~~.
\label{eq:8.18}
\eea
and the superfield action for the multiplet is simply 
\bea
{\cal S}_{Spng~Part.}^{N=1ex} &=~ \int d \t ~ d \z ~ \Big[\, i \,  
\fracm 1{2} \, {\bf U} \, D \, {\bf U} \, \Big] ~=~ \int d \t \, d 
\z ~ \Big[ \, \fracm 12 \, {\bf U} \, {\bf \Psi} \, \Big] ~~~. 
\label{eq:8.19}
\eea

For $N$ = 3 , we introduce the prepotential superfield
\bea
 {\bf U}{}^{{\,}^{\rm {I_1 I_2 I_3}}}(\z, \, \t)
~~~,
\label{eq:8.20a}
\eea
along with two other prepotential superfields,
\bea
(\, {\bf {{\bf V}}}(\z, \, \t), \, {\bf {{\bf V}}}{}^{{\,}^{\rm 
{I_1 I_2}}}(\z, \, \t)\,) ~~~, 
\label{eq:8.21a}
\eea
For $N ~ >$ 4, we introduce the $2{}^{N - 1} - 1$ prepotential superfields
\bea
(\, {\bf U}{}^{{\,}^{\rm {I_1 I_2 I_3}}}(\z, \, \t), \, \dots , \, 
{\bf U}{}^{{\,}^{\rm {I_1 I_2 \cdots I_{2p + 1}}}}(\z, \, \t) \, ) 
~~~,
\label{eq:8.20}
\eea
along with another set of $2{}^{N - 1}$ prepotential superfields,
\bea
(\, {\bf {{\bf V}}}(\z, \, \t), \, {\bf {{\bf V}}}{}^{{\,}^{\rm 
{I_1 I_2}}}(\z, \, \t), \, \dots , {\bf {{\bf V}}}{}^{{\,}^{\rm 
{I_1 \cdots I_{2p }}}}(\z, \, \t) \,) ~~~, 
\label{eq:8.21}
\eea
whose Grassmann coordinate expansions contain all of the component fields 
in (\ref{eq:8.2}) {\em {plus}} many more.  

The number of component fields contained in (\ref{eq:8.20}) and
(\ref{eq:8.21}) is $2{}^{N} (2{}^{N} -1)$.  This should be compared
with the 2${}^{N + 1}$ total number of Wess-Zumino gauge component fields
in (\ref{eq:8.2}).   The results in (\ref{eq:8.20}) and (\ref{eq:8.21})
follow from the fact that highest engineering dimension component fields 
of the USPM correspond to spinorial auxiliary fields $\L{}_{\hat k} {}^i$ 
and for the conjugate momentum multiplet these are the bosonic fields 
$P$ and ${\cal G}{}_i {}^j$.

In other words, the superfields in (\ref{eq:8.20}) and (\ref{eq:8.21}) are
the unconstrained pre-potentials for the  off-shell spinning particle with
$N$-extended supersymmetries on its  worldline. The nature of the
superfields changes drastically according to the value of $N$.  If $N$ is
even, then these superfields are bosonic quantities.  If $N$ is odd, then
these superfields are fermionic. The position coordinate is contained in
the ``smallest'' $\bf U$ pre-potential.  In the ordinary member of the
generic sequence, the 0-brane coordinate, canonical conjugate momentum,
NSR fermion and spinorial momentum can be defined by 
\be \eqalign{
{\rm X}(\t) &\propto~ \fracm 1{~(N -3)! 3! \,} \, \e^{{\,}^{{\rm I}_1
\,\cdots
\, {\rm I}_N }}   D^{{\,}^{\rm I_1}} \, \cdots \, D^{{\,}^{{\rm I}_{N -3}}} 
\, {\bf U} {}^{{\, }^{{\rm I}_{N -2} {\rm I}_{N -1} {\rm I}_{N} }}\,  
{\Big |}~~~, \cr
{\rm P}(\t) &\propto~ \fracm 1{N!} \, \e^{{\,}^{{\rm I}_1 \,\cdots \, {\rm
I}_N }}   D^{{\,}^{\rm I_1}} \, \cdots \, D^{{\,}^{{\rm I}_{N}}} 
\, {\bf V} \, {\Big |}~~~,\cr
\Psi{}^{\rm I}(\t) &\propto~ \fracm 1{N!} \, \e^{{\,}^{{\rm I} \,\cdots \,
{\rm I}_N }}   D^{{\,}^{\rm I}} \,D^{{\,}^{\rm I_1}} \, \cdots \,
D^{{\,}^{{\rm I}_{N -3}}} \, {\bf U} {}^{{\, }^{{\rm I}_{N -2} {\rm I}_{N
-1} {\rm I}_{N} }}\,   {\Big |} ~~~, \cr
\p^{\rm I}(\t) &\propto~ \fracm 1{~(N -1) ! \,} \, \e^{{\,}^{{\rm I} \,{\rm 
I}_2 \,\cdots \, {\rm I}_N }}  \, D^{{\,}^{\rm I_2}} \, \cdots \,
D^{{\,}^{{\rm I}_{N }}} \, {\bf V} \,
{\Big |}~~~, 
} \label{eq:8.22}
\ee
and the remaining component fields in (\ref{eq:8.2}) can easily be assigned
as the $\z \, \to$ 0 limit of the $D$'s acting on the superfields in
(\ref{eq:8.20}) and (\ref{eq:8.21}).  Note that since the engineering
dimensions of $X$, $P$, $\psi{}_{\rm I}$ and $\pi{}_{\rm I}$ are fixed,
as $N$ increases the engineering dimensions of $\bf U {}^{{\rm I}_1 
{\rm I}_2 {\rm I}_3 }$, \dots , ${\bf U} {}^{{\rm I}_1 \cdots {\rm I}_{2p +
1}}$,  ${\bf V}$, \dots  ,${\bf V} {}^{{\rm I}_1 \cdots {\rm I}_{2p}}$ become
increasingly negative in order to compensate for the numbers of spinorial
superderivatives.

From the expression in (\ref{eq:8.22}) it can be seen that {\bf X}, {\bf P},
${\bf {\Psi}} {}^{\rm I}$ and ${\bf {\p}}{}^{\rm I}$ are expressed in terms
of the unconstrained pre-potential superfields.  This implies that the
superfields {\bf X}, {\bf P}, ${\bf {\Psi}} {}^{\rm I}$ and ${\bf {\p}}
{}^{\rm I}$ must satisfy some set of differential equations. Such differential
equation on superfields are called ``constraints.''  This observation
leads us to our third conjecture in this work.

${~~~~~}$ {\it {The constraints to which all irreducible superfields 
in all dimensions  \newline ${~~~~~~~~~}$ are
subjected insure that irreducible supermultiplets
are also irreduc- \newline ${~~~~~~~~~}$ ible representations of the}} ${\cal
GR}($d$, N)$ {\it {algebra.}}

In passing, it is worth mentioning that after all the component fields in
(\ref{eq:8.2}) are expressed in terms of the pre-potentials in
(\ref{eq:8.20}) and (\ref{eq:8.21}), the action (\ref{eq:8.1}) can be
written as the integral of a superfield Lagrangian
\be \eqalign{ {~~}
{\cal S}_{Spng~Part.} &=~ \int d \t \, \int d^N \z ~ {\cal L}_{Spng~Part.
} ({\bf U}{}^{{\,}^{\rm {I_1 I_2 I_3}}},\dots, {\bf U}{}^{{\,}^{\rm {I_1
\cdots I_{2p +1}}}}; \, {\bf V} , \, \dots , \, {\bf V}{}^{{\,}^{\rm 
{I_1 \cdots I_{2p }}}}) \,   {\Big |}~~~,
\cr
&{~~~}\int d^N \z ~\equiv~  \fracm 1{N!} \, \e^{{\,}^{{\rm I}_1 \, \cdots 
\, {\rm I}_N }} D^{{\,}^{\rm I_1}} \, \cdots \, D^{{\,}^{{\rm I}_{N}}} 
~~~, } \ee
which describes the spinning particle for arbitrary values of $N$ (except
1,2 and 4).  For $N~ \ge$ 3, all the actions suggested in this chapter
to describe spinning particles are superfield {\em {gauge}} theories.
The existence of these superfield actions is a direct consequence of
the existence of the pre-potential superfields $\bf U {}^{{\rm I}_1 
{\rm I}_2 {\rm I}_3 } $, \dots , ${\bf U}
{}^{{\rm I}_1 \cdots {\rm I}_{2p + 1}}$,  ${\bf V}$, \dots , ${\bf V}
{}^{{\rm I}_1 \cdots {\rm I}_{2p}}$.

We end this section by pointing out that the ${\cal GR}$(d, $N)$ Clifford 
algebra approach has gone well beyond the naive use of a Salam-Strathdee 
superspace.  By its use we have; (a.) identified a set of superfield
prepotentials (\ref{eq:8.20}), (\ref{eq:8.21}) and (b.) identified the 
component fields that remain in the WZ gauge of the pre-potentials 
(\ref{eq:8.2}).  It is now also obvious that had we simply begun with 
the action in (\ref{eq:8.6}) together with the transformation laws after 
substitution of (\ref{eq:8.5}) into (\ref{eq:8.2}), the ``spinorial'' 
indices of ${\cal GR}$(d, $N)$ do not explicitly appear anywhere in the 
formulation!  In other words, the ${\cal GR}$(d, $N)$ origin of the 
multiplets of pre-potentials is totally hidden.

\subsection{On-shell 3D, $N$-extended Vector Multiplets} 

~~~~Simultaneous with our initial exposition on the role of the ${\cal 
GR}$(d, $N)$ algebras in 1D $N$-extended systems, it was also noted that 
this same algebraic structure plays a role in the construction of
on-shell $N$-extended supersymmetrical vector multiplets in three
dimensions.

The following supersymmetry variations close up to terms involving
the Dirac equations of the spinor fields
\bea
\d_Q {\cal B}_i {}^j &=& \e^{{\,}_{\a \, \rm I}} \, ({\rm L}_{{\,}_{\rm
I}})_k  {}^{\hat k} \, \left[ ~ \d_i {}^k \l_{\a \, \hat k} {}^j ~-~ 
{\rm d}^{-1} \d_i {}^j \l_{\a \, \hat k} {}^k   ~ \right] ~~~, \cr
\d_Q \l_{\a \, \hat k} {}^k &=& i \e^{{\,}_{\b \, \rm I}}  \, ({\rm 
R}_{{\,}_{\rm I}} )_{\hat k}{}^j (\g^{\un d} )_{\a \b} ~\left[ \, \pa_{\un 
d} {\cal B}_j {}^k ~+~ \fracm 14 \, {\rm d}^{-1} \, \d_j {}^k \e_{\un d} 
{}^{\un b \un c} \, F_{\un b \un c}  \, \right] ~~~, \cr
\d_Q A_{\un a} &=&  i \, \e^{{\,}_{\a \, \rm I}} \, ({\rm L}_{
{\,}_{\rm I}})_k {}^{\hat k} \, (\g_{\un a} )_{\a \b} \, \l^{\b
}{}_{\hat k} {}^k   ~~~. 
\label{eq:8.01}
\eea
The conventions used here to describe 3D Lorentz spinors and 
vectors are given by
\be  \eqalign{      {~~~~~~~~~}
 \eta_{ \un a \un b}  \,&=~ diag(1 , -1 , -1)  ~~~,~~~ \e_{{\un
a} {\un  b} {\un  c}} \, \e^{{\un d} {\un e} {\un f}} ~=~ 
\d_{[\un a}{}^{\un d} \d_{\un b}{}^{\un e} \d_{\un c]}{}^{\un f} ,
~~~ \e^{0 1 2} =  +1 ~~~,   \cr
(\g^{\un a})_{\a}{}^{ \g} (\g^{\un b})_{\g}{}^{ \b} \, &=~   
\eta^{ \un a \un b} \, \d_{\a}{}^{\b} + i \e^{ {\un a}
{\un  b} {\un  c}} \,  (\g_{\un c})_{\a}{}^{\b} ~~.  
} \label{eq:3D1} \ee
In these expressions, $\e_{{\un a} {\un  b} {\un  c}}$ is the
Levi-Civita tensor.  Some useful Fierz identities are:
\be \eqalign{
(\g^{\un a})_{\a \b} (\g_{\un a})^{\g \d} \, &=~ - \d_{(\a}{}^\g  
\d_{\b)}{}^\d ~=~ - (\g^{\un a})_{(\a}{}^\g (\g_{\un a})_{\b)}{}^\d  
~~, ~~~~~~~   \cr
e^{{\un a} {\un b} {\un  c}} \, (\g_{\un b})_{\a \b} (\g_{\un 
c})_\g {}^\d  \, &=~ - i C_{\a \g} (\g^{\un a})_\b 
{}^\d - i (\g^{\un a})_{\a \g} \d_{\b} {}^\d ~~~.~~~~~
}  \label{eq:3D2}
\ee

where $C_{\a \, \b} $ = $i \, \e_{\a \, \b}$ and $\e_{\a \, 
\b}$ is also a Levi-Civita tensor.

The fields ${\cal B}_i{}^j$ and $\l_{\a \, {\hat k}}{}^l$ may be regarded
as DKP fields in the normal part of the enveloping algebra (as in the 
case of the spinning particle).  Consequently, each of these field operators 
can be expanded as
\bea
{\cal B}_i {}^j (x) &=& \fracm 12 (f^{{\,}^{\rm {{\rm I}_1 {\rm I}_2}}})_i 
{}^j \, \varphi_{{\,}_{\rm {{\rm I}_1 {\rm I}_2}}} (x) ~+~ \fracm 1{4!}
(f^{{\,}^{\rm {{\rm I}_1 {\rm I}_2 {\rm I}_3 {\rm I}_4}}})_i {}^j \,
\varphi_{{\,}_{\rm {{\rm I}_1 {\rm I}_2 {\rm I}_3 {\rm I}_4}}} (x) ~+~ 
\cdots \cr   
&\,&~+~ \fracm 1{~(2{\rm p})!}(f^{{\,}^{ \rm{{\rm I}_1 \cdots {\rm
I}_{2p}}}})_i {}^j \, \varphi_{{\,}_{\rm {{\rm I}_1 \cdots {\rm I}_{2p}
}}}(x) ~~~, \cr  
{\l}_{\a \, {\hat k}} {}^{k} (x) &=& ({\rm R}^{{\,}^{\rm{{\rm I}_1}}}
)_{ \hat k}{}^{k} \, \l_{{\,}_{\a \, {\rm {{\rm I}_1}}}}(x) ~+~ \fracm
1{3!} (f^{{\,}^{\rm {{\rm I}_1 {\rm I}_2  {\rm I}_3}}})_{\hat k} {}^{k} 
\, \l_{{\,}_{\a \, {\rm {{\rm I}_1 {\rm I}_2 {\rm I}_3 }}}} (x) ~+~ 
\cdots \cr    
&\,&~+~ \fracm 1{~(2{\rm q} +1) !}(f^{{\,}^{\rm {{\rm I}_1 \cdots 
{\rm I}_{2q + 1}}}} )_{\hat k} {}^{k} \, \l_{{\,}_{\a \, {\rm {{\rm
I}_1 \cdots {\rm I}_{2q + 1 }}}}}(x)  ~~~.
\label{eq:8.02}
\eea

All of the results above are cast in the form of component formulations.
We now switch to the superfield viewpoint by the introduction of
connection superfields $\G_{\un A}{}^{\hat \a}$.  These superfields
are expanded over the three bosonic coordinates and $N$ (three
dimensional) spinor coordinates and are used to introduce superspace
gauge covariant derivatives
\be  \eqalign{
\nabla_{\un A} \, &\equiv~ (\,  \nabla_{{}_{\a \, {\rm I}}} , ~
\nabla_{\un a}  \,)   ~~~, \cr
 \nabla_{{}_{\a \, {\rm I}}}  &\equiv~  D_{{}_{\a \, {\rm I}}} ~+~  i \, g \,
\G_{{}_{\a \, {\rm I}}} {}{}^{\hat \a} \, t_{\hat \a}   ~~~, ~~~
\nabla_{\un a}    ~\equiv~  \pa_{\un a} ~+~  i \, g \,
\G_{\un a} {}^{\hat \a} \,  t_{\hat \a}  
}  \label{eq:8.03a}
\ee
where $ t_{\hat \a} $ denote a set of generators for a set of Abelian
symmetries and satisfies $( t_{\hat \a} )^*$ = $-$   $( t_{\hat \a} )$. 
The supersymmetry transformation laws above are consistent with a 
superspace covariant derivative $\nabla_{{\,}_{\a {\rm I}}}$ associated
with an Abelian group and which satisfies the restrictions
\bea
[ \, \nabla_{{}_{\a \, {\rm I}}} ~,~ \nabla_{{}_{\b \, {\rm J}}} \, 
\} &=& i  \, 2 \, \d_{{\,}_{{\rm I}\, {\rm J}}} \, (\g^{\un c})_{\a \b}
\nabla_{\un c} ~+~ i  \,  4 \, g  \, C_{\a \, \b} \, (f_{{\,}_{\rm {I \,
J}}})_j{}^i \, {\cal B}_i{}^j {}^{\hat \a} \,  t_{\hat \a}   ~~~, \cr 
[ \, \nabla_{{}_{\a \, {\rm I}}} ~,~ \nabla_{\un b} \, \} &=&  2 \, g \, 
 (\g_{\un b})_{\a}{}^{\b} ({\rm L}_{{\,}_{\rm I}})_{k}{}^{\hat k} 
\, \l_{\b}{}_{\hat k}{}^{k} {}^{\hat \a} \,  t_{\hat \a}  ~~~, \cr  
[ \, \nabla_{\un a} ~,~ \nabla_{\un b} \, \} &=& i \, g \, F_{\un a \, 
\un b} {}^{\hat \a} \,  t_{\hat \a}  ~~~. 
\label{eq:8.03}
\eea

In order to show that these constraints satisfy the usual superspace
Bianchi identities, it is required to note the identities
\be  \eqalign{
(f_{{\,}_{\rm {I \, J}}})_p {}^r  \,  ({\rm L}_{{\,}_{\rm K}})_{r}{}^{
\hat q} \, &=~ - \,  \d_{{\,}_{{\rm J}\, {\rm K}}} \, ({\rm L}_{{\,}_{
\rm I}})_{p}{}^{\hat q} ~+~  \d_{{\,}_{{\rm I}\, {\rm K}}} \, ({\rm L
}_{{\,}_{\rm J}})_{p}{}^{\hat q} ~+~ (f_{{\,}_{\rm {I \, J \, K}}})_p 
{}^{\hat q}   ~~~, \cr
(f_{{\,}_{\rm {I \, J}}})_i{}^k  \, &=~  \frac 12 \, [ ~   ({\rm L}_{
{\,}_{\rm I}})_{i}{}^{\hat q}  \,  ({\rm R}_{{\,}_{\rm J}})_{\hat q}{
}^{k}  ~-~  ({\rm L}_{{\,}_{\rm J}})_{i}{}^{\hat q}  \,  ({\rm R}_{{
\,}_{\rm I}})_{\hat q}{}^{k}  ~] 
}   \label{eq:8.03A}
\ee
where $(f_{{\,}_{\rm {I \, J \, K}}})$ is a 3-form element as discussed in
section (2.2).

An interesting point about the equations (\ref{eq:8.03}) is that they demonstrate 
that {\it {only}} $\varphi_{\rm {I \, J}}$ and $\l_{{\,}_{\a {\rm I}}}$
appear  as components of the field strength superfield.  The remaining
fields in (\ref{eq:8.02}) appear via derivatives at higher orders in the $
\q$-expansions of the superfields $\varphi_{{\,}_{\rm {I \, J}}} (x)$ 
and $\l_{{\,}_{\a {\rm I}}}(x)$. The component fields $\varphi_{{\,}_{\rm {
I_1 \, I_2 \, I_3  \, I_4 }}} (x), \dots  , \, \varphi_{{\,}_{\rm {I_1 \, \cdots 
\, I_{2p}}}}(x)$ and $\l_{{\,}_{\a  {\rm {I_1 \,I_2 \, I_3}}}}(x) , \dots , 
\, \l_{{\,}_{\a {\rm {I_1 \, \cdots \, I_{2q + 1}}}}} (x)$ provide examples 
of a phenomenon that has {\it {not}} been observed in superspace formulations
previously.   We may call these ``exo-field strength components'' 
because they play the same roles at the usual field strength components
$\varphi_{{\,}_{\rm {I \, J}}}(x)$ and $\l_{{\,}_{\a {\rm I}}} (x)$ but
they do not occur within the set of conventional field strength
superfields common to all known Salam-Strathdee superspace constructions.

There is one other interesting implication of the presence of the 
exo-field strength components.  Due to the fact that they only occur
via their spacetime derivatives in (\ref{eq:8.03}) implies that there
is a huge space of both bosonic and fermionic moduli associated with
these theories in general.  This follows from the invariance of the 
results in (\ref{eq:8.03}) with respect to transformations of the forms
\be  \eqalign{ {~~~~~~}
&\d_m \, \varphi_{{\,}_{\rm {I_1 \, I_2 \, I_3  \, I_4 }}}(x) ~=~ (c_0
)_{{\,}_{\rm {I_1 \, I_2 \, I_3  \, I_4 }}} , \cdots  , \,  \d_m \, 
\varphi_{{\,}_{\rm {I_1 \, \cdots \,I_{2p}}}} (x) ~=~ (c_0)_{{\,}_{\rm 
{I_1 \, I_2 \, I_3  \, I_{2p} }}}  \cr
&\d_m \, \l_{{\,}_{\rm {I_1 \, I_2 \, I_3   }}} (x) ~=~ (\a_0)_{{\,}_{
\rm {I_1 \, I_2 \, I_3   }}} , \cdots  , \,  \d_m \, \l_{{\,}_{\rm {I_1 
\, \cdots \, I_{2q + 1}}}} (x) ~=~ (\a_0)_{{\,}_{\rm {I_1 \, I_2 \, I_3  
\, I_{2q + 1} }}}
} \label{eq:8.03Z}
\ee
where the quantities $(c_0)_{{\,}_{\rm {I_1  \, I_2 \, I_3  \, I_4 }}} 
,\,  \dots  , \, (c_0)_{{\,}_{\rm {I_1  \, I_2 \, I_3  \, I_{2p} }}}$ 
are bosonic constants and $(\a_0)_{{\,}_{\rm {I_1  \, I_2 \, I_3  }}} ,$ 
$ \dots  , \, (\a_0)_{{\,}_{\rm {I_1  \, I_2 \, I_3  \, I_{2q + 1} }}}$ 
are fermionic constants.  The dimensions of the bosonic and fermionic space 
of moduli are respectively given by
\be
{\rm {dim}} \Big( \,   {\cal U}_L \,  \Big/ \,  {\bf I}  \, \oplus \, 
f_{{\,}_{\rm {I \, J}}} \, \Big)  ~~~\& ~~~ {\rm {dim}} \Big( \,   
{\cal M}_L \,  \Big/ \,  f_{{\,}_{\rm {I }}} \, \Big)
\label{eq:8.03ZZ} 
\ee
and these formulae are only evaluated over the normal parts of the
spaces.

There are two obvious actions that are invariant under the supersymmetry
variations in (\ref{eq:8.01}).  The first of these is the usual kinetic 
energy of a vector supermultiplet
\bea
{\cal S}_{3{\rm D}\, VM} &=& \int d^3 x ~\Big[ ~ - \fracm 14 F^{\un a \, \un b}(A) 
\, F_{\un a \, \un b}(A) \,+\, i  \, {\rm d}^{-1} \, \l^{\a}{}_{\hat k} {}^{k} (\g^{\un a})_{\a \b} 
\pa_{\un a} \, \l^{\b} {}_{\hat k}{}^{k} {~~~~~~~} \cr
&\,& ~~~~~~~~~~
\,+\,   {\rm d}^{-1}  \, (\, \pa^{\un a} {\cal B}_i{}^j \,) \, (\, 
\pa_{\un a} {\cal B}_j{}^i \,  ) ~\Big] ~~~.
\label{eq:8.04}
\eea
From its form, the space of moduli described in (\ref{eq:8.03Z}) is seen
to leave this action invariant.

As well, it is possible to construct the supersymmetric BF-theory.
In order to do this, it is first necessary to introduce the supersymmetrical 
dual multiplet.  The fields of this supermultiplet can be defined by that 
fact that their transformation laws are such that the BF-action 
\bea
{\cal S}_{3{\rm D}\, BF} ~=~ \int d^3 x ~\Big[ ~  \fracm 12 \, \e^{\un 
a \, \un b  \, \un c} \, B_{\un a} \, F_{\un b \, \un c}(A)  ~+~ {\cal 
X}^{\a}{}_i {}^{\hat k} \, \l_{\a}{}_{\hat k} {}^i  ~+~ {\cal H}_j{}^i 
\, {\cal B}_i {}^j ~\Big]
~~~.
\label{eq:8.05}
\eea
is invariant under the supersymmetry variations. The variations of the 
dual multiplet components that accomplish this are
given by
\bea
\d_Q B_{\un a} &=& i {\rm d}^{-1}\, \e^{\a \, {\rm I}} \, (\g_{\un a})_{\a 
\b} \, {\cal X}^{\b}{}_{k}{}^{\hat k} \, ({\rm R}_{{\,}_{\rm I}})_{\hat k}{
}^{k} ~~~ , \cr
\d_Q {\cal X}_{\a \, i}{}^{\hat k} &=& \e^{\b \, {\rm I}} \, \Big[~ C_{\a \b}
{\cal H}_i {}^l \,+\, i \fracm 14 \d_i {}^l (\g^{\un a} )_{\a \b} \, \e_{\un a}
 {}^{\un b \, \un c} \,  F_{\un b \, \un c}(B) ~\Big] \, ({\rm L}_{{\,}_{\rm 
I}})_{l}{}^{\hat k}  ~~~, \cr
\d_Q {\cal H}_i {}^j &=& - i \, \e^{\a \, {\rm I}} \, (\g^{\un a})_{\a \b} \, 
\pa_{\un a} \, \left[ ~ {\cal X}^{\b}{}_{i} {}^{\hat k} \d_k {}^j ~-~ {\rm d}^{
-1} \d_i  {}^j {\cal X}^{\b}{}_{\hat k} {}^k ~ \right] \, ({\rm R}_{{\,}_{\rm 
I}})_{\hat k}{}^{k} ~~~. 
\label{eq:8.06}
\eea
Finally it is interesting to note that the action in (\ref{eq:8.05}) is
a superconformal action.  Via the AdS/CFT correspondence there
should exist a 4D, $N$-extended AdS supergravity theory that is
closely related to this action.  

In closing this section, it should pointedly be noted that we have not
attempted to construct the non-Abelian extensions of the models 
discussed.  The non-linearities due to the presence of non-trivial
commutators can be expected to place stringent restrictions on
$N$.  This is a topic for possible future study.

\section{The $N$ = 8 Spinning Particle - Supergravity \\Surprise}

~~~~Before ending our recitation on the relation between ${\cal {GR}}$(${\rm 
d}, N$) Pauli algebras and supersymmetric algebras, there is a surprising
observation to be made.  We return to the case of the ${\cal 
{GR}}$(8, 8) enveloping algebras where we found
\bea
{\cal U}_{L} &=& \{~ {\bf I}, ~ {f}_{{\,}_{\rm {I \, J}}}, ~ {f}^{-}_{{\,
}_{\rm {I_1 I_2 I_3 I_4}}} ~\} ~~~,~~~ {\cal U}_{R} ~=~ \{~ {\bf I}, ~ 
{\Hat f}_{{\,}_{\rm {I \, J}}}, ~ {\Hat f}^{+}_{{\,}_{\rm {I_1 I_2 I_3 
I_4 }}} ~\} ~~~, \cr
{\cal M}_L &=& \{~ f_{{{\,}_{\rm {I}}}} \, , ~ f_{{\,}_{\rm {I \,J \, 
K}}} ~\} ~~~,~~~ {\cal M}_R~=~ \{~ {\Hat f}_{{{\,}_{\rm {I}}}} \, , ~
{\Hat f}_{{\,}_{\rm {I \,J \, K}}} ~\} ~~~. 
\label{eq:9.1}
\eea

One of our other on-going lines of investigation \cite{curto} has been a 
model-independent formulation of super Virasoro algebras.   There we have
seen that the co-adjoint of the totality of generators required to form a
closed super Virasoro algebra  naturally leads to the appearance of
fields that bare a striking resemblance to  the spectrum of supergravity
theories. In particular, the spin of the co-adjoint fields is determined
by the relation
\bea
s ~\equiv~ (\, 2 \,-\, \fracm 12 p \,)  ~~~,
\label{eq:9.1b}
\eea
where $p$ is the rank of the generator (all such generators are forms in 
this approach) associated with the co-adjoint field.  If we now simply apply
this observation to the forms that appear in the enveloping algebra
above, we are led to the spins and degeneracies indicated in the
following table.
\begin{center}
\renewcommand\arraystretch{1.2}
\begin{tabular}{|c|c|c| }\hline
 $~~{\rm {Algebraic~ element}}~~$ & ${\rm Spin}$  & ${\rm Degeneracy}$ 
\\ \hline  \hline
$~~{\cal E}_L({\bf I})~~$ &  $~~2~~$ & $~~1~~$  \\ \hline
$~~{\cal E}_R({\bf I})~~$ &   $~~2~~$ & $~~1~~$  \\ \hline
$~~{\cal O}_L(f_{{}_{\rm I}})~~$ &  $~~3/2~~$ & $~~8~~$  \\ \hline
$~~{\cal O}_R({\Hat f}_{{}_{\rm I} })~~$ &   $~~3/2~~$ & $~~8~~$  \\ \hline
$~~{\cal E}_L(f_{{}_{\rm I} J})~~$ &  $~~1~~$ & $~~28~~$  \\ \hline
$~~{\cal E}_R({\Hat f}_{{}_{\rm I} J})~~$ &   $~~1~~$ & $~~28~~$  \\ \hline
$~~{\cal O}(f_{{}_{{\rm I}_1} {}_{{\rm I}_2} {}_{{\rm I}_3}})~~$ & 
$~~1/2~~$ & $~~56~~$  \\ \hline
$~~{\cal O}_R({\Hat f}_{{}_{{\rm I}_1} {}_{{\rm I}_2} {}_{{\rm I}_3} })~~$
&  
$~~1/2~~$ & $~~56~~$  
\\ \hline
$~~{\cal E}_L(f^{-}_{{}_{{\rm I}_1} {}_{{\rm I}_2} {}_{{\rm I}_3} {}_{{\rm
I}_4}})~~$ &  $~~0~~$ & $~~35~~$  
\\ \hline
$~~{\cal E}_R({\Hat f}^{+}_{{}_{{\rm I}_1} {}_{{\rm I}_2} {}_{{\rm I}_3}
{}_{{\rm I}_4}})~~$ &  $~~0~~$ & $~~35~~$  
\\ \hline
\end{tabular}
\end{center}
\vskip.2in
\centerline{{\bf Table {II}}} 
These spins and degeneracies are exactly those of on-shell 4D, $N$ = 8
supergravity.  We thus assert that each state of the on-shell supergravity
theory is in one-to-one correspondence with the elements of ${\cal
{EGR}}$(8,  8).  It is our suspicion that the relationship we have elucidated
here is  no accident but instead is hinting at a {\it {new}} deep
relationship  between the ${\cal GR}$(d, $N)$ Pauli-Clifford algebras and
${\cal {GR}}$  super Virasoro algebras on one side and supergravity and 
superstring/M-theory on the other.  Stated another way, upon choosing
$N$ = 8, each of the ``fibers'' of the directed links in PpG diagram can
apparently be associated with one of the states in $N$ = 8 supergravity.

\section{Conclusion}

~~~~With this paper, we hope to have provided the reader with convincing
arguments that show that there exists a deep and largely overlooked 
connection between supersymmetrical theories and the theory of a special 
class of real Clifford algebras.  Evidence for these connections first 
began to emerge from investigations of spinning particles \cite{jim1,jim2}.  
It appears within the context of the 1D spinning particle theories that 
the component fields of superfields have two {\em {simultaneous}} 
interpretations.  First, as is widely known, the component fields are 
the coefficients of superfields when expanded over a set of Grassmann 
coordinates.  Our work shows that these components may also be
interpreted as the coefficients of the expansions of a set of linear
operators acting on the vector space ${\cal V}_L \oplus {\cal V}_R$. In
the former approach the field operators are characterized by a Grassmann
algebra while in the latter they are defined through a Clifford algebra. 

One other intriguing possible application of our work is to strengthen
the relation of KO-theory to supersymmetrical representation theory.
Some years ago \cite{AGaume}, it was shown that topological indices
can be related to supersymmetrical quantum mechanical models closely
related to the action in (\ref{eq:SUSYact}).  In turn, the present
work suggests how the representation theory of the supersymmetrical model
is itself related to the ${\cal GR}$(d, $N)$ Pauli-Clifford algebras.
It is also well known that KO-theory is related to real Clifford algebras.
So we see a nexus involving supersymmetry, ${\cal GR}$ algebras and
KO-theory.  The results in section (3.5) show that the usual component
fields of spacetime supersymmetrical theory seem to possess root
superfield representations in 1D, $N$ = 4 theories and that these
component fields correspond in a definite way to geometrical structures
associated with real Clifford algebras.  

It has been observed by Landweber \cite{GLand}  that our construction may 
be linked to KO-theory by noting that we  construct KO${}^{-k}(X)$ by looking 
at $Z_2$-graded $Cl(k)$ bundles  over $X$, and identifying ones which admit 
a $Cl(k+1)$ action.   Our present work takes $X$ to simply be $R^1$ parametrized 
here by $\t$.  Upon the imposition of boundary conditions, in other words
deal with  K-theory with compact supports or take the one point compactification, 
KO($R$) shifts degree by one.  So it may be possible to view space-time locally  
as a bundle over the timelike direction.

We wish to comment based upon an accumulating amount of evidence that
it is conceivable that all aspects of supersymmetrical theories in all
dimensions are encoded in some manner in 1D, $N$-extended theories.
This thought has been in the background of this line of research every
since we began it some time ago \cite{jim1,jim2}.  Two of the main
results in this present paper add more such evidence.  One of these
is the discussion that was given in sections (3.5-3.6).  There it 
was seen in a rather precise way, how the fields of the 4D, $N$ 
= 1 chiral multiplet seem possess an alternate interpretation of being
associated with certain irreducible tensor operators that act on the spaces
${\cal V}_L$ and ${\cal V}_R$ (as well as Cartesian products of these
spaces).  In equation (86) we have seen that the 4D, $N$ = 1 chiral
multiplet seems to possess a ``root superfield'' representation.
We have also outlined how other representations of 4D, $N$ = 1 
supersymmetry likely also possess such representations among the higher
rank representations analogous to that used for the chiral multiplet.
A second striking piece of evidence is the unexpected apparent link
between some representations of the ${\cal {EGR}}(8,8)$ algebras and
the states that appear in 4D, $N$ = 8 supergravity.

All of this suggests that there exist some new type of ``holography'' at
work here.

Finally, we have seen that a Clifford-algebraic based construction of
supersymmetrical theories exist independent of the more traditional
Salam-Strathdee superspace based constructions.  However, as explicitly
seen in sections (4.1-4.2), the Clifford-algebraic based construction
is perfectly compatible with off-shell and on-shell Salam-Strathdee 
superspace based constructions.  We ultimately expect this to be universally
true.  Exploring these relations further will be a primary purpose of
future studies along these lines.

${~~~}$ \newline
${~~~~~~~~~}$``{\it {We share a philosophy about Linear Algebra: We
think basis-free, 
we  \newline
${~~~~~~~~~}$ write basis-free, but when the chips are down we close the
office door \newline
${~~~~~~~~~}$ and compute 
with matrices like fury.}}'' -- I. Kaplansky.


$$ \eqalign{
~~& ~~ \cr  ~~& ~~ \cr  
~~& ~~ \cr  
}
$$

\noindent{\bf Acknowledgments}

\noindent The authors would like to acknowledge, John H. Schwarz and H. Tuck 
for the hospitality extended during their visit to the California Institute 
of Technology, where some of this research was undertaken.  Additionally, 
S.J.G. wishes to recognize the support rendered by the Caltech administration 
during this visit.   We also thank Prof. V.\ G.\ J.\ Rodgers for assistance with this
paper.  Finally, we wish to thank Prof. Gregory Landweber for his critical 
reading of an earlier version of this work and making numbers of
valuable comments.

\newpage
\section{Appendix: Trace Relations and Inner Product Structure 
on ${\cal EGR}$(d, $N)$}

~~~In this section we present a symmetric bilinear form on the 
enveloping algebra ${\cal EGR}(8)$. This will induce, via the reduction 
procedure described above, an inner  product on the algebras 
${\cal EGR}(N)$ for $5\le N \le 8$.  
In order to define the inner product on the enveloping algebra, it is 
necessary to investigate the traces which may be defined on the generators. 
Due to the mixed markings on the elements of ${\cal EGR}(8)  \cong \{ 
{\cal M}_L \} \oplus \{ {\cal M}_R \} \oplus \{ {\cal U}_L \} \oplus \{ {\cal U}_R 
\}$, only the traces on the following subspaces may be defined properly: 
$ \{ {\cal U}_R \} \oplus \{ {\cal U}_R \} $, $\{ {\cal U}_L \} \oplus \{
{\cal U}_L \}$, and $\{ {\cal M}_L \}\oplus\{ {\cal M}_R \}$.   The relevant 
traces are given by
\bea
\label{eq:traces}
{\rm tr}\Big[ {\Tilde f}_{{}_{{\rm I}_1} \cdots {}_{{\rm I}_p}}{\Tilde
f}^{~{}^{{\rm J}_1} \cdots {}^{{\rm J}_q} }  \Big] &=& (-)^q \cdot d 
\cdot {\rm sgn} \left( {}^{12\cdots q}_{q\cdots 21} \right) \cdot 
\d_{p,q} \cdot \d_{{}_{{\rm I}_1}}{}^{{}^{[ ~{\rm J}_1}}\cdots \d_{
{}_{{\rm I}_q} }{}^{{}^{{\rm J}_d ~ ]}}
\cr 
&~&+~\ell \cdot d \cdot \d_{p,N-q} \cdot \e_{{}_{{\rm I}_1} \cdots 
{}_{{\rm I}_p} }{}^{~{}^{{\rm J}_1} \cdots {}^{{\rm J}_q} } ~~,
\eea
with all other traces vanishing.  Here ${\Tilde f}_{[p]}:={\Tilde f}_{
{}_{{\rm I}_1} \cdots {}_{{\rm I}_p} }=f_{[p]}\in \{{\cal U}_R \}$ or ${\Hat
f}_{[p]}\in \{{\cal U}_L \}$ if $p$ is even and ${\Tilde f}_{[p]}\in \{ 
{\cal M}_L \}$ or $\{ {\cal M}_R \}$  if $p$ is odd.  Also, d here is the
dimension of the representation, $\ell$ gives the duality of the ${\fracm
N2}$-form and ${\rm sgn} \left( {}^{12\cdots q}_{q\cdots 21} \right) =
(-)^{{\sum^{ q-1}_{n=1}} n}=(-)^{{\fracm q2}(q-1)}$ denotes the sign of the
permutation reversing the order of the $q$ indices.

Using these traces we may proceed to define the following inner product.
Let 
\bea
\f_{i}{}^j&=&\sum_{p=0}^{N/2} \f_{{}_{{\rm I}_1} \cdots {}_{{\rm 
I}_{2p}}}(f_{ {}_{{\rm I}_1} \cdots  {}_{{\rm I}_{2p}} })_i{}^j 
~\in~ \{{\cal U}_L \} ~~~,\cr
{\Hat \F}_{\hat k}{}^{\hat l}&=&\sum_{p=0}^{N/2} {\Hat \f}_{
{}_{{\rm I}_{1}} \cdots {}_{{\rm I}_{2p}}}({\Hat f}_{{}_{{\rm 
I}_{1}} \cdots {}_{{\rm I}_{2p}}})_{\hat k}{}^{\hat l} ~\in~ 
\{ {\cal U}_R \} ~~~, \cr 
\J_k{}^{\hat l}&=&\sum_{p=0}^{N/2-1} \j_{{}_{{\rm I}_{1}} \cdots 
{}_{{\rm I}_{(2p + 1) }} }( f_{{}_{{\rm I}_{1}} 
\cdots {}_{{\rm I}_{(2p + 1) }}})_k{}^{\hat l} ~\in~ \{{\cal M}_L \}
~~~, \cr 
{\Hat \J}_{\hat l}{}^k&=&\sum_{p=0}^{N/2-1} {\Hat \j}_{{}_{{\rm I}_{1}}
\cdots {}_{{\rm I}_{(2p + 1)}}}({\Hat f}_{{}_{{\rm I}_{1}}
\cdots {}_{{\rm I}_{(2p+ 1)}} })_{\hat
l}{}^k ~\in~ 
\{ {\cal M}_R \} ~~~.
\eea
We define $ \left<\cdot , \cdot \right>$ by
$$  \eqalign{
(\F^{(1)},\F^{(2)})&\mapsto + {\fracm 1d} \sum_{p,q=0}^{N/2} {\fracm
1{(p+q)!}} \cdot \f^{(1)}_{{}_{{\rm I}_{1}} \cdots {}_{{\rm I}_{2p 
}} }\f^{(2)}_{ {}_{{\rm J}_{1}}  \cdots {}_{{\rm J}_{2q}} }
\cdot {\rm tr}\Big[f_{ {}_{{\rm I}_{1}} \cdots {}_{{\rm I}_{2p}}
}f_{{}_{{\rm J}_{2q}} \cdots {}_{{\rm J}_{1}} }\Big] ~~, \cr
({\Hat \F}^{(1)},{\Hat \F}^{(2)})&\mapsto +{\fracm 1d}\sum_{p,q=0}^{N/2}
{\fracm 1{(p+q)!}} \cdot {\Hat \f}^{(1)}_{{}_{{\rm I}_{1}} \cdots
{}_{{\rm I}_{2p}} }{\Hat \f}^{(2)}_{ {}_{{\rm J}_{1}} \cdots 
{}_{{\rm I}_{2q}} }\cdot {\rm tr}\Big[{\Hat f}_{ {}_{{\rm I}_{1}}
\cdots {}_{{\rm I}_{2p}} }{\Hat f}_{ {}_{{\rm J}_{2q}} \cdots 
{}_{{\rm J}_{1}} }\Big] ~~,  
} $$
\be  \eqalign{
(\J^{(1)},\J^{(2)})&\mapsto -{\fracm 1d}\sum_{p,q=0}^{N/2-1} 
{\fracm 1{(p+q+1)!}} \cdot \j^{(1)}_{ {}_{{\rm I}_{1}} \cdots 
{}_{{\rm I}_{(2p + 1 )}} }\j^{(2)}_{ {}_{{\rm J}_{1}} \cdots 
{}_{{\rm J}_{(2q + 1) }} } \cdot {\rm tr}\Big[f_{ {}_{{\rm I}_{1}}
\cdots {}_{{\rm I}_{(2p + 1)}} }f_{ {}_{{\rm J}_{(2q + 1)}}
 \cdots {}_{{\rm J}_{1}} } \Big] ~~, \cr 
({\Hat \J}^{(1)},{\Hat \J}^{(2)})&\mapsto -{\fracm 1d}\sum_{p,q=0
}^{N/2-1} {\fracm 1{(p+q+1)!}}\cdot {\Hat \j}^{(1)}_{{}_{{\rm I}_{1}}
\cdots {}_{{\rm I}_{(2p + 1) }} }{\Hat \j}^{(2)}_{{}_{{\rm J}_{1}}
\cdots {}_{{\rm J}_{(2q + 1)}} } \cdot {\rm tr} \Big[{\Hat f}_{
{}_{{\rm I}_{1}}  \cdots {}_{{\rm I}_{(2p + 1)}} }{\Hat f}_{
{}_{{\rm J}_{(2q + 1) }} \cdots  {}_{{\rm J}_{1}} }\Big] ~~.\cr
}  \ee
Note that the adjoint indices on the second generator under the trace
stand in reverse order w.r.t. the indices on its accompanying component
field.

The following property of this inner product is crucial to the
construction of $N$-extended supersymmetry Lagrangians: The non-vanishing
piece of these traces which are not proportional to the $\e$ tensor are 
{\it positive definite}. e.\ g.
\bea
\left<\F^{(1)},\F^{(2)} \right>&=&+ {\fracm 1d} \sum_{p,q=0}^{N/2} 
{\fracm 1{(p+q)!}} \cdot \f^{(1)}_{ {}_{{\rm I}_{1}} \cdots
{}_{{\rm I}_{2p}} }\f^{(2)}_{{}_{{\rm J}_{1}} \cdots {}_{{\rm J}_{2q}} 
} \cdot {\rm tr}\Big[f_{{}_{{\rm I}_{1}} \cdots {}_{{\rm I}_{2p}}}
f_{{}_{{\rm J}_{2q}} \cdots {}_{{\rm J}_{1}} }\Big]  \cr
&=&+ {\fracm 1d} \sum_{p,q=0}^{N/2} {\fracm 1{(p+q)!}} \cdot \f^{(1)
}_{{}_{{\rm I}_{1}} \cdots {}_{{\rm I}_{2p}}}\f^{(2)}_{
{}_{{\rm J}_{1}} \cdots {}_{{\rm J}_{2q}}} \cdot (-)^{2q}
\cdot d
\cdot \Big[{\rm sgn} \left( {}^{12\cdots q}_{q\cdots 21} \right) 
\Big]^2\cr
&~&~~~~~~~~~~~~~~~~~~~~~~~~~\cdot \d_{p-q} \cdot \d_{{}_{{\rm I}_{1}}
}{}^{{}^{[ {\rm J}_{1}} } \cdots \d_{ {}_{{\rm I}_{p}} }{}^{ {}^{{\rm 
J}_{q} ]} } +\e{\rm -terms} \cr 
&=&+ \sum_{p=0}^{N/2}
\f^{(1)}_{{}_{{\rm I}_{1}} \cdots {}_{{\rm I}_{2p}} }\f^{(2)}_{
{}_{{\rm I}_{1}}  \cdots {}_{{\rm I}_{2p}}}~~+~\e{\rm  -terms}.
\eea
Without this result, the would-be kinetic terms in a generic lagrangian
would introduce classical ghosts into the particle spectrum.


\end{document}